\documentclass[fleqn,usenatbib]{mnras}

\usepackage{newtxtext,newtxmath}

\usepackage[T1]{fontenc}
\usepackage{ae,aecompl}
\usepackage[normalem]{ulem}
\usepackage[utf8x]{inputenc}

\usepackage{graphicx}	
\usepackage{amsmath}	
\usepackage{amssymb}	
\usepackage{array}
\usepackage{multirow}
\usepackage{xcolor}
\usepackage{soul}

\newcolumntype{P}[1]{>{\centering\arraybackslash}p{#1}}
\definecolor{cadmiumred}{rgb}{0.89, 0.0, 0.13}

\definecolor{scc}{rgb}{0.0, 0.26, 0.15}
\definecolor{mbc}{rgb}{0.8, 0.58, 0.46}
\definecolor{mpc}{rgb}{0.6, 0.4, 0.8}
\definecolor{agc}{rgb}{0, 0, 0}
\definecolor{rda}{rgb}{1,0,1}
\definecolor{glo}{rgb}{0.31, 0.78, 0.47}
\definecolor{eims}{rgb}{1.0, 0.5, 0.0}
\definecolor{nnc}{rgb}{0, 0.82, 0}
\definecolor{mcc}{rgb}{0, 0.37, 0.38}
\definecolor{gri}{rgb}{0.19, 0.55, 0.91}

\newcommand*\mathinhead[2]
{\texorpdfstring{$\boldsymbol{#1}$}{#2}}
\title[AstroInformatics Globular Clusters in Fornax region]{Astroinformatics based search for globular clusters in the Fornax Deep Survey}

\author[G. Angora et al.]{
G. Angora$^{1,2}$\thanks{E-mail: gius.angora@gmail.com},
M. Brescia$^{2}$, 
S. Cavuoti$^{2,3,4}$, 
M. Paolillo$^{2,3,4}$,
G. Longo$^{3,4,5}$,  \and
M. Cantiello$^6$, 
M. Capaccioli$^{2,3}$, 
R. D'Abrusco$^{7}$, 
G. D'Ago$^8$, 
M. Hilker$^{9}$, \and
E. Iodice$^2$, 
S. Mieske$^{10}$,  
N. Napolitano$^{11}$, 
R. Peletier$^{12}$, 
V. Pota$^2$, 
T. Puzia$^{8}$, \and
G. Riccio$^{2}$ and
M. Spavone$^2$
\\
\\ \\
$^{1}$ Department of Physics and Earth Science of the University of Ferrara, Via Saragat 1, I-44122 Ferrara, Italy\\
$^{2}$ INAF - Astronomical Observatory of Capodimonte, via Moiariello 16, I-80131 Napoli, Italy\\
$^{3}$ University of Naples Federico II - Dept. of Physics ``E. Pancini'', via Cinthia 21, I-80126 Napoli, Italy\\
$^{4}$ INFN - Napoli Unit, via Cinthia 21, I-80126 Napoli, Italy\\
$^{5}$ Merle Kingsley distinguished visitor, California Institute of Technology, Pasadena, 90125-CA, USA\\
$^{6}$ INAF - Astronomical Observatory of Abruzzo, Via Mentore Maggini snc, Loc. Collurania, 64100 Teramo, Italy\\
$^{7}$ Center for Astrophysics | Harvard \& Smithsonian, 60 Garden St, 02138 Cambridge - MA (US)\\
$^{8}$ Institute of Astrophysics, Pontificia Universidad Cat\'{o}lica de Chile, Av. Vicu\~{n}a Mackenna 4860, 7820436 Macul Santiago, Chile\\
$^{9}$ European Southern Observatory, Karl-Schwarzschild-Str. 2, D-85748 Garching, Germany\\
$^{10}$ European Southern Observatory, Alonso de Cordova 3107, Vitacura, Santiago, Chile\\
$^{11}$ School of Physics and Astronomy, Sun Yat-sen University, Guangzhou 519082, Zhuhai Campus, P.R. China\\
$^{12}$ Kapteyn Astronomical Institute, University of Groningen, PO Box 800, NL-9700 AV Groningen, the Netherlands
}

\date{Accepted 2019 September 30. Received 2019 September 2019; in original form 2019 May 7}

\pubyear{2019}

\begin{document}
\label{firstpage}
\pagerange{\pageref{firstpage}--\pageref{lastpage}}
\maketitle

\begin{abstract}

In the last years, Astroinformatics has become a well defined paradigm for many fields of Astronomy. In this work we demonstrate the potential of a multidisciplinary approach to identify globular clusters (GCs) in the Fornax cluster of galaxies taking advantage of multi-band photometry produced by the VLT Survey Telescope using automatic self-adaptive methodologies.
The data analyzed in this work consist of deep, multi-band, partially overlapping images centered on the core of the Fornax cluster. 
In this work we use a Neural-Gas model, a pure clustering machine learning methodology, to approach the GC detection, while a novel feature selection method ($\Phi$LAB) is exploited to perform the parameter space analysis and optimization. We demonstrate that the use of an Astroinformatics based methodology is able to provide GC samples that are comparable, in terms of purity and completeness with those obtained using single band HST data \citep{brescia:2012} and two approaches based respectively on a morpho-photometric \citep{cantiello2018} and a PCA analysis \citep{dabrusco2015} using the same data discussed in this work. 
\end{abstract}



\section{Introduction}
In modern observational Astronomy, the amount of data collected by an instrument in a single day is often more than enough to keep occupied an entire community of scientists for long time; LSST, for instance, will produce 20 trillion bytes of raw data per night\footnote{\url{https://www.lsst.org/about/dm}}.  These huge datasets are further enlarged by the possibility to combine data obtained at different wavelengths and epochs by different instruments. Astronomy, in fact, is going to enter the big data era not only for the sheer size of its data, but also for the high dimensionality and complexity of the parameter spaces to be explored. These spaces are composed by a variable mixture of photometry, spectroscopy, structural and morphological features, depending on the specific context of the problem under investigation.
A complexity which allows, on the one hand, to answer long standing questions with higher accuracy and, on the other hand, to address completely new and more difficult problems. In such scenario a new paradigm is required, mostly based on a multi-disciplinary approach, through a virtuous integration of Astrophysics, Data Science and Informatics.
A symbiosis which is at the very heart of the relatively new discipline of Astroinformatics, or Knowledge Discovery in astrophysical data \citep{Borne2009,Brescia2013,Feigelson2014,Brescia2018}. 
Astroinformatics, however, is just a label which summarizes the emerging awareness that complex problems  can be tackled only by heterogeneous groups of experts, and that multi-disciplinary approach is not a presumptuous ambition, but rather an unavoidable and precious quality. 
In the last decade, in many different fields, it has been clearly demonstrated that the emulation of the mechanisms underlying natural intelligence, if translated into efficient algorithms and supplied to super computers, is fully and rapidly able to analyze, correlate and extract huge amounts of heterogeneous information \citep{baron2019, Brescia2018}.

When dealing with high dimensionality parameter spaces, it appears evident 
the crucial importance of an optimal choice of the parameter space (i.e. feature selection, hereafter FS) adopted to represent the data to be explored in the context of a specific problem. 

The selection of an optimal set of features strictly depends on the concept of \textit{feature importance}, based on the quantification of its \textit{relevance}. 

Formally, the importance of a feature is the relevance of its informative contribution to the solution of a learning problem. 
There is plenty of FS solutions proposed in literature \citep{Guyon2003}, such as Principal Component Analysis (PCA, \citealt{Jolliffe2002}), \textit{filter} techniques \citep{Gheyas2010}, \textit{wrapper} \citep{Kohavi1997} and \textit{embedded} methods \citep{Lal2006}, among which a typical example is the Random Forest model \citep{Breiman2001}. These methods are basically oriented to find the smallest (best) parameter space able to solve a given problem \citep{Jain:1997,Guyon:2006,hastie2009}.

Such multidisciplinary paradigms have been concretely followed in the present work by exploiting automatic self-adaptive methodologies (e.g. feature selection and the two machine learning paradigms, respectively supervised and unsupervised learning, \citealt{Russell2010}) with the main goal of identifying globular clusters (hereafter GCs) in the Fornax cluster of galaxies.

GCs represent an important category of widely studied astronomical sources. Since GCs harbour a wide variety of stellar types of the same age, each single GC acts as a stellar laboratory, suitable to observe and analyze the formation, behavior and evolution of stellar systems concentrated within just $\sim 10$ parsec. As a population, on a galactic scale, they trace the dynamics, the kinematics and the chemistry of their host galaxy, behaving like a sort of a footprint left by the galactic evolution \citep{ashman:2008, brodie2006}.

It is now well established that GC can be split in populations \citep[e.g.][]{geisler1990, ashman:1995, brodie2006, pota:2013}: \textit{(i)} a red, metal-rich, spatially concentrated sub-population, \textit{(ii)} a blue, metal-poor, spatially extended sub-population. 

The data analyzed in this work consist of deep, multi-band images, partially overlapped, centered on the core of the Fornax cluster. The extracted catalogue is composed by several thousands of sources, each one characterized by a large set of features (i.e. parameters), such as luminosity, colours and morphological information, for a total of more than $60$ features. 
Given the high number of dimensions involved, the difficulty to disentangle different types of objects (e.g. foreground stars, background galaxies and GCs), together with the fact that spectroscopic confirmation was available only for a quite limited number of sources, it was decided to tackle the task of recognizing and classifying GCs (against a variety of background and foreground sources) by investigating both the parameter space optimization and the classification capabilities of specific Machine Learning (ML) methods. 

The work has therefore focused on the Growing Neural Gas model (GNG, \citealt{martinez:1991,martinez:1993,fritzke:1994}), a pure clustering category of ML methods, together with a novel feature selection method, named $\Phi$LAB  \citep{brescia:2018b}. 
In order to compare the performance of the Neural-Gas based model, a Multi Layer Perceptron with Quasi-Newton approach (MLPQNA, \citealt{Byrd1994, bortoletti, Cavuoti:2012, brescia:2012}) and a K-Means \citep{Bishop:2006} have been used as a test benchmark. While, in order to evaluate the feature selection performances, Growing Neural Gas and Random Forest (RF, \citealt{Breiman2001}) methods have been compared on several datasets.
To evaluate the accuracy and in particular the efficiency in identifying secure GC candidates, a direct comparison of these methods has been performed with other techniques as well as with very promising results obtained with other types of machine learning methods applied on single-band HST data of NGC1399, the giant elliptical galaxy at the center of the Fornax cluster.
An important corollary aspect of this work was, in fact, to evaluate the level of accuracy in GC classification within two different contexts: multi-band ground-based data and the single-band high spatial resolution data obtained from space.

The paper is structured as follows: in Sec.~\ref{sec:data} we describe the data used in this work. In Sec.~\ref{sec:models} we present the methods employed for the experiments. In Sec.~\ref{sec:exp} we describe and discuss the experiment results. Sec.~\ref{sec:comp} is dedicated to a comparison with similar ML experiments performed on HST data and with other approaches. In Sec.~\ref{ss:densitymap} we estimate the density maps of the GCs spatial distribution as further validation method. Finally, in Sec.~\ref{sec:conc} we draw our conclusions.     

\section{Data}\label{sec:data}

The data used in this work cover the central region of the Fornax cluster and were obtained with the OmegaCam \citep{Kuijken2011} camera, installed on the VLT Survey Telescope (VST \citealt{schipani:2012}) as part of the Fornax Deep Survey (FDS, \citealt{iodice2016}, Peletier et al. in prep.)\footnote{FDS is a ESO joint program, based on two Guaranteed Time Observation surveys, FOCUS (P.I. R. Peletier) and VEGAS (P.I. E. Iodice; Capaccioli et al. 2015), having as main goal the study of the whole Fornax cluster out to the its viral radius}.
The images were obtained through $76$ exposures of $150$s in the \emph{u}-band, $54$s in the \emph{g} and \emph{r} bands, and $35$s in the \emph{i}-band, reaching a $S/N$ $\sim 10$ at, respectively, $23.8$, $24.8$, $24.3$, $23.3$ magnitudes in the \emph{u}, \emph{g}, \emph{r}, \emph{i} \citep{dabrusco:2016}. 
The average seeing was $1.17\arcsec\pm0.08\arcsec$ in the \emph{g}-band and $0.87\arcsec\pm0.07\arcsec$ in the \emph{r} band (\textit{u} and \textit{i} bands show similar variations over the observed field.)

The catalogue\footnote{The catalogues for the full FDS survey will be presented in a forthcoming paper \citep{cant2019}. Here we adopted the catalogues used in \cite{dabrusco:2016} and \cite{cantiello2018}}, extracted using SExtractor \citep{bertin:1996}, consists of $94,067$ sources whose right ascension (RA) and declination (DEC) are inside the celestial square of limits $\sim[54.02, 55.38]\times[-34.91, -36.03]$ (measured in degrees). The catalogue does not contain the same number of sources in each band, due to the different depth of observations in different filters: there are $15,095$ sources in the \emph{u}-band, $73,497$ sources in \emph{g}-band, $72,385$ in \emph{r}-band and $49,207$ in \emph{i}-band. For each band and for each source SExtractor was used to derive the following information \citep{bertin:1996}:

\begin{itemize}
\item the automatic aperture magnitudes with error (\textit{MAG\_AUTO}), i.e. an estimation of the total magnitude;
\item the fixed aperture magnitudes (\textit{MAG\_APER}): an estimation of the flux above the background within different circular apertures ($4$, $6$, $8$, $16$ and $32$ pixels, respectively), with the related errors;
\item the peak surface brightness above background (\textit{MU\_MAX});
\item the average FWHM of the image assuming a gaussian core (\textit{FWHM\_IMAGE}). It is the average, due to the various overlaps, by considering the small variations among the fields; 
\item the semi-major and semi-minor axis lengths (\textit{A\_WORLD}, \textit{B\_WORLD}) with the errors; 
\item the position angle between the major axis and the x-axis of the image (\textit{THETA\_WORLD}); 
\item the ratio between the semi-major and semi-minor axes lengths (\textit{ELONGATION});
\item the fraction-of-light radii. It measures the radius of the circle centered on the barycenter that encloses about half of the total flux (\textit{FLUX\_RADIUS}); 
\item Kron apertures (\textit{KRON\_RADIUS}), within $2.5\times$\textit{FLUX\_RADIUS}; 
\item The Petrosian apertures (\textit{PETRO\_RADIUS}), i.e. the apertures defined by the petrosian radius, i.e. the radius limit of the ratio between the local surface brightness and the mean interior surface brightness of the source.   
\end{itemize}

\begin{figure}
\includegraphics[width=\columnwidth]{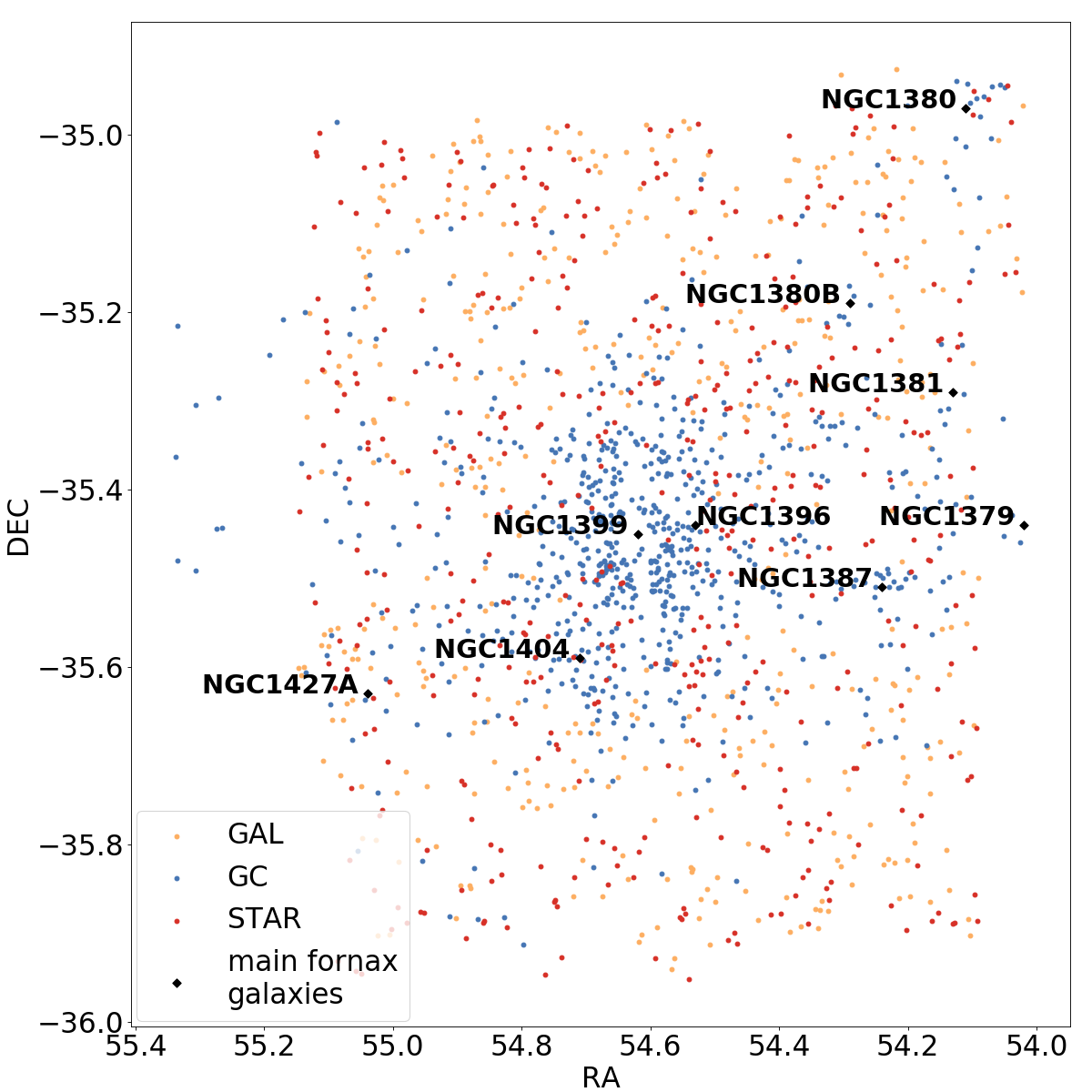}
\caption[Fornax cluster central region]{Distribution of spectroscopically sources: GCs (blue), foreground stars (red), background galaxies (yellow) and bright Fornax cluster galaxies (black diamonds).}
\label{fig:fornax_data}
\end{figure}

By adding colours (\emph{u-g}, \emph{g-r} and \emph{r-i}) and by excluding the two larger 
apertures to minimize contamination from nearby sources and to limit the magnitude errors induced by the background contamination, the final parameter space consists of $64$ features: $16$ magnitudes, $36$ photometric parameters and $12$ colours.

To build the Knowledge Base (KB) needed for the training of the ML network, we used a set of spectroscopically confirmed sources obtained by combining the catalogues from \cite{pota:2018}, consisting of newly confirmed GCs and previous datasets from \cite{Wittmann:2016}, which is mostly based on \cite{schub:2010}. In addition, the foreground stars were provided by \cite{pota:2018} and the background galaxies by \cite{dago2019}. The sky distribution of the various objects is illustrated in Fig.~\ref{fig:fornax_data} for both the spectroscopically confirmed objects and the unclassified sources. 

Since these catalogues were derived with different instruments and methodologies, we applied our in-house cross-matching method \citep{Riccio:2017} between GCs and galaxies and the FDS catalogue, imposing a matching tolerance of  $0.25\arcsec$. Stars were not cross-matched because they are already available in the FDS catalogue.

After the cross-matching procedure, $1,627$ sources were labeled: $706$ GCs, $464$ foreground stars and $457$ background galaxies.  
However, not all these labeled objects turned out to be suitable to construct the KB, due to the presence of missing data (i.e. missing values in some feature columns). In particular the missing data for the $1,627$ cross-matched sources were: (u) $509$ ($31.3$\%), (g) $6$ ($0.4$\%), (r) $8$ ($0.5$\%), (i) $5$ ($0.3$\%). While the missing data for the whole catalogue ($94,067$ sources) were: (u) $78,971$ ($84.0$\%), (g) $20,531$ ($21.8$\%), (r) $21,637$ ($23.0$\%), (i) $44,822$ ($47.6$\%). 

Most of these missing data are in the \textit{u}-band (i.e. the less deep). However, in this case, missing information is mostly due to the sensibility limit of the instrument, rather than to the presence of \textit{holes} in the data distributions, causing an intrinsic difficulty to test any imputation method. In fact, although there are numerous imputing techniques \citep{yoon:2018, zhang:2018, camino:2019, poulos:2016}, able to predict missing values within the sample features, the prediction of feature values outside the training distribution is a more tricky and complex problem, beyond the goals of this work.
We excluded all of them from the final sample due to the known negative impact of such missing information on the performances of machine learning models \citep{BatistaMonard2003,marlin2008,parker2010,brescia:2018b}. We did not introduce any  further error-based cuts in order to avoid any additional reduction of the KB.


\section{The Methods}\label{sec:models}
In this work we make use of an optimized implementation of the Growing Neural Gas (GNG) network \citep{fritzke:1995} obtained using the Theano programming environment \citep{theano:2016}, and a novel feature selection method, $\Phi$LAB, to optimize the parameter space. Moreover we briefly introduce the three methods used as test benchmark. Such models are described in the following sections. For all these networks the hyper-parameters have been set by following a heuristic pruning process. 

\subsection{Growing Neural Gas}\label{ss:GNG}
The Growing Neural Gas (GNG) model was introduced in \cite{fritzke:1994} as a variant of the Neural Gas algorithm \citep{martinez:1991}, which combines the Competitive Hebbian Learning (CHL, \citealt{martinez:1993}) with a vector quantization technique, to achieve a learning that retains the topology of the dataset. This is an important property, since the vector quantization introduces an order relationship between the data parameter space and the internal architecture of the network. In fact, Vector quantization techniques \citep{martinez:1993}, encode a data manifold, e.g. $V\subseteq R^m$, using a finite set of reference vectors $w=w_1\dots w_N, w_i \in R^m, i=1\dots N$. Every data vector $v\in V$ is described by the best matching unit (BMU), i.e. the neural unit whose reference vector $w_{i(v)}$ minimizes the distortion error $d(v,w_{i(v)})$. This procedure divides the manifold $V$ into a number of subregions: 
$V_i=\{v\in V : ||v-w_i||\le||v-w_j||\, \forall j\}$, called Voronoi polyhedra \citep{montoro:1993}, within which each data vector $v$ is described by the corresponding reference vector $w_i$. 
The BMU and the second-BMU develop a connection that, if not energized again during learning, tends to decay and then to be removed \citep{fritzke:1994}. 
The GNG network is characterized by a variable number of neurons during the learning phase: new units are added to an initially small number of units through the estimation of a statistical local measure obtained during the previous adaptation steps, while isolated units are removed.
The insertion mechanism has to be able to find the location in the parameter space where to introduce a new neuron, in order to reduce the reconstruction error. In other words, the insertion mechanism finds subregions of the data manifold whose reconstruction is more complex, i.e. the subregions characterized by a relatively high density.\\
Each neuron has an attribute defined as the local error $E_i$, whose value is updated at each iteration only for the BMU $i_0$: $\Delta E_{i_0} = ||w_{i_0}-v||^2$, where $v$ is the extracted input vector. After a certain number of iterations, $E_i$ represents a local reconstruction error for the neural unit $i$. Units characterized by high values of $E_i$ are associated to large Voronoi polyhedra, and these regions require better sampling to be correctly reconstructed.\\
The adaptation rule is applied only for the BMU $i_0$ and for its topological neighbours:
\begin{equation}\label{eqn:update_w}
\begin{split}
&\Delta w_{i_0} = \epsilon_w \cdot(v-w_{i_0})\\
&\Delta w_j = \epsilon_n \cdot(v-w_j) \quad \forall j \in \text{Neighbours}(i_0)
\end{split}
\end{equation}

The advantage of this method is that the learning is completely determined by the input data, i.e. it is not necessary to superimpose a structure to the network as, for instance, the expected number of clusters. The downside is the single input extraction at each iteration, which leads to an extra computational cost on large dataset. 
For this reason we optimized the GNG implementation using Theano \citep{theano:2016}, an open source Python library allowing an efficient computation of tensor mathematical expressions and an easy exploitation of the Graphical Processing Unit (GPU). Furthermore, we revised the adaptation rule of eq.~(\ref{eqn:update_w}), by introducing a gradient descent method with respect to the cost function, represented by the quantization error: 
\begin{equation}
\begin{split}
&\Delta W = -\eta\nabla_W (\mathcal{QE})\\
&\mathcal{QE}=\frac{1}{2|V|}\sum_{i=1}^p\sum_{n\in BMU_i}||v_n-w_i||^2
\end{split}
\end{equation}
where we have assumed that: $v_n$ is the $n$-th input vector mapped by the BMU $i$ whose reference vector is $w_i$, $V$ is the data manifold composed by $|V|$ records, $p$ is the number of BMUs.\\
Finally, we added also a batch extraction criteria, i.e. at each iteration a subset of sources has been extracted from the data, whose dimension is between $1$ (equivalent to the original case) and the full dataset. 

\subsection{\mathinhead{\Phi}{Phi}LAB: a novel feature selection method}\label{ss:PhiLAB}
We recently investigated the possibility to find a parameter space optimization method able to cope with the all-relevant feature selection requirements and to infer knowledge within the data-driven analysis domain, hence particularly suitable for astrophysical problems. The designed method is called $\Phi$LAB (PhiLAB, Parameter handling investigation LABoratory), which aims at identifying the exact parameter space (PS), by solving the so-called \textit{all-relevant} feature selection problem \citep{Delliveneri:2019,brescia:2018b}. The method is a hybrid approach, including properties of both wrappers and embedded feature selection categories \citep{Tangaro:2015}. 
It is, in fact, based on two joined concepts: \textit{shadow features} \citep{Kursa2010} and \textit{Na\"{\i}ve LASSO} statistics (Least Absolute Shrinkage and Selection, (\citealt{Tibshirani2012,Hara2016,Hara2017,hastie2012}), by using the Random Forest (RF, \citealt{breiman:2003}) as feature importance computing engine. 
By joining the two concepts, $\Phi$LAB is able to determine a threshold to filter the most relevant candidate features and to refine the final selection by determining the additional weak relevant features through a $L_{1}-norm$ regularization of a ridge regression \citep{Tikhonov1998}, retaining only the non-zero features representing the optimal solution.

\subsection{Benchmark methods}\label{ss:benchmethod}

 In order to compare the classification capability of the GNG and to explore the features selection performed by $\Phi$LAB, we used three methods, described in the following. Having the possibility to use both supervised and unsupervised models, we tried also to perform a comparison between the two categories, by taking into account that, although supervised paradigm is generally preferred whenever a knowledge base is available, we were particularly interested to evaluate the performances of the GNG model in a complex astrophysical problem. 
\begin{itemize}
    \item[-] Multi-Layer-Perceptron with Quasi-Newton Approach (MLPQNA): it is a very robust supervised machine learning model, as it has been already demonstrated by its capability to achieve high performances on a variety of astrophysical problems \citep{Cavuoti:2012, Cavuoti:2015, brescia:2012, brescia:2013}. For this reason we have chosen this model as upper limit benchmark method. Its architecture is similar to an MLP \citep{Bishop:2006}, with a Quasi-Newton algorithm used as optimizer furthermore it makes use of the known L-BFGS algorithm (Limited memory - Broyden Fletcher Goldfarb Shanno, \citealt{Byrd1994}). This network has been applied as test benchmark of GNG performances (results shown in Sec.~\ref{ss:GNGclass}). The network is composed by two hidden layers, respectively with $2N+1$ and $N-1$ neural units, where $N$ is the number of input dimensions (i.e. the number of features). The neuron activation function is a hyperbolic tangent. Furthermore, the network weight updating is based on the L2-norm regularization term \citep{Bishop:2006}, with a decay factor of $0.01$. 
    \item[-] Random Forest (RF): it is a widely known supervised machine learning ensemble method that uses a random subset of features to build a collection of decision trees \citep{Breiman2001}. The method is characterized by an intrinsic absence of training overfitting (i.e. excess of training data fitting with consequent poor fitting of blind test data). RF has been applied in order to verify the sensitivity of GNG to noisy and redundant data parameters and to investigate regarding the efficiency of $\Phi$Lab to individuate the best set of features. The results are shown in appendix~\ref{app:FI} and discussed in Sec.~\ref{ss:FS}. The method has been trained with $500$ trees, the gini index is used in order to evaluate the quality of the split \citep{Breiman2001}, while the maximum number of features, required to search the best split, coincides with the involved number of parameters. Furthermore the minimum number of samples necessary to split a node has been set equal to $2$ and no limit was imposed to the depth of the tree growth.
    \item[-] K-means: it is a clustering method able to partition the dataset into $K$ clusters, minimizing the distortion measure. At the end of the training phase the dataset has been divided into $K$ Voronoi polyhedra \citep{montoro:1993}. Such method provides a benchmark lower limit: although the model is able to perform a vector quantization process, it is necessary to over-impose a structure on the data, i.e. the number of cluster is not automatically determined. Thus, in order to compare the unsupervised networks, we trained the model by setting $K$ to the number of clusters found by the GNG. Finally, the Expectation Maximization (EM) algorithm was used to train the model.
\end{itemize}

\section{Experiments}\label{sec:exp}
We generated two different datasets to train our models, one including the \textit{u}-band and the second excluding it. This partition was imposed by the fact that in FDS the \textit{u}-band is much shallower than the others (among the $\sim 94,000$ sources in the catalogue, only $15\%$ of the objects were detected in \textit{u}-band, against higher percentages for \textit{g}, \textit{r} and \textit{i} bands, respectively, $78\%$, $77\%$ and $53\%$).
The resulting datasets can be summarized as it follows: \textit{(a)} \emph{ugri} dataset with $1,113$ objects, of which $357$ are GCs, $416$ galaxies and $340$ stars; \textit{(b)} \emph{gri} dataset with $1,618$ objects, including $699$ GCs, $457$ galaxies and $462$ stars. For both datasets we performed three different classification experiments: 
\begin{itemize}
\item a $3$-class problem, i.e. stars, GCs and galaxies (named as \textit{3CLASS}); 
\item a $2$-class problem by grouping in the same class stars and galaxies against GCs (named as \textit{GCALL}); 
\item a $2$-class problem, namely stars versus GCs (named as \textit{GCSTAR}).  
\end{itemize}
Hence, we performed a total of six experiments. Due to the limited amount of labeled samples available (i.e. sources with a known spectroscopic classification), a canonical splitting of the KB into training ($\sim 80\%$) and blind test set ($\sim 20\%$) could not be applied. In order to circumvent this problem and to balance the samples for each class during the learning phase, the training-test experiments involved an approach based on the stratified k-fold cross validation \citep{hastie2009, Kohavi1995}: the KB is split into $5$ non-overlapping subsets.
In this way, by iteratively taking each time $4$ of these subsets as training set, and using the fifth as blind test set, an overall blind test on the entire KB available can be performed.
As it was already mentioned, in order to perform the optimal choice of the parameter space, we applied the method $\Phi$LAB, by identifying a proper subset of features for each of the six experiments. We use a RF method, together with the GNG, to analyze the selection achieved by $\Phi$LAB: we measured the model performances by varying the parameter space (see next section).

The statistical estimators used to measure both the feature selection and classification performances are: \textit{(i)} the average efficiency (\textit{AE}), the ratio between the number of correctly classified objects and the total number of objects, averaged over all involved classes; \textit{(ii)} the purity (\textit{pur}, also called \textit{precision}), i.e. the ratio between the number of correctly classified objects of a class and the total number of objects classified in that class (the dual of the contamination); \textit{(iii)} the completeness (\textit{compl}, also called \textit{recall}), i.e. the ratio between the number of correctly classified objects of a class and the total amount of objects of that class; \textit{(iv)} the F1 score (\textit{F1}), defined as two times the ratio between the product of purity and completeness and their sum, for each class \citep{stehman:1997}. 

\subsection{Feature selection}\label{ss:FS}

\begin{figure*}
\includegraphics[width=\textwidth]{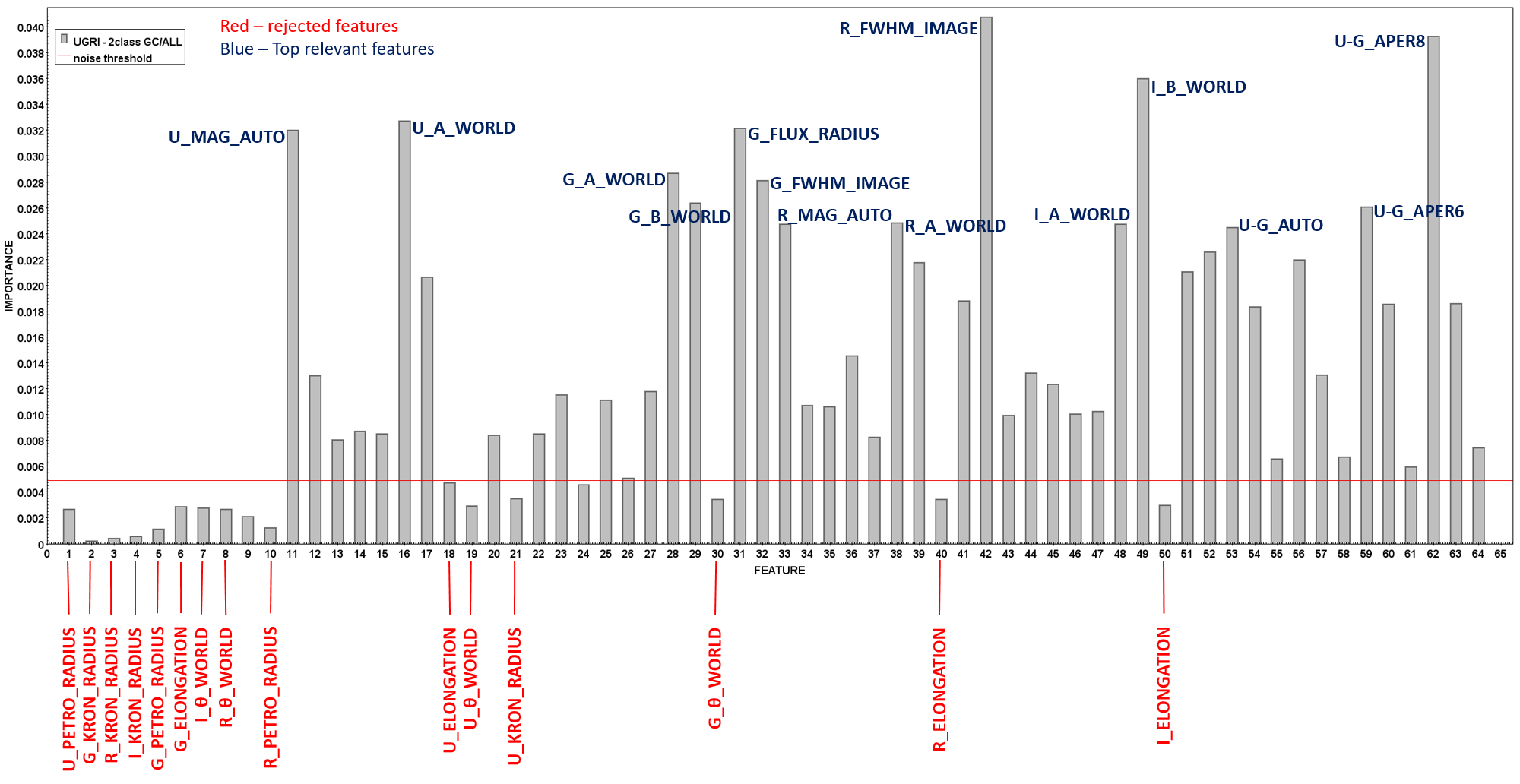} 
\caption[Example of feature selection results related to the experiment $2$-class \emph{ugri}, GCs versus the others]{Example of feature selection results for the experiment $2$-class \emph{ugri} (GCs-vs-stars+galaxies). The red line is the \textit{shadow feature} noise threshold, defining the separation between \textit{best} and \textit{weakly relevant} features. The rejected features are in red, while the top relevant features are in blue. Table~\ref{tab:FS}, in appendix~\ref{app:FI}, reports the feature importance values estimated by $\Phi$LAB for all the six experiments. Features 9 (\textit{i}-band \textit{PETRO\_RADIUS}) and feature 24 (\textit{g}-band \textit{MAG\_APER6}), although under the shadow feature noise threshold, have not been considered as rejected, because retained as weak relevant after the application of the LASSO statistics (see Sec.~\ref{ss:PhiLAB} for details).}\label{fig:ugri:GCALL:FS}
\end{figure*}

In order to analyze and optimize the parameter space suitable for classification purposes, the six datasets were first processed by $\Phi$LAB. In this work, our main interest was to obtain the most simplified parameter space able to predict a good classification on new objects (i.e. outside the data used for training+test) based on the training on the full dataset available. That is why, along the reported experiments, we took in consideration the results of feature 
selection applied always to the whole dataset. As an example, the feature importance related to the \textit{ugri-GCALL} case is shown in Fig.~\ref{fig:ugri:GCALL:FS}, where the parameter space is partitioned into a \textit{rejected} set of features (considered to be irrelevant in terms of information contribution) and a retained set of \textit{all relevant} features, composed by both \textit{best} and \textit{weakly relevant} features. 

The results of the six FS experiments are summarized in Table~\ref{tab:FS} (appendix \ref{app:FI}) for both the selected and the rejected subsets of features. By analyzing such results, the \textit{FWHM}, \textit{FLUX\_RADIUS} and colours appear relevant in all the six classification experiments (these three sets of features, with respect to the involved set of relevant features, have an informative contribution ranging in $9\%-31\%$, $4\%-24\%$ and $12\%-26\%$, respectively), thus confirming the higher relevance of the colours over magnitudes, as well as the intrinsic importance of \textit{FWHM} and \textit{FLUX\_RADIUS}, indispensable to disentangle the extended objects from foreground point-like sources and the unresolved GCs. Concerning the \textit{FWHM}, its informative contribution remains high, although being averaged to take into account small variations among the FDS fields.
The features \textit{MU\_MAX}, achieving a large informative contribution (ranging in $5\%-18\%$), mark a slight difference between the experiments that include galaxies and those in which only the GC/star separation is required.

Moving from the \emph{ugri} type to \emph{gri} and from the \emph{3CLASS} experiments to \emph{GCSTAR}, there is a flattening of the informative contribution among the features, mainly due to the exclusion of the \textit{u}-band and by grouping together stars and galaxies, thus increasing the complexity of the classification problem.

Regarding the removed parameters: \textit{THETA} features have a negligible informative contribution; \textit{KRON\_RADIUS} carries a very weak contribution (the sum over all the involved filters does not reach $1\%$ of the whole informative contribution).
The improvement due to the \textit{ELONGATION} features is always much smaller than the information shared by the \textit{SEMI-AXIS} features (sometimes about one order of magnitude). Although it may appear as a bit surprising that the elongation does not show a particular relevance, this can be due to the information already carried by the semi-axes that are an absolute quantity depending on the shape and distance of the object and its extension. In particular, the combination of the semi-axes and extension (e.g. \textit{FLUX\_RADIUS}) may embed the elongation information.

The information carried by the \textit{PETRO\_RADIUS} oscillates among the experiments and its informative contribution disuniformity is mostly due to \emph{i}-band. Thus, according to the \textit{all-relevant feature selection} approach, this feature has not been rejected. From all these considerations, the resulting optimized parameter space extracted for the six classification experiments consists of $49$ features (listed in upper Table~\ref{tab:FS} in appendix~\ref{app:FI}). 

In order to verify that the PS extracted by $\Phi$LAB, is the best suitable set of representative features for the GC classification, we performed a test, based on the following training+test classification experiments, involving all six datasets:
\begin{itemize}
\item \textit{BEST PS}: using the optimized PS, composed by the $49$ features extracted by the FS method, hence representing the best solution to the \textit{all-relevant} FS problem; 
\item \textit{FULL PS}: by using the full PS, composed by all $64$ features available;
\item \textit{MIXED PS}: by altering the BEST PS, replacing a group of $15$ randomly selected features with the $15$ rejected features;
\item \textit{BEST+REJECTED PS}: by replacing the $15$ least relevant features of the best PS with the $15$ rejected features.
\end{itemize}
In all these tests, for each of the six datasets involved, the same training and test sets have been used, as well as the same configuration setup for the model GNG, in order to avoid any spurious effect on the classification statistics induced by the change of internal model parameters and by the data used for training and test. The resulting parameter spaces are summarized in Table~\ref{tab:PSs}, while the related GNG performances are reported in Table~\ref{tab:PSs:performance}, showing that:
\begin{itemize}
    \item[-] the selected set of features (named as BEST) allows the GNG to achieve high performances, reaching an increase of $50\%$ in terms of average efficiency (AE), i.e. the GNG trained on the BEST dataset always reaches better scores in terms of statistical estimators, whereas the performance degradation is due only to the removal of the \textit{u}-band ($\sim8\%$ in terms of AE), although always remaining well above the other PSs; 
    \item[-] the separation between GCs and notGCs (star and galaxies) appears to be the least complex problem for the GNG (showing an increase of the AE between $\sim3\%$ and $\sim45\%$);
    \item[-] concerning the other PSs (FULL, MIXED, BEST+REJECTED), the additional information carried by a greater number of sources (\textit{gri}) predominates on the information represented by the \textit{u}-band (\textit{ugri}) only in the \textit{GCALL} experiments (separation between GCs and notGCs), gaining up to $10\%$ in terms of AE. This trend is reversed when the classification involves the separation of the stars and galaxies (\textit{3CLASS} and \textit{GCs versus stars}): in these cases there is a greater dependence on the absence of the \textit{u}-band rather than on the number of sources (with differences of AE $\lesssim 9\%$). The only exception occurs in the most complete and complex case, i.e. the \textit{ugri} \textit{3CLASS} with FULL parameter space, where the amount of objects has a greater impact (with $\lesssim 7\%$ gain in terms of AE);  
\end{itemize}


This analysis seems to support the idea that GNG is particularly sensitive to noisy or redundant features and, in order to verify this hypothesis, we repeated the same training/test process (i.e. the same dataset for the same experiments) replacing the GNG with a Random Forest method (RF, \citealt{Breiman2001}). 

The RF method, by generating a random ensemble of decision trees, is very robust to parameter variations. The results are shown in \ref{tab:PSs:RFperformance}, from which:
\begin{itemize}
    \item[-] in all cases, RFs trained on the dataset BEST achieve the highest scores, showing an increasing of $\lesssim8\%$ in terms of AE;
    \item[-] concerning the \textit{3CLASS} and \textit{GCALL} problem, the \textit{u}-band improves the AE ($\lesssim8\%$) but, without this additional photometric information, BEST is always more robust than others ($\lesssim6\%$);
    \item[-] the MIXED dataset is always the worst PS, with a decrease larger than $5\%$ in terms of AE;
    \item[-] in the \textit{GCSTAR} case, the \textit{u}-band seems to loose partially its positive role (with an AE decreasing of $\sim2\%$) and, without the \textit{u}-band, the dataset \textit{FULL} is quite often more robust than the others ($\lesssim3\%$);
    \item[-] Purity is almost stable in all PSs, while completeness presents large fluctuations (the largest purity deviation is $\sim2\%$, whereas the largest completeness is $\sim7\%$).
    \item[-] it appears that the absence of galaxies in the sample (i.e. \textit{GCSTAR} cases) makes the experiments less sensitive to the presence of the \textit{u}-band data (showing a decrease of AE of $0.5\%$, i.e. the results show fluctuations around a mean value). While the presence of the \textit{u}-band affects more significantly the classification capability in presence of galaxies, making the \textit{BEST} case the most powerful dataset. The role of the \textit{u}-band can be derived from Fig.~\ref{fig:gngmlp:col} and Fig.~\ref{fig:RUN:colours}, according to what was widely discussed in \cite{munoz:2014}, where the authors showed (see, for instance, their figures $13$ and $16$) how the bluer Spectral Energy Distribution (SED) of star forming galaxies and passive galaxies at moderate redshift is well identified in the colour-colour diagrams. The \textit{u}-band, in this case based only on optical colours, shows much less discriminant power for GCs and stars, becoming more relevant in identifying galaxies.
\end{itemize}

Given these premises, in order to explore the impact of the \textit{u}-band on the classification, we performed two further experiments: (i) a dataset extracted from the \textit{ugri} set without the \textit{u}-band, named as \textit{gri*}; and (ii) by using only the features related to the \textit{u}-band. Such test has the role to disentangle the \textit{u}-band contribution from the effect carried by the increase of the samples. Results are shown in Table~\ref{tab:roleofu} (in appendix~\ref{app:FI}), from which we conclude that:
\begin{itemize}
    \item[-] the use of the single \textit{u}-band information still allows the separation among classes with a slight decrease of AE for both GNG ($\lesssim 2\%$) and RF ($\sim{3}\%$);
    \item[-] concerning the \textit{gri} dataset the average efficiency reduction appears higher if compared to the \textit{ugri} dataset ($\gtrsim 5\%$);
    \item[-] RF performances significantly decrease with respect to the results achieved in the case where all the available samples are used (rising up an AE difference of $\sim 10\%$ in the same cases), showing the known dependence of ML methods from the training dimension \citep{Brescia2013}.
\end{itemize}

Hence, the experiments performed by the two models confirm: (i) the strong dependence of GNG on the structure of the parameter spaces and thus support the need for a robust method of features selection, (ii) the capability of $\Phi$LAB to identify the set of relevant features, (iii) the high impact of the \textit{u}-band on the classification capability for both models.

\subsection{GC classification}\label{ss:GNGclass}
As introduced in Sec.~\ref{ss:benchmethod} the GC classification experiments have been performed by comparing the GNG model with MLPQNA and K-means. For the latter model the performances have been reported in Table~\ref{tab:kmeanscompare} and represent a lower limit for the GNG. In terms of average efficiency, the GNG shows better results, gaining from $1.0\%$ to $6.6\%$. For this reason we focused more on the comparison between MLPQNA and GNG.
The classification results on the blind test sets, obtained by GNG and MLPQNA using the k-fold technique for the six datasets, are summarized in Table~\ref{tab:results}. In Fig.~\ref{fig:GNG:ROC} the ROC curves \citep[Receiver Operating Characteristic, ][]{hanley:1982} have been reported to study the purity-completeness trade-off. In order to compare the results, it is important to remark that the unsupervised model (GNG) does not take into account the knowledge of the source labels during the learning, entrusting the weights adaptation to the minimization of the quantization error, while the supervised model (MLPQNA) uses labels to guide learning, allowing also the identification of the minimum with a very high efficiency \citep{brescia:2012}. Therefore, in principle, some performance differences are expected.

In order to identify the GCs, the most interesting measure is the purity, i.e. the fraction of true GCs within the set of objects that are classified by the method as GCs \citep{brescia:2012}, although we are interested to find the best trade-off with completeness, as usual in any classification scheme but crucial in  astrophysical problems \citep{D'Isanto20163119}. 

Concerning the \emph{ugri} dataset, both models show comparable performances in terms of trade-off between purity and completeness of the order of $80$-to-$95\%$ (see Table~\ref{tab:results}). However, in order to explore the classification differences between the two methods, we counted the number of times a model exceeds the other for more than $3\%$ in terms of statistical measurements. Concerning the \textit{ugri} dataset, MLPQNA overcomes the GNG results in the \textit{GCSTAR} experiment (showing an average improvement of $\sim5\%$), while GNG performs better than MLPQNA in the \textit{GCALL} experiment (where the GNG average increase is $\sim3\%$). Regarding the \textit{3CLASS} experiments, both methods show very similar efficiency (in fact, the MLPQNA average increase is less than $0.4\%$). Furthermore, GNG seems to identify GCs better than other classes (showing an average improvement of $\sim2\%$ in the case of the \textit{ugri} experiments), while MLPQNA shows a greater detecting capability for stars and galaxies (with respect to which the average gain related to \textit{ugri} experiments is $\sim3\%$). 

The scenario is different for the \emph{gri} datasets, where MLPQNA outperforms GNG, by disentangling the source classes with very few losses (by reaching a purity and a completeness on GCs higher than $86\%$), although without the \textit{u}-band. Only in the \textit{2CLASS} experiment \textit{GCs vs ALL} the GNG network achieves similar (although minor) performances. In this case the AE difference between model classification capabilities is $\sim4\%$, whereas in \textit{3CLASS} and \textit{GCSTAR} experiments the differences are $\sim10\%$ and $\sim8\%$, respectively.

By comparing the results between the \textit{GCSTAR} and the other experiments, it appears that the exclusion of galaxies from the train set makes both models more efficient to identify the GCs, although both methods are able to identify galaxies with a good trade-off between purity and completeness ($\gtrsim92\%$ in the \textit{ugri} case, $\gtrsim88\%$ for the \textit{gri} experiment). This behaviour appears more pronounced for the MLPQNA.

The GNG performance gaps between \textit{ugri} and \textit{gri} dataset can be visually deduced also from the ROC curves (Fig.~\ref{fig:GNG:ROC}) studying the trend of the Area Under the Curve (AUC), which represents the probability that a classifier correctly predicts the membership of a sample, although the ``positive" probability thresholds are higher than the ``negative" ranks \citep{fawcett:2006}. Concerning the GC classification, the AUCs gain up to $9.9\%$ by moving from \textit{gri} to \textit{ugri} dataset; the best result is obtained by the \textit{GCALL-ugri} experiments ($93\%$ and $94\%$, respectively for GCs and \textit{not}GCs). The performances drop down for the \textit{GCSTAR-gri} experiment (AUC $\sim80\%$ for both classes), where the photometric similarity between sources and the lack of information makes the GNG less performing.

\begin{table}\centering\caption[Classification results in terms of statistical estimators]{Classification results in terms of statistical estimators: average efficiency (\emph{AE}), purity (\emph{pur}), completeness (\emph{compl}) and F1-score (\emph{F1}) for both \emph{ugri} and \emph{gri} dataset types and for both GNG and MLPQNA models. Top Table reports the results for the \textit{3CLASS} experiment, middle Table for the $2$-class experiment between GCs and \textit{not} GCs (stars + galaxies), and bottom Table shows the results concerning the $2$-class experiment between GCs and stars. Regarding the \emph{ugri} dataset case, GNG and MLPQNA have similar performances, although MLPQNA shows an optimal trade-off between purity and completeness; while in the case of \emph{gri} dataset the GNG cannot reach MLPQNA performance, particularly in the $3$-class and stars VS GCs cases. All these experiments are performed using the \textit{BEST} parameter space. The values higher than $90$\% are marked in bold.}\label{tab:results}
\resizebox{\columnwidth}{!}{
\begin{tabular}{lcccc}
\hline
\multicolumn{1}{c}{\textbf{3CLASS}} & \multicolumn{2}{c}{\textbf{ugri}} & \multicolumn{2}{c}{\textbf{gri}} \\
\hline
Estimator [\%] & GNG  & MLPQNA & GNG  & MLPQNA \\\hline
AE         & 86.5 & 88.2   & 79.4 & 88.8\\\hline\hline
pur STAR   & 85.8 & 84.8   & 71.9 & 85.6 \\\hline
compl STAR & 80.3 & 85.6   & 66.9 & 82.0 \\\hline
F1 STAR    & 83.0 & 85.2   & 69.3 & 83.8 \\\hline\hline
pur GCs    & 80.0 & 83.2   & 78.2 & 87.2 \\\hline
compl GCs  & \textbf{90.8} & 83.2   & 79.6 & 89.4 \\\hline
F1 GCs     & 85.1 & 83.2   & 78.9 & 88.3\\\hline\hline
pur GAL    & \textbf{92.5} & \textbf{95.4}   & 88.3 & \textbf{94.5} \\\hline
compl GAL  & \textbf{95.4} & \textbf{94.7}   & \textbf{92.1} & \textbf{94.7} \\\hline
F1 GAL     & \textbf{93.9} & \textbf{95.0}   & \textbf{90.2} & \textbf{94.6} \\\hline\hline
\hline
\multicolumn{1}{c}{\textbf{GCs vs ALL}} & \multicolumn{2}{c}{\textbf{ugri}} & \multicolumn{2}{c}{\textbf{gri}} \\
\hline
Estimator [\%]        & GNG & MLPQNA & GNG & MLPQNA \\\hline
AE          & 88.7 & 87.1 & 84.0 & 88.4\\\hline\hline
pur notGC   & 85.1 & \textbf{91.2} & 81.3 & \textbf{90.1} \\\hline
compl notGC & 88.0 & 89.6 & 85.5 & 89.4 \\\hline
F1 notGC    & 86.5 & \textbf{90.5} & 83.4 & 89.8 \\\hline\hline
pur GCs     & \textbf{91.3} & 78.9 & 86.2 & 86.3 \\\hline
compl GCs   & 89.1 & 81.8 & 82.1 & 87.1 \\\hline
F1 GCs      & \textbf{90.2} & 80.3 & 84.2 & 86.7 \\\hline\hline
\hline
\multicolumn{1}{c}{\textbf{GCs vs STARs}} & \multicolumn{2}{c}{\textbf{ugri}} & \multicolumn{2}{c}{\textbf{gri}} \\
\hline
Estimator [\%]       & GNG & MLPQNA & GNG & MLPQNA \\\hline
AE         & 86.8 & \textbf{90.3} & 78.2 & 87.9\\\hline\hline
pur STAR   & 87.0 & 84.3 & 77.3 & 81.8 \\\hline
compl STAR & 83.2 & 80.7 & 84.9 & 83.2 \\\hline
F1 STAR    & 85.1 & 82.5 & 81.1 & 82.5 \\\hline\hline
pur GCs    & \textbf{91.6} & \textbf{92.6} & 79.7 & \textbf{91.2} \\\hline
compl GCs  & 80.3 & \textbf{94.1} & 70.3 & \textbf{90.4} \\\hline
F1 GCs     & 85.6 & \textbf{93.3} & 75.0 & \textbf{90.8}\\\hline\hline
\end{tabular}
}
\end{table}

\begin{table}\centering\caption{Intersection between predictions performed by GNG and MLPQNA for both \textit{ugri} and \textit{gri} dataset types. Further, in order to disentangle the influence on performances due to the amount of samples in the train set, we evaluate the intersection between the predictions performed on the \textit{gri} dataset using only the \textit{ugri} indices (i.e. \textit{gri} features with \textit{ugri} samples), named as \textit{gri*}. Top Table refers to the \textit{3CLASS} experiment, middle Table to the \textit{GCALL} experiment, while bottom Table to the \textit{GCSTAR} experiment. Row \textit{commons} reports the intersection between predictions regardless of whether they are correct or not. Row \textit{corrected} specifies the common objects correctly classified. The values are expressed: (i) as percentage with respect to whole set (third and fifth columns), (ii) as percentage with respect to the number of objects in the corresponding class (fourth and sixth columns). }\label{tab:common}
\resizebox{\columnwidth}{!}{
\begin{tabular}{llcccccc}
\hline
\multicolumn{8}{c}{\textbf{3CLASS}}\\\hline
&&\multicolumn{2}{c}{\textbf{ugri}[\%]} & \multicolumn{2}{c}{\textbf{gri}[\%]} & \multicolumn{2}{c}{\textbf{gri*}[\%]} \\\hline
\multirow{3}{*}{commons} 	& STAR & \multirow{3}{*}{85.3}  & 78.8 & \multirow{3}{*}{80.8} & 67.3 & \multirow{3}{*}{78.9} & 72.1 \\
		   					& GCs  &                        & 79.8 &  				   & 82.4 &                       & 67.6 \\
		   					& GAL  &                        & 95.2 &  				   & 92.1 &                       & 92.0 \\\hline
\multirow{3}{*}{corrected} 	& STAR & \multirow{3}{*}{80.1}  & 75.9 & \multirow{3}{*}{75.0}   & 63.0 & \multirow{3}{*}{75.6} & 77.6 \\
		   					& GCs  &                        & 71.1 &  				   & 74.4 &                       & 58.0 \\
		   					& GAL  &                        & 91.3 &  				   & 88.2 &                       & 88.9 \\\hline\hline
\multicolumn{8}{c}{\textbf{GCALL}}\\\hline
&&\multicolumn{2}{c}{\textbf{ugri}[\%]} & \multicolumn{2}{c}{\textbf{gri}[\%]} & \multicolumn{2}{c}{\textbf{gri*}[\%]} \\\hline
\multirow{2}{*}{commons} 	& notGCs& \multirow{2}{*}{86.0}  & 88.3 & \multirow{2}{*}{82.2}   & 81.1 & \multirow{2}{*}{83.0} & 85.6\\
		   					& GCs  &                       & 82.6 &  					    & 83.4 &                     & 78.7\\\hline
\multirow{2}{*}{corrected} 	& notGCs& \multirow{2}{*}{80.6}  & 84.6 & \multirow{2}{*}{77.0}   & 77.1 & \multirow{2}{*}{77.0} & 81.9\\
		   					& GCs  &                       & 75.7 &  					    & 77.1 &                     & 71.1\\\hline\hline
\multicolumn{8}{c}{\textbf{GCSTAR}}\\\hline
&&\multicolumn{2}{c}{\textbf{ugri}[\%]} & \multicolumn{2}{c}{\textbf{gri}[\%]} & \multicolumn{2}{c}{\textbf{gri*}[\%]}\\\hline
\multirow{2}{*}{commons} 	& STAR & \multirow{2}{*}{90.3}  & 78.6 & \multirow{2}{*}{80.0} & 70.4 & \multirow{2}{*}{80.6} & 82.8\\
		   					& GCs  &                      & 95.0 &  					 & 86.2 &                     & 74.7\\\hline
\multirow{2}{*}{corrected} 	& STAR & \multirow{2}{*}{84.7}  & 74.3 & \multirow{2}{*}{73.1} & 66.0 & \multirow{2}{*}{72.3} & 77.8\\
		   					& GCs  &                      & 88.8 &  					 & 77.8 &                     & 66.7\\\hline\hline
\end{tabular}
}
\end{table}

\begin{figure*}\centering
\includegraphics[width=\columnwidth]{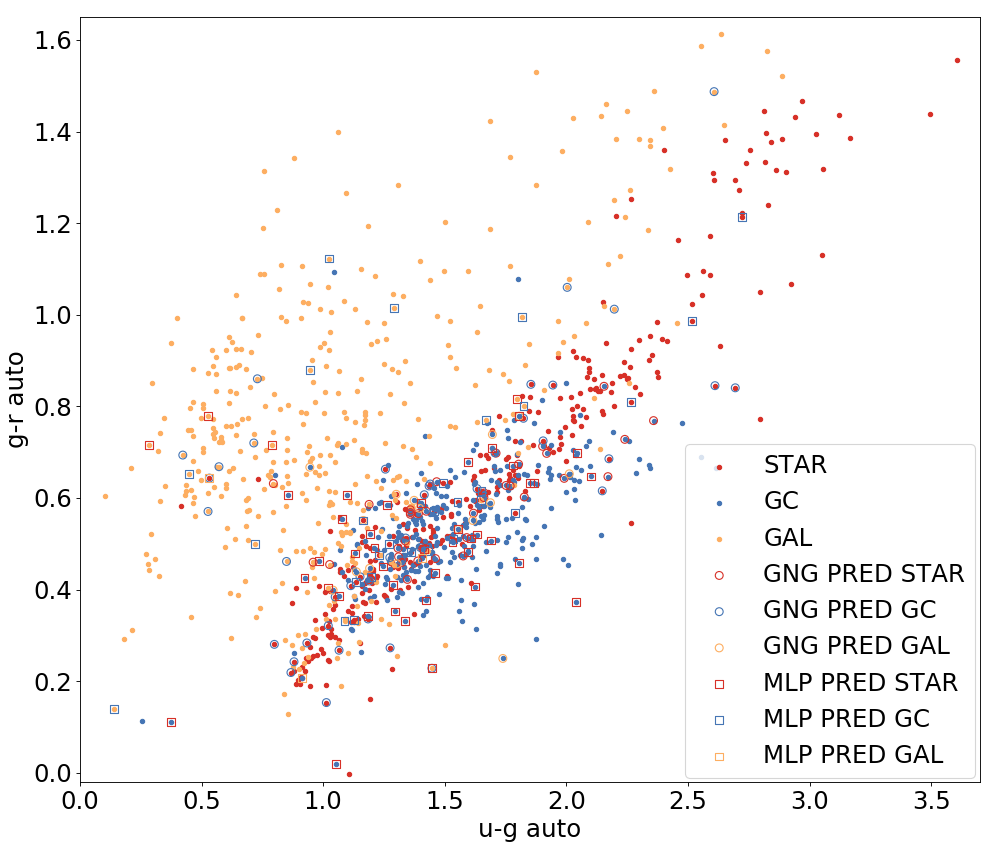}
\includegraphics[width=\columnwidth]{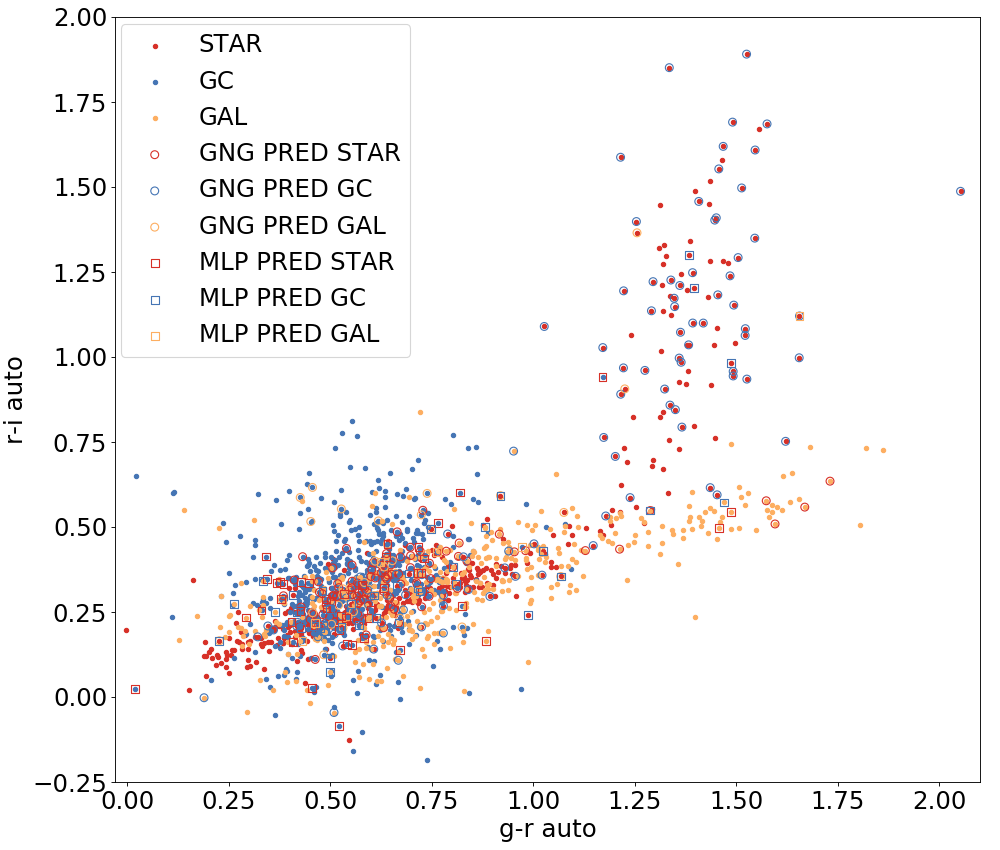}
\includegraphics[width=\columnwidth]{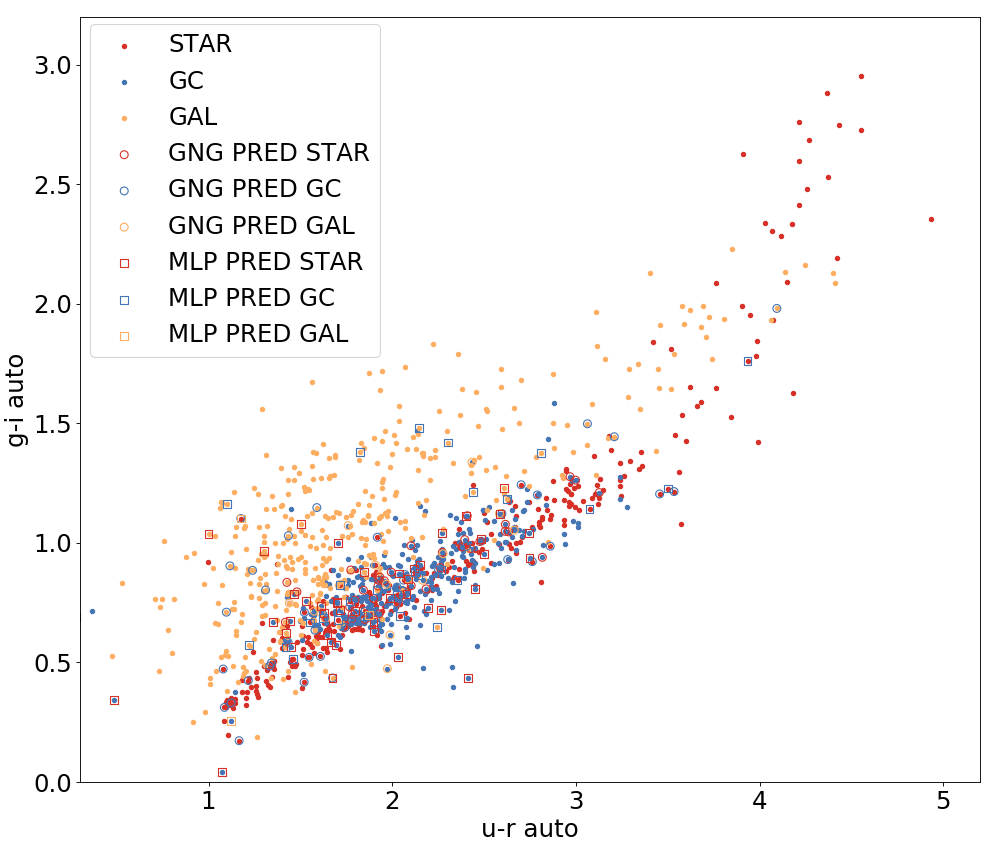}
\includegraphics[width=\columnwidth]{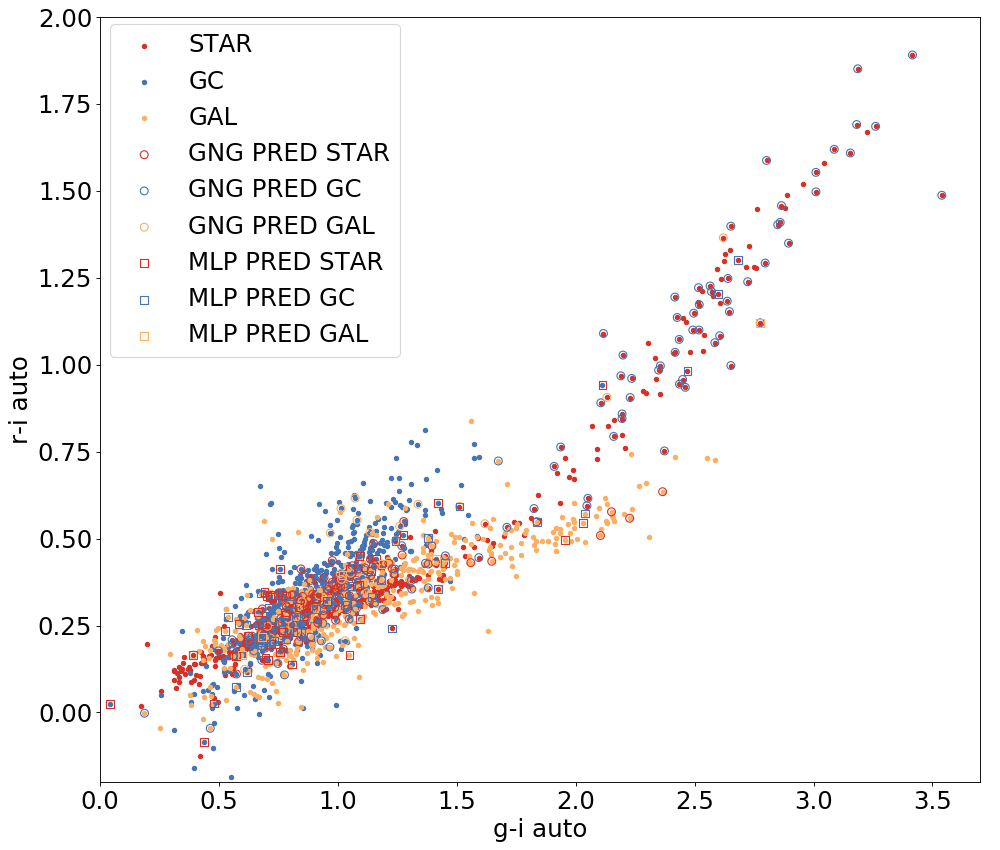}
\caption{ Misclassified sources plotted on colour-colour diagrams related to the \textit{ugri} dataset (left top panel, \textit{g-r} vs. \textit{u-g} and left bottom \textit{g-i} vs. \textit{u-r}) and related to the \textit{gri} dataset (right top panel, \textit{r-i} vs. \textit{g-r}, right bottom panel, \textit{r-i} vs. \textit{g-i}), together with the spectroscopic set. In all figures train GCs are plotted with blue dots, train stars with red dots and train galaxies with orange dots. The incorrect predictions made by GNG are plotted with open circles, while MLPQNA misclassified sources are plotted with open squares. For both of them the incorrect classifications are colored in red, blue and orange, respectively for sources predicted as stars, GCs and galaxies.}\label{fig:gngmlp:col}
\end{figure*}

\begin{figure*}\centering
\includegraphics[width=\columnwidth]{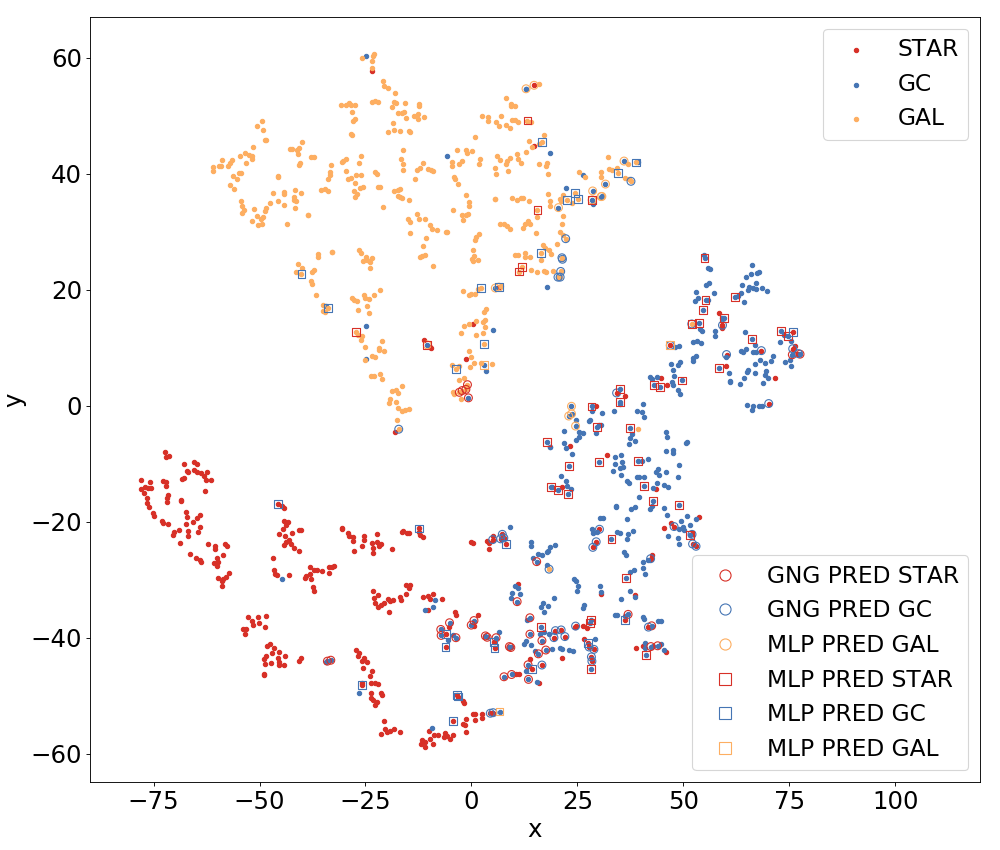}
\includegraphics[width=\columnwidth]{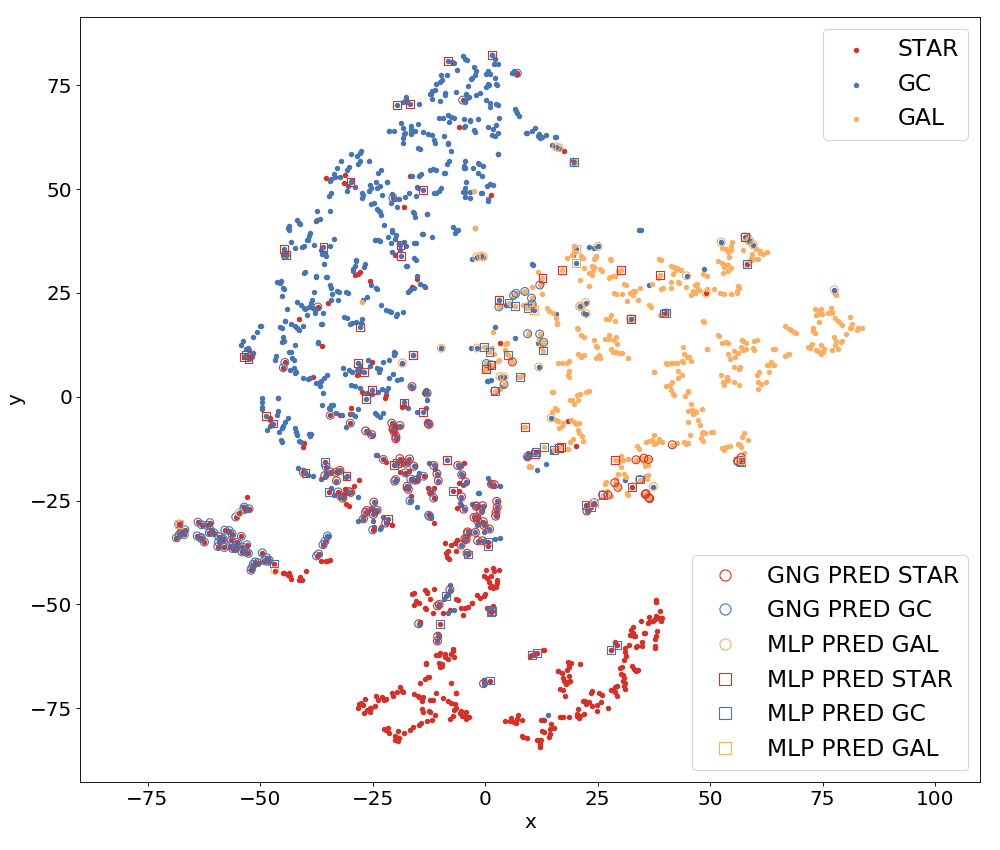}
\caption{Bi-dimensional projection performed by \textit{ECODOPS} of the \textit{ugri} (top panel) and \textit{gri} (bottom panel) train set together with the misclassified sources predicted by GNG (open circles) and MLPQNA (open squares). In both figures GCs (spectroscopic and predicted) are in blue, stars (spectroscopic and predicted) are in red and galaxies (spectroscopic and predicted) are in orange.}\label{fig:gngmlp:tsne}
\end{figure*}

In Table~\ref{tab:common} we distinguish the commonalities between GNG and MLPQNA referred to their predictions and corrected ones. Such analysis was computed for all the six kinds of experiments and evaluated on a parameter space composed by the \textit{gri} features, but restricting the data samples only to the objects available in the \textit{ugri} case (named as \textit{gri*}). Again, the underlying idea is to disentangle the contribution of features from the increasing of sample size in the train set. As expected, from the estimated performances (see Table~\ref{tab:results}), the largest sets of common predictions are related to the \textit{ugri} experiments ($80$-to-$90\%$). As previously discussed, the inclusion of galaxies in the training set reduces the capabilities of both methods to detect GCs (a $5\%$ drop in term of commonalities). Concerning the \textit{gri*} case, there is a negligible reduction in terms of global common classification fraction (i.e. regarding all the involved source types), but the fraction of identified (and correctly identified) GCs is considerably decreased (up to $~20\%$). 
In order to understand the origin of the misclassified objects, we plot on colour-colour diagrams the spectroscopic sources together with the incorrect predictions with respect to all classes of objects, for both \textit{ugri} and \textit{gri} training sets, shown in Fig.~\ref{fig:gngmlp:col}. The incorrect sources could be due to the photometric similarity between the sources, for instance GCs and stars, accentuated when the \textit{u}-band is removed. However, given the high number of dimensions involved, the diagrams describe a small portion of the whole space of features (each one of them represents less than the $5\%$ of total information contribution, estimated as sum of the feature importances related to the involved colours).
Thus, we used our model \textit{ECODOPS}\footnote{\url{http://dame.dsf.unina.it/ecodops.html}} (Efficient Coverage of Data On Parameter Space), a Python based system wrapping a high-dimensions data visualization technique: t-distributed Stochastic Neighbor Embedding \citep[tSNE][]{vanDerMaaten2008, VanDerMaaten:2014}. Our tSNE implementation is based on the object imported from the library \textit{sklearn} \citep{sklearn:2011}.\\  
Such method, already applied in other astrophysical contexts \citep{Nakoneczny:2019}, guarantees the preservation of the significant structures of the high-dimensional data visualized in a low-dimensional map. Therefore it converts similarities between data points to joint probabilities and tries to minimize the Kullback-Leibler \citep{Kullback} divergence between the embedding space and the high-dimensional data. In this way, the tSNE maps the multi-dimensional data to a lower dimensional space and attempts to find patterns in the data by identifying observed clusters based on similarity of data points with multiple features. However, after this process, the input features are no longer identifiable, and any inference based only on the output of the method cannot be done. Hence it must be considered as mainly a data exploration and visualization technique. The resulting 2D representation is obtained as a bunch of data points scattered on a 2D space (Fig.~\ref{fig:gngmlp:tsne}), where the underlined concept is that two close data points in the 2D space have similar properties in the high-dimensional space.

The embedding maps, illustrated in Fig.~\ref{fig:gngmlp:tsne} for both \textit{ugri} and \textit{gri} sets, together with the misclassified objects, show the great separability between the class types when the information carried by the \textit{u}-band is added to the train set. Most of the incorrect predictions are located in the border regions, particularly in the \textit{ugri} case (top panel in Fig.~\ref{fig:gngmlp:tsne}), while other false predictions are surrounded by contaminants. The embedded space shows a correspondence to the colour-colour plane: the similarity between stars and GCs is still noticeable, although their separation is larger than in any other feature combinations, i.e. the whole ensemble of features computed by $\Phi$LAB maximizes the separation capabilities of both methods.

\subsection{Photometric search for new GCs in Fornax core}
It is now crucial to verify the capability of GNG to identify GCs by testing it against a set of unlabeled sources (hereafter, \textit{run} process). This should allow not only to analyze the performances of the classifiers, but also to provide an additional set of GCs, suitable for advances in the astrophysical studies of the GC population and their connection with host galaxies.

As preliminary step, the fainter sources are excluded from the dataset, by applying the cuts summarized in Table~\ref{tab:runcuts}, due to their low $S/N$ ratio and in order to cut unclassified data at the same limit of the KB. The \emph{u}-band magnitudes were excluded from the cut, since the training and the run magnitude distributions share the same range of values. Moreover, cuts on magnitude errors have been applied on all the available bands, in order to limit the presence of noisy sources (Table~\ref{tab:runcuts}). Furthermore, samples affected by missing data were excluded from the catalogue. At the end of this selection process, two datasets have been produced, one including the \emph{u}-band, the other excluding it, in a similar way to what was done for the KB data. The \emph{ugri} dataset consists of $5,562$ sources ($\sim45\%$ of the available run set), while the \emph{gri} dataset counts $6,884$ sources ($\sim17\%$ of the available run set).

\begin{table}\centering\caption[Magnitude cuts to build the run dataset]{Magnitude and magnitude error cuts adopted for the run dataset, for the \textit{u}, \textit{g}, \textit{r} and \textit{i} bands, deduced from error trends.}\label{tab:runcuts}
\begin{tabular}{p{1.8cm}P{1.1cm}P{1.1cm}P{1.1cm}P{1.1cm}}
\hline
Magnitudes& \emph{u}-band & \emph{g}-band & \emph{r}-band & \emph{i}-band \\\hline
\textit{AUTO} & & 23.7 & 23.0 & 23.0 \\\hline
\textit{APER4}& & 25.2 & 25.0 & 24.6 \\\hline
\textit{APER6}& & 24.2 & 24.1 & 23.6 \\\hline
\textit{APER8}& & 24.3 & 23.6 & 23.0 \\\hline\hline
Errors& \emph{u}-band & \emph{g}-band & \emph{r}-band & \emph{i}-band \\\hline
\textit{AUTO}& 0.18 & 0.040 & 0.040 & 0.050 \\\hline
\textit{APER4}& 0.04 & 0.050 & 0.050 & 0.070 \\\hline
\textit{APER6}& 0.05 & 0.035 & 0.034 & 0.055 \\\hline
\textit{APER8}& 0.18 & 0.033 & 0.030 & 0.050 \\\hline
\end{tabular}
\end{table}

We have performed the \textit{run} process with the GNG whose learning had involved all the three source types, i.e. galaxies, stars and GCs. Thus the GNG models trained with the \textit{GCSTAR} have been excluded from the \textit{run} process. 
Indeed the \textit{run} set is composed by all sources detectable from the instrument, so, when a galaxy is presented to this network, the model tries to assign a label to the source, i.e. \textit{star} or \textit{GC}, making a mistake in both cases. The purpose of the \textit{GCSTAR} experiments is to test the effectiveness of the network to photometrically disentangle GCs from stars, which is the most complex among the proposed problems, due to the morphological and photometric similarity of both source types. 

Since a \textit{leave-k-out} approach has been adopted, we used the five available trained networks to analyze the GNG performance fluctuations.
Table~\ref{tab:intersection} shows the results of the intersection between the different results produced by the GNG in terms of common predictions among stars, GCs, and galaxies sources: as expected from the blind test performance (Table~\ref{tab:results}), the common percentages reveal a gap between the \emph{ugri} and \emph{gri} \textit{3CLASS} experiments; nevertheless the other three run experiments reach more than $90\%$ of common predictions, finding about $500$ GCs candidates.

\begin{table}\centering\caption[Common predictions among the GNG trained networks performed on the run datasets]{Common predictions among the GNG trained networks performed on the unlabeled sources for the 4 dataset experiments involving the three class types: upper rows refer to \textit{3CLASS} experiments, while bottom rows refers to \textit{GCALL} experiments.}\label{tab:intersection}
\resizebox{\columnwidth}{!}{
\begin{tabular}{cccccc}
\hline
\textit{3CLASS} & COMMON & \% & GCs & stars & galaxies \\\hline
\textit{ugri} & 5115 & 92.0 & 522 & 2022 & 2571 \\\hline
\textit{gri} & 5861 & 85.1 & 425 & 2601 & 2835 \\\hline\hline
\textit{GCALL} & COMMON & \% & GCs & \multicolumn{2}{c}{notGC} \\\hline
\textit{ugri} & 5228 & 94.0 & 472 & \multicolumn{2}{c}{4756} \\\hline
\textit{gri} & 6437 & 93.5 & 790 & \multicolumn{2}{c}{5647} \\\hline
\end{tabular}
}
\end{table}

After having verified the robustness of the method with respect to the dataset variations and that the results seem to reflect the performances achieved on the blind test set, we trained the GNG on the whole KB. In order to quantify the overall performances of the network, we used samples of \textit{bona fide} Hubble space Telescope (HST) GCs in the central region of \textit{NGC1399} \citep{brescia:2012, cavuoti2013b, puzia2014}. After a cross-match between VST and HST catalogues, we found $~100$ HST sources (GC and \textit{not}GC) within our run dataset. The resulting classification statistics are shown in Table~\ref{tab:hstresults}. Despite the reduced number of samples, the measures reflect what was obtained on the blind tests: an increase of the classification accuracy for the \textit{ugri}-band dataset with respect to the \textit{gri} case, and for the \textit{GCALL} experiments with respect to the \textit{3CLASS} case. 

We want to emphasize the result obtained with the experiment \textit{ugri GCALL}, which outperforms the others, reaching an excellent purity-completeness trade-off and a very low GC-\textit{not}GC contamination ($\lesssim 4\%$). Thus, given these measures and the performances achieved on the blind test set, we used the GCs identified by the GNG trained with \textit{ugri GCALL} dataset. Furthermore, since we were also interested in other source types, we defined as stars (or galaxies) the sources classified as \textit{not}GCs trained with the \textit{ugri GCALL} experiments, which have been predicted as stars (or galaxies) by GNG trained with the \textit{ugri 3CLASS} experiment, i.e. the common prediction among the \textit{ugri} experiments.

\begin{table}\centering\caption[Classification results in term of statistical estimators using HST samples as bona fide]{Classification results in terms of statistical estimators using HST samples as \textit{bona fide} for both \emph{ugri} and \emph{gri} datasets. The estimator nomenclature is the same as that adopted in Table~\ref{tab:results}. 
The columns \textit{3CLASS} shows the results concerning the $3$-class experiments, while the columns \textit{GCALL} shows the results concerning the $2$-class. Despite the limited amount of labeled sources within the HST sample, the statistical estimators reflect the performance obtained with the blind test (Table~\ref{tab:results}).}\label{tab:hstresults}
\resizebox{\columnwidth}{!}{
\begin{tabular}{lcccc}
\hline
\multicolumn{1}{c}{} & \multicolumn{2}{c}{\textbf{ugri}} & \multicolumn{2}{c}{\textbf{gri}} \\
\hline
Estimator [\%]& \textit{3CLASS} & \textit{GCALL} & \textit{3CLASS} & \textit{GCALL} \\\hline
AE & 90.1 & 96.7 & 80.4 & 85.5\\\hline
pur \textit{not}GCs & 92.7 & 97.7 & 86.2 & 89.1 \\\hline
compl \textit{not}GCs & 86.4 & 95.6 & 64.1 & 81.7 \\\hline
F1 \textit{not}GCs & 89.4 & 96.7 & 75.1 & 85.4 \\\hline
pur GCs & 88.0 & 95.6 & 77.8 & 82.3 \\\hline
compl GCs & 93.6 & 97.8 & 92.5 & 89.1 \\\hline
F1 GCs & 90.8 & 95.6 & 85.1 & 85.7 \\\hline
\end{tabular}
}
\end{table}

\begin{figure}
\includegraphics[width=0.98\columnwidth]{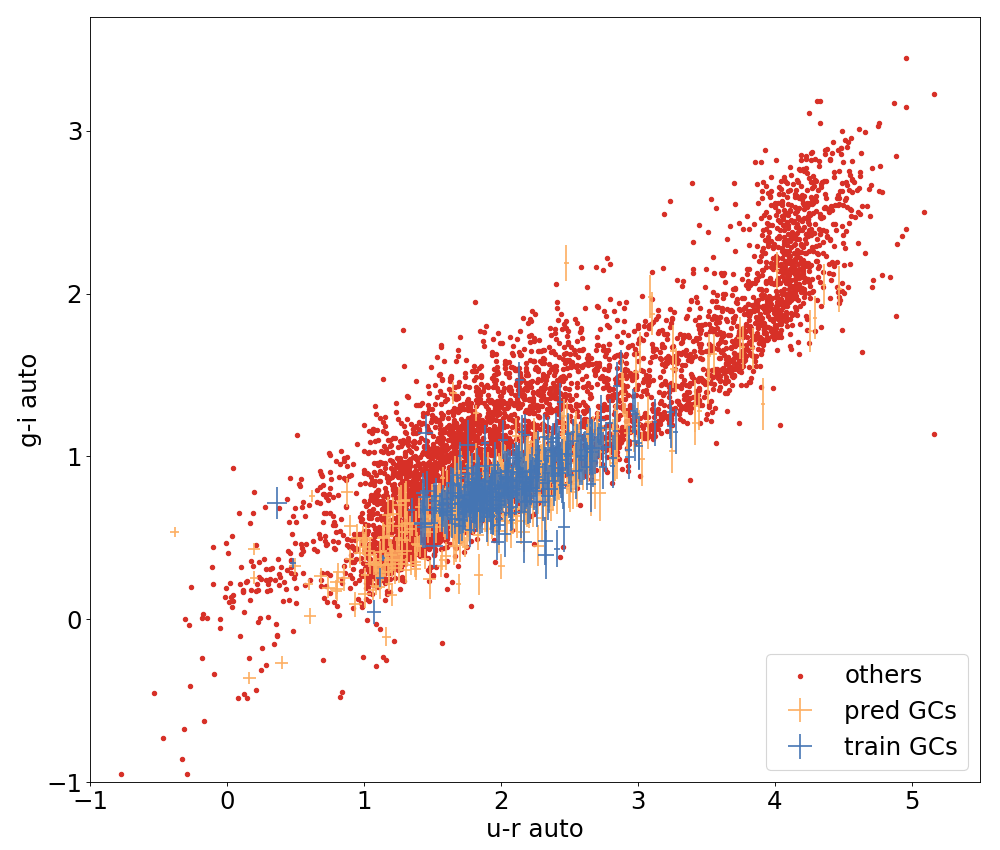}
\includegraphics[width=0.98\columnwidth]{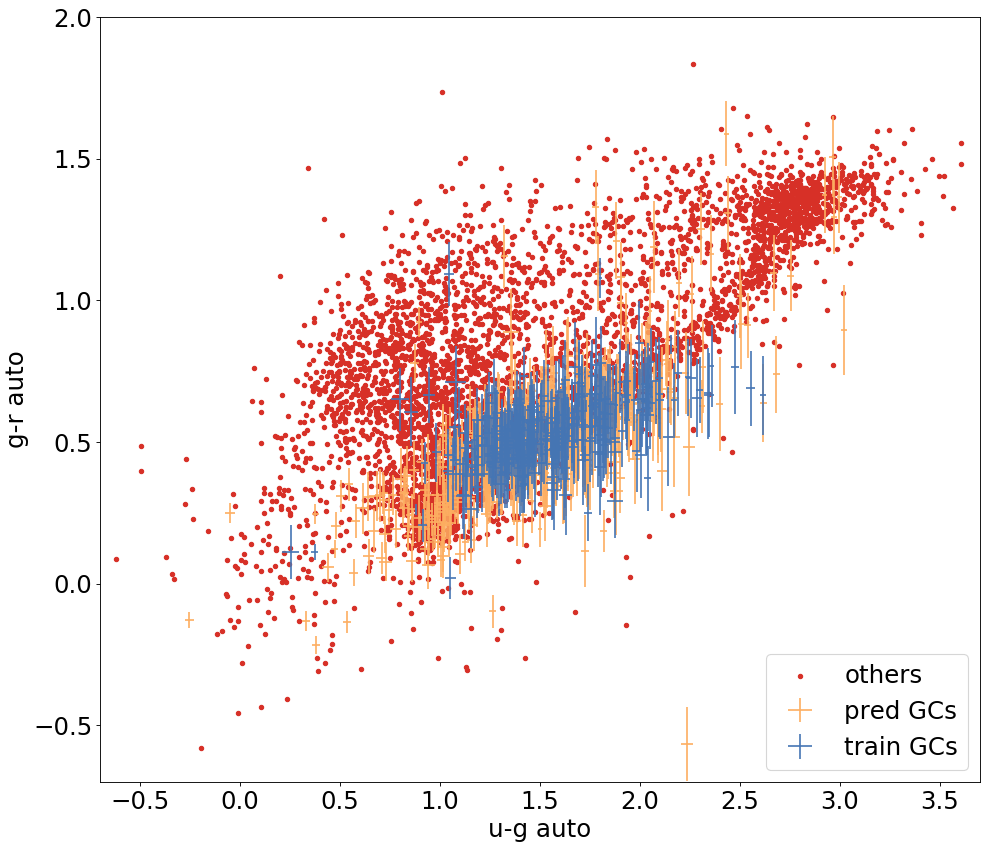} 
\includegraphics[width=0.98\columnwidth]{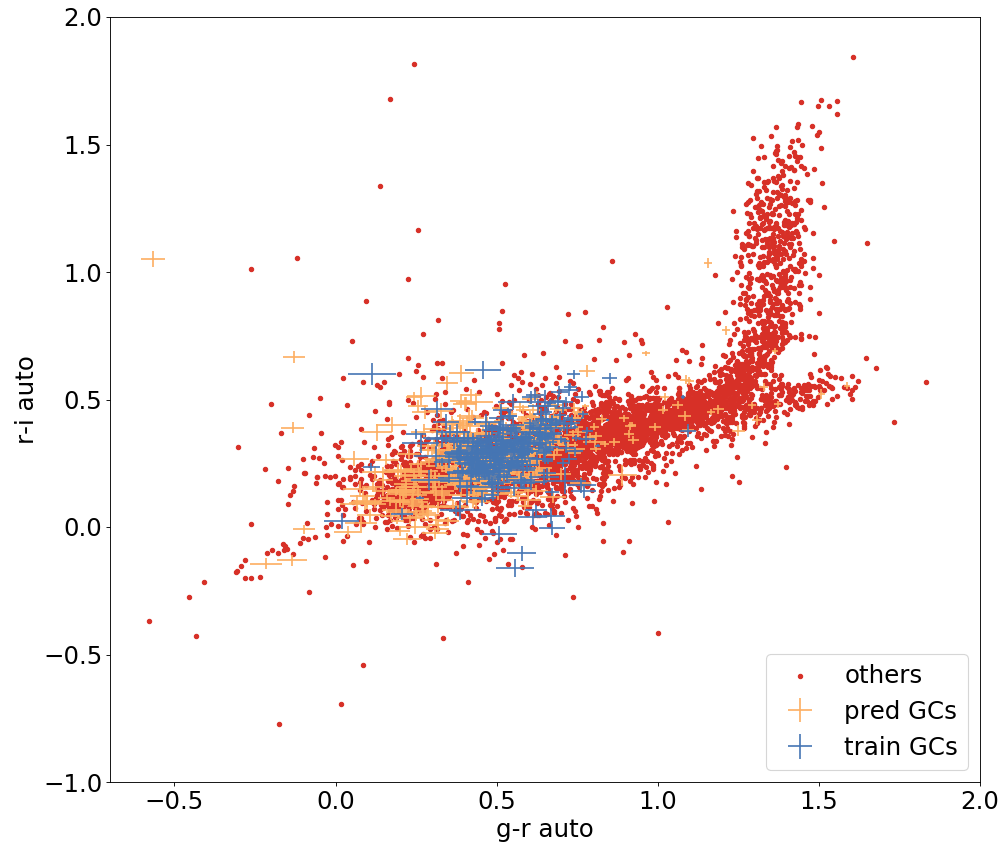} 
\caption[Colours for the run predicted sources compared with the train sources]{Colour-colour diagrams for the run predicted sources (orange) compared with the train sources (blue), overlapped to the color-color distribution related to the other sources (i.e. galaxies and stars, red dots in figures). From the top panel to the bottom the figures refers to the colours, respectively, \textit{g-i} vs \textit{u-r}, \textit{g-r} vs \textit{u-g} and \textit{r-i} vs \textit{g-r}.}\label{fig:RUN:colours}
\end{figure}

The resulting colour-colour diagrams for the predicted GCs are illustrated in Fig.~\ref{fig:RUN:colours} together with the other sources (i.e. galaxies and stars). These panels clearly show a large overlap between predicted and training GCs probing the network capability to extract the GC population, photometrically indistinguishable from the background and foreground sources. The results confirm the capability of our method, able to identify the sources without any pre-selection and in spite of the limited number of labelled sources. The only requirement is the correspondence, in term of photometric coverage, between the training and the run datasets.

\begin{figure*}\centering
\includegraphics[width=2\columnwidth]{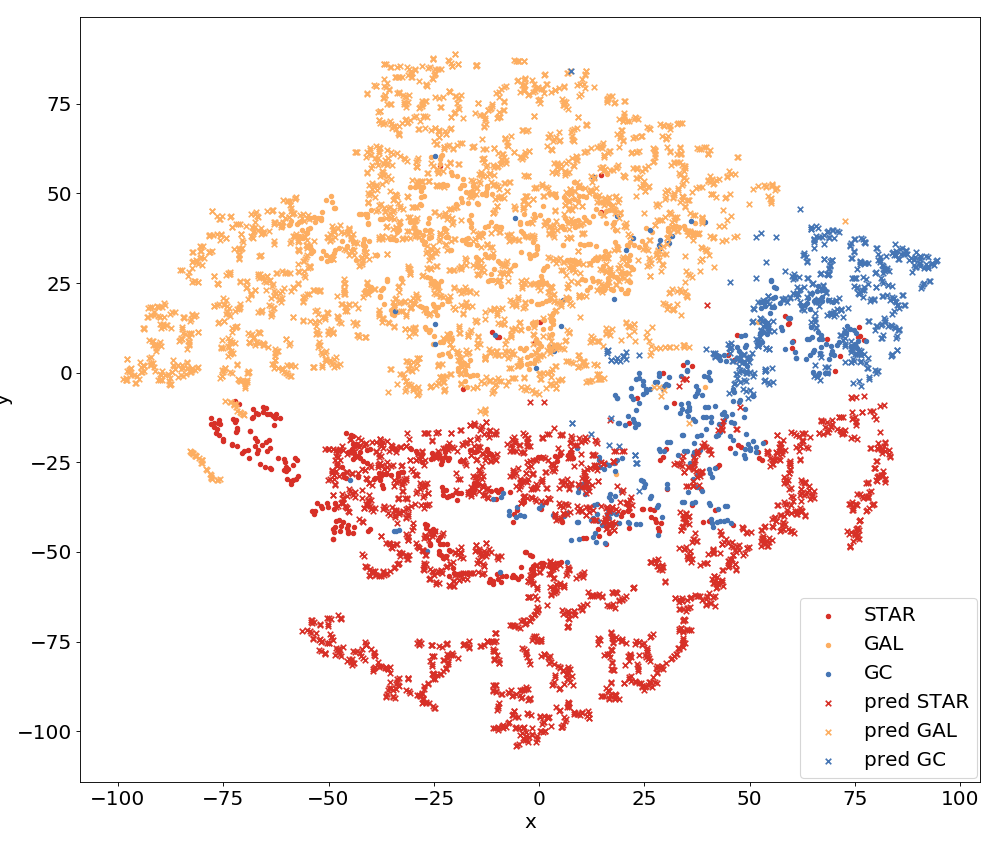}
\caption{Bi-dimensional projection performed by tSNE of the \textit{ugri} set both for spectroscopic sources (blue, red and orange dots, respectively for GCs, stars and galaxies) and for predicted candidate objects (blue, red and orange crosses, respectively for GCs, stars and galaxies).}\label{fig:RUN:tsne}
\end{figure*}

Concerning the residual misclassified objects, some of these could be false positives (FPs). 
This misclassification could be due to the exiguous number of training sources or a non-uniform sampling of the parameter space. In the top panel of Fig.~\ref{fig:gngmlp:tsne} we have shown a visualization of the \textit{ugri} train set into a bi-dimensional space through the \textit{ECODOPS} tool. Although this is a projection, it is evident that the space is not uniformly sampled and characterized by the presence of several contaminants. Such two factors, together with the exiguous number of training sources, could cause the presence of the outliers. Thus, in order to visualize the result of the run process, we estimate the bi-dimensional projection of the run set, analogous to what was already done in Sec.~\ref{ss:GNGclass}. In Fig.~\ref{fig:RUN:tsne} we show this same projection by overlapping the training set objects. Most of the predicted sources seem to populate well-defined regions, predominantly occupied by spectroscopic objects, although stars and GCs show a large overlap, as expected. This aspect, together with the already discussed problems of the misclassification at the border of class regions, may cause the presence of redder and bluer outliers.\\
Given such premises, by considering also the low fraction of spectroscopic sources available, a larger fraction of FPs could be expected. However the resulting exiguous number of contaminants is a consequence of the approached PS optimization process.

\begin{figure}\centering
\includegraphics[width=\columnwidth]{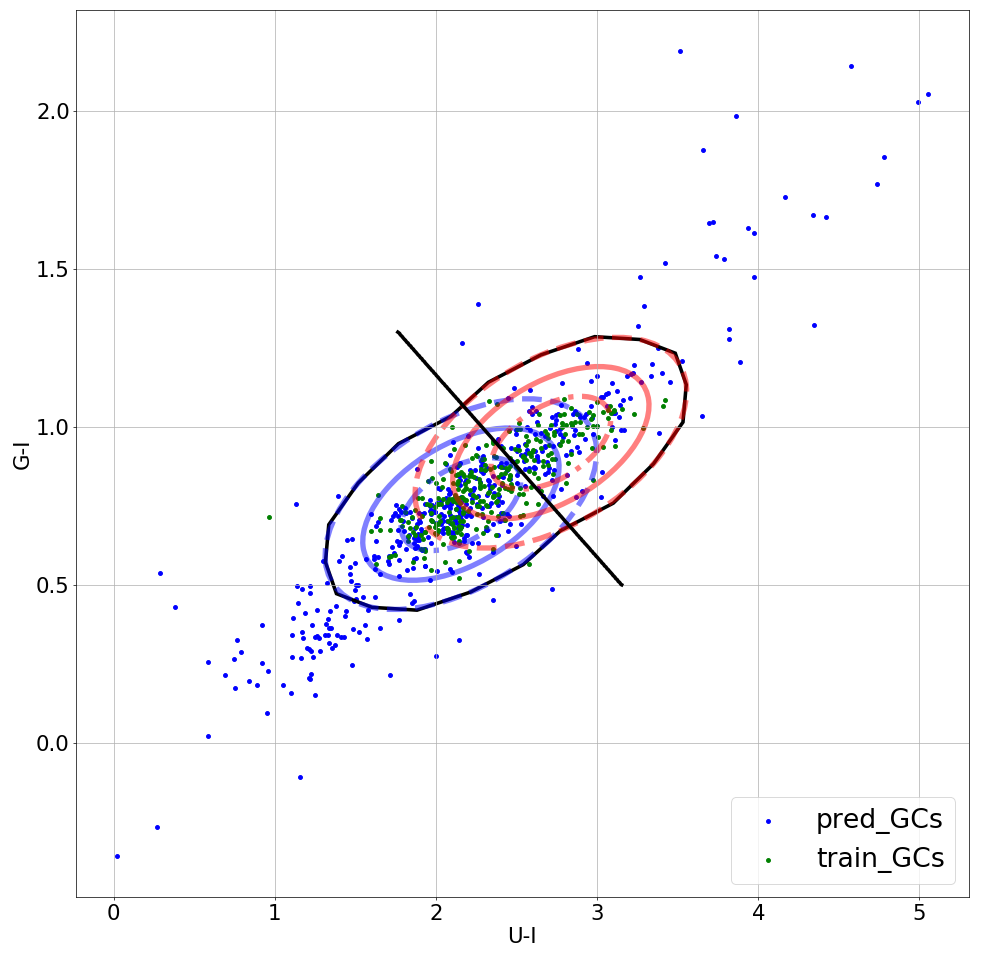}
\includegraphics[width=\columnwidth]{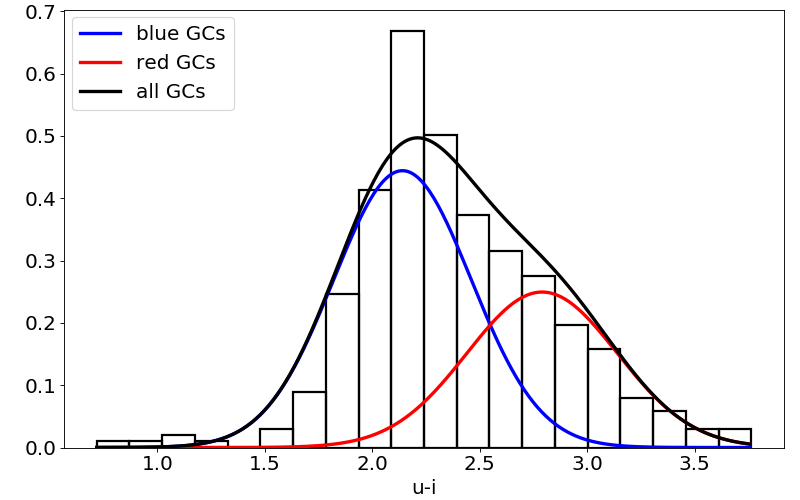}
\includegraphics[width=\columnwidth]{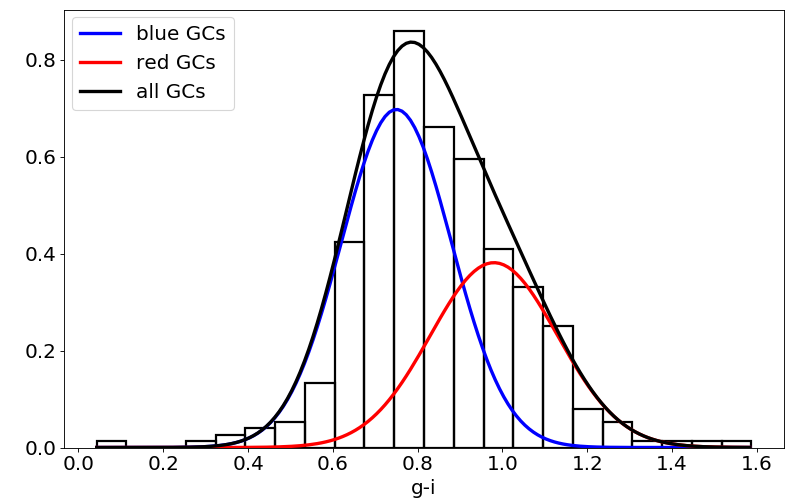}
\caption[Projection of the bivariate bimodal Gaussian distribution on the colour-colour plane and the colour distribution]{Top panel: projection of the bivariate bimodal Gaussian distributions on the colour-colour plane fitting the training GCs (in green). The predicted GCs are scattered in blue. The ellipses are the contours levels related to $1\sigma$, $2\sigma$ and $3\sigma$, respectively, for the red and blue population. All sources within the union of the $3\sigma$ ellipses are assumed to be \textit{true} GCs. The black line is the projected intersection between the bivariate Gaussians. Middle and bottom panels: colour distribution (\textit{u-i} and \textit{g-i}) for the bimodal GC population.}\label{fig:GMM:result}
\end{figure}

Clearly these outliers must be excluded from the set of the identified GCs. A simple approach could be excluding sources from the colour-colour diagrams. However given the arbitrariness of such procedure, we neglected this solution. In order to exclude such candidate FPs, we took advantage from the existence of the two GC populations (namely red and blue, \citealt{kundu1998}), to fit a bimodal bivariate Gaussian distribution deduced from the spectroscopic GCs. We used a Gaussian Mixture Model (GMM, \citealt{muratov2010}) implemented through the library \textit{sklearn} \citep{sklearn:2011}, which is a generalization of the K-Mixture Model \citep{KMM1994}. The method maximizes the likelihood of the dataset using the expectation-maximization (EM) algorithm, which allows to derive explicit equations for the maximum likelihood estimate of the parameters. 
The projection of this surface on the colour-colour plane is illustrated in the top panel of Fig.~\ref{fig:GMM:result}. The red and blue ellipses symbolize the contour levels matching, respectively, $1\sigma$, $2\sigma$ and $3\sigma$ of the bivariate bimodal Gaussian distribution of the underlying spectroscopic GC population. The black line crossing the ellipses is the projection on the colour-colour plane of the intersection between the two bivariate Gaussian surfaces. Training (i.e. spectroscopic) and predicted sources above such line and within the $3\sigma$ levels are considered components of the red population, while those below and within the $3\sigma$ levels are assumed to be members of the blue population. Finally, we assume as FPs the predicted sources outside the union of the $3\sigma$ ellipses ($110$ objects, $\sim 23.3\%$). Middle and bottom panel of Fig.~\ref{fig:GMM:result} show the evident bimodal colour distributions of the GC population. Once the FPs have been excluded from the GC set, the intracluster error \citep{Floudas:2006, Murtagh2014}, defined as the measure of the overlap between training and predicted GCs, decreases by $\sim5\%$ while variance drops down by about $80\%$.

\begin{figure}\centering
\includegraphics[width=0.95\columnwidth]{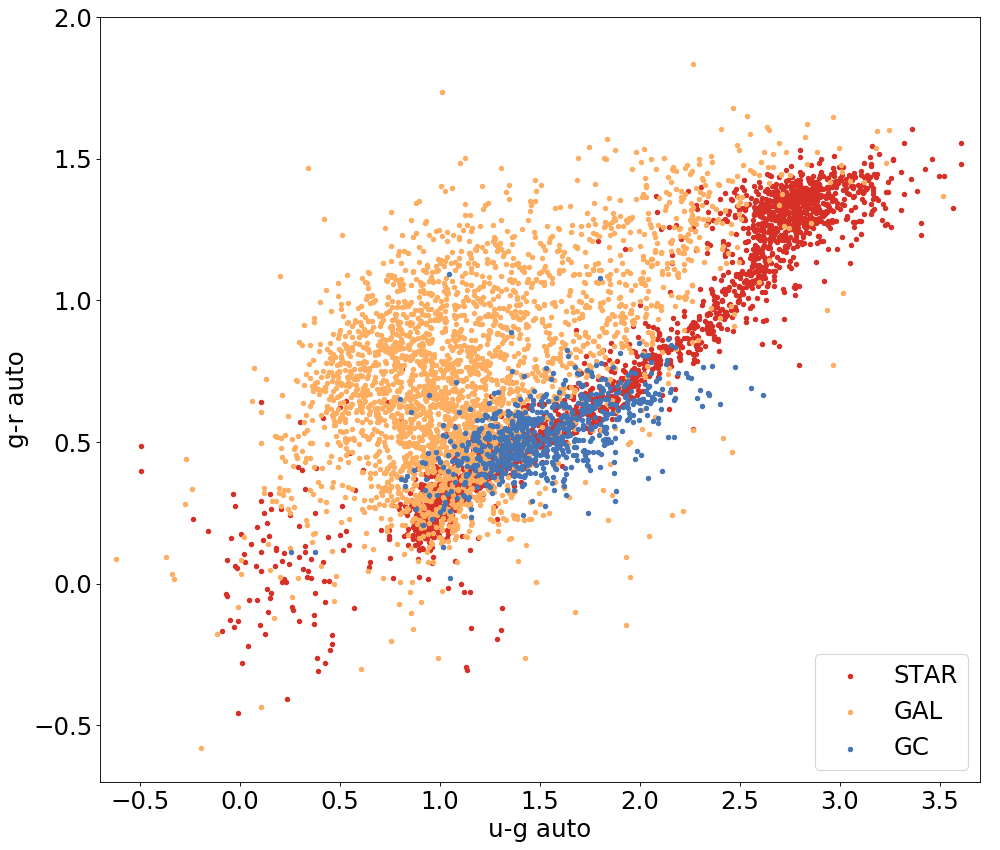}
\includegraphics[width=0.95\columnwidth]{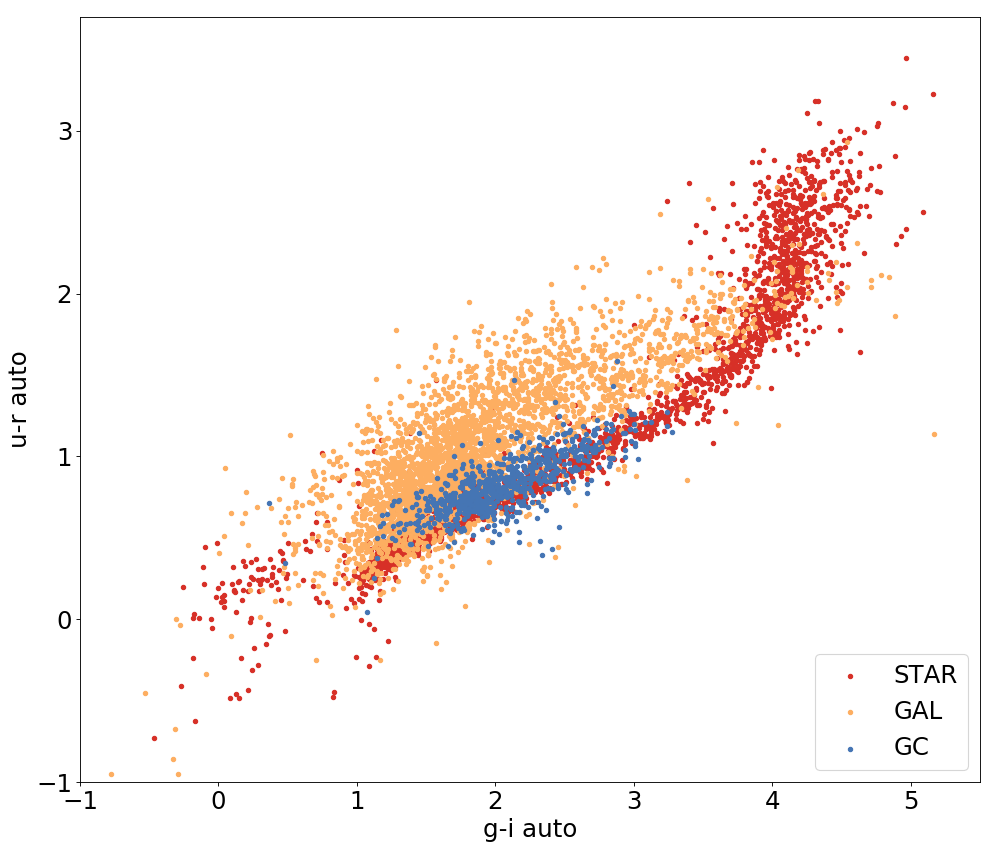}
\includegraphics[width=0.95\columnwidth]{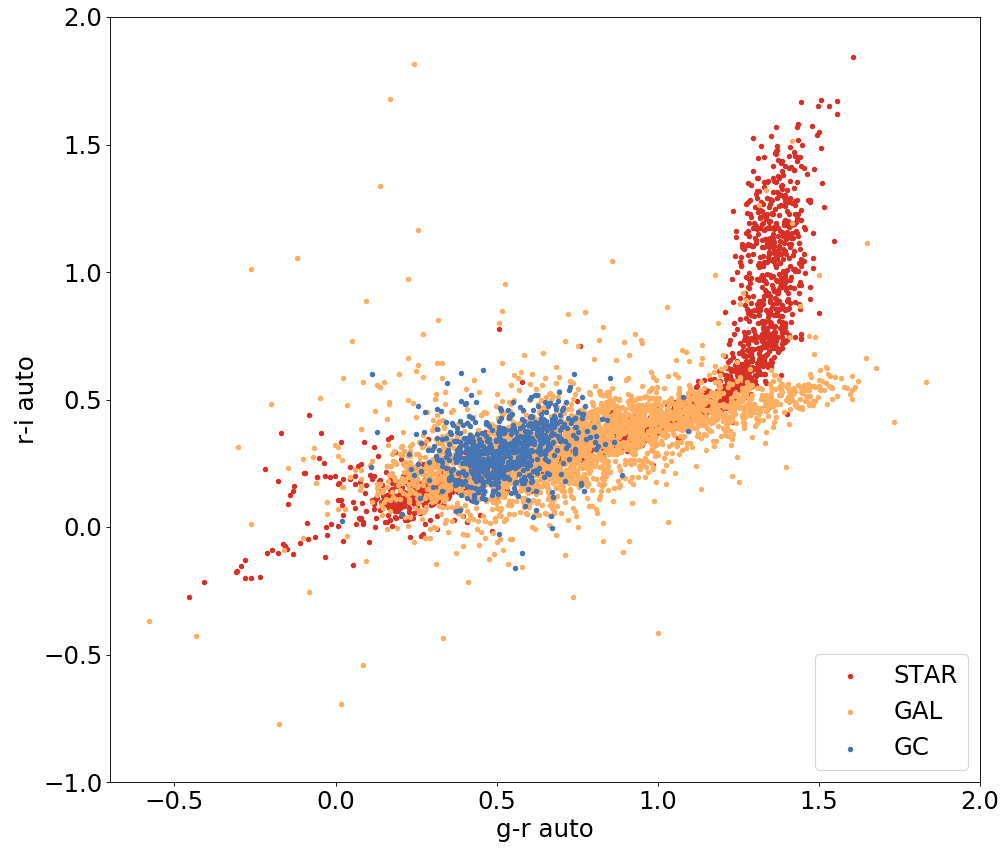}
\caption[Colour diagrams for stars, galaxies and GCs]{Colour diagrams for stars, galaxies and GCs downstream the GMM false positive exclusion process. From the top panel to the bottom one the figures show: \textit{g-r} versus \textit{u-g}, \textit{r-i} versus \textit{u-g}, \textit{r-i} versus \textit{g-r}. In these figures the GCs are blue, the stars are red and the galaxies are orange. The diagrams recall the results obtained in others works \citep[e.g.][]{can2018}.}\label{fig:finalcolours}
\end{figure}

Panels in Fig.~\ref{fig:finalcolours} illustrate the colour-colour diagrams for the selected GCs, galaxies and stars. Concerning the branch of the stars, some bluer sources could be false positives, since their distribution appears to be the extension of the galaxies trend. However, the stellar branch goes through the diagrams following the expected shapes and, above all, the GCs occupy the restricted region well known in literature (e.g. \citealt{can2018}), in which GCs are particularly difficult to be separated from stars or galaxies through colours.

\section{Comparison with external data}\label{sec:comp}
In order to validate the GCs identification through the GNG-GMM approach, we compare our selection with other similar works: 
\begin{itemize}
    \item[-] Machine Learning experiments performed by \cite{brescia:2012} and more recently 
    by \cite{angora2017} on single-band Hubble Space Telescope (HST) data of \textit{NGC1399}. This comparison, presented in Sec.~\ref{ss:hstcomp}, analyses the performances achieved by the same methods (GNG and MLPQNA) varying the instruments (VST vs HST);   
    \item[-] Other experiments carried out by \cite{dabrusco:2016} and \cite{cantiello2018}, which exploit techniques different from Machine Learning methods on the same VST dataset. In this case we compare performances obtained by different approaches (ML vs not-ML) using the same instruments (discussed in Sec.~\ref{ss:tradcomp}.
\end{itemize}

\subsection{Comparison with HST data}\label{ss:hstcomp}
The catalogue used by \cite{brescia:2012} and \cite{angora2017} was extracted from single-band Hubble Space Telescope (HST) images of \textit{NGC1399}, reaching $7\sigma$ at m$_V=27.5$, that is $\sim3.5$ magnitudes fainter than the GC luminosity function turnover point, thus it allows the sampling of nearly the entire GC population \citep{puzia2014}. The parameter space is composed by seven photometric features, respectively, four magnitudes (isophotal and three different apertures), \textit{FWHM}, \textit{central surface brightness}, \textit{Kron radius} and four structural parameters, respectively, \textit{King's tidal}, \textit{effective radius}, \textit{core radius} and \textit{ellipticity}.         

\begin{table}\centering\caption[Comparison between classification performed with HST and VST data]{Comparison between classification performed with HST and VST data. The results related to the HST data obtained by the MLPQNA and GNG are derived, respectively, from \cite{brescia:2012} and \cite{angora2017}. The estimators are referred to the GCS. In order to measure the amount of correctly classified GCs, the  contamination (dual estimator of the purity) has been used.}
\begin{tabular}{P{1.2cm}p{2.0cm}P{1.4cm}P{1.4cm}}
\hline
&Estimator [\%] &MLPQNA&GNG\\\hline
\multirow{3}{*}{HST}& AE & 98.3 & 86.8 \\\cline{2-4}
&completeness& 97.8 & 83.8 \\\cline{2-4}
&contamination& 1.6 & 15.9\\\hline\hline
\multirow{3}{*}{VST}& AE & 87.1 & 88.7\\\cline{2-4}
&completeness& 81.8 & 89.1\\\cline{2-4}
&contamination& 21.1 & 8.7\\\hline
\end{tabular}\label{tab:HSTVST}
\end{table}

Table~\ref{tab:HSTVST} reports a comparison between the best results obtained by the MLPQNA and the GNG networks on the HST and VST data (the latter extracted from Table~\ref{tab:results}). Instead of the purity, the contamination (i.e. the complementary of purity, $1-$\textit{purity}) was used to evaluate the capability of the ML models to correctly classify the GCs. 
MLPQNA achieves a remarkable result on the HST data, with a contamination of $\sim1.8\%$ (thus corresponding to a purity of $\sim98.2\%$). 
The GNG performances appear to be similar by increasing the accuracy in the case of VST data. The statistics suggest the capability of ML models to disentangle the GCs from the background and foreground sources also with high-quality single-band photometry.
In order to investigate such result, the k-fold based training/test procedure, described in Sec.~\ref{sec:exp}, has been reproduced for VST data using all possible filter combinations and varying the involved number of bands for all the classification experiments.

\begin{figure}\centering
\includegraphics[width=\columnwidth]{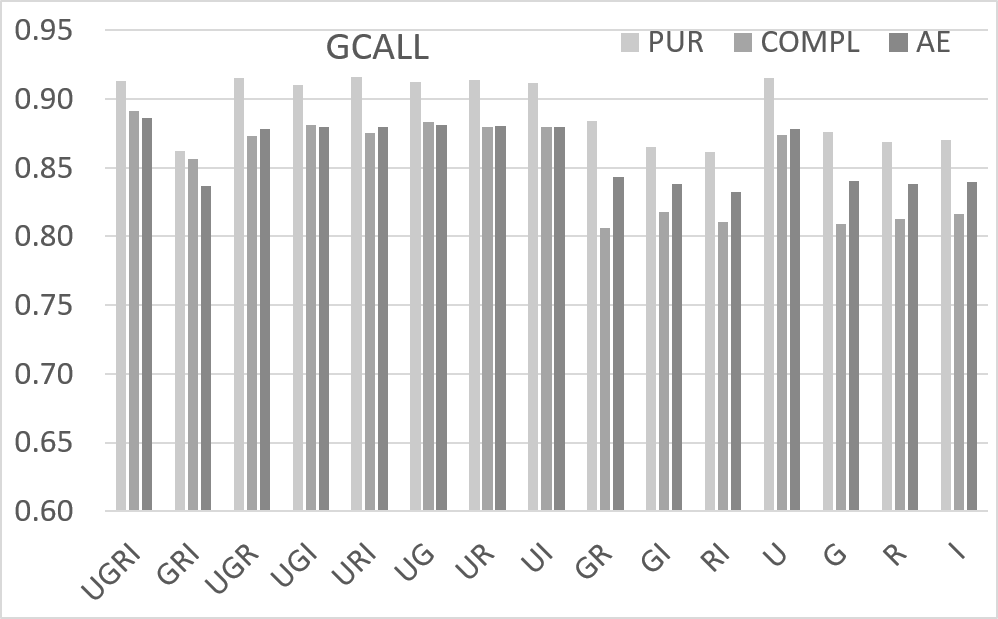}
\includegraphics[width=\columnwidth]{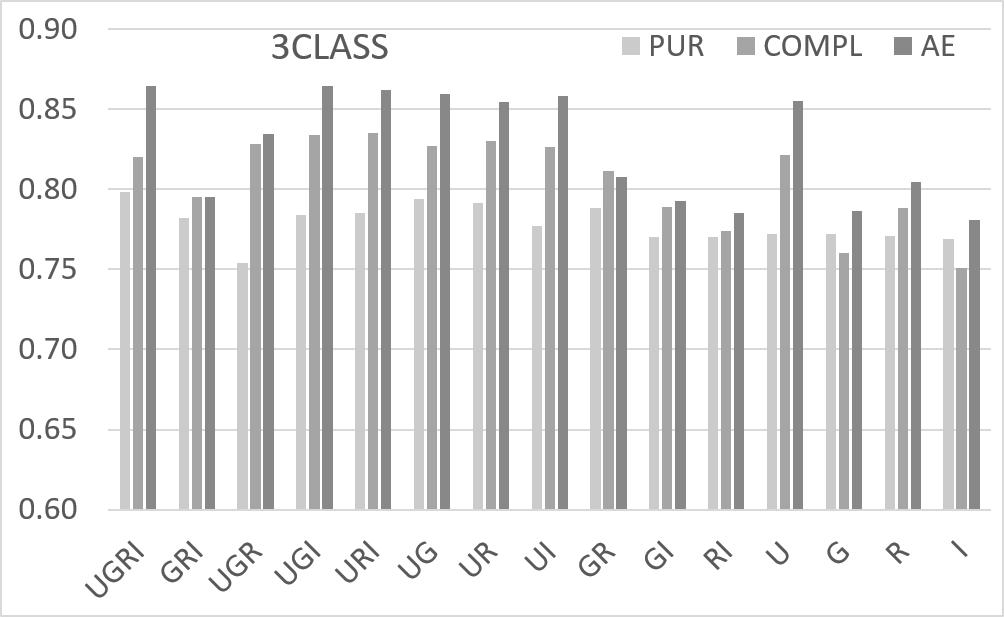}
\includegraphics[width=\columnwidth]{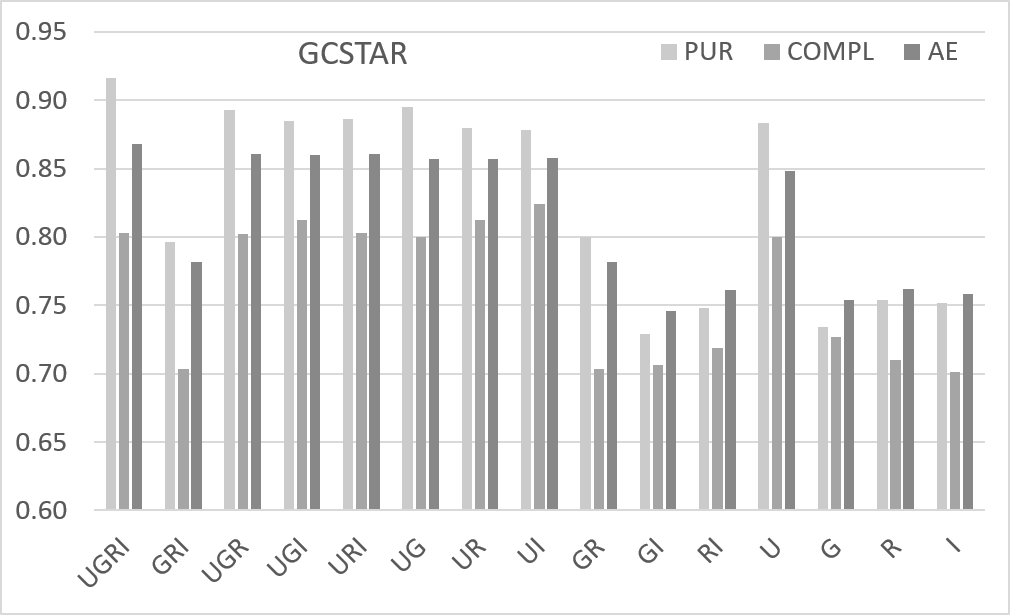}
\caption[Comparison among blind test performances, obtained by the GNG network, as function of different bands involved, in terms of purity, completeness and average efficiency]{Comparing among GNG network blind test performances as function of the involved filter numbers in terms of GCs purity (\textit{pur}, light grey), completeness (\textit{compl}, grey), and average efficiency (\textit{AE}, dark grey) for the experiments \textit{GCALL} (top panel), \textit{3CLASS} (middle panel) and \textit{GCSTAR} (bottom panel). Experiments involving the \emph{U}-band improve significantly their purity (more than $5\%$ in the \textit{GCALL} experiment), their completeness (more than $7\%$ in the \textit{GCALL} experiment) and their average efficiency. This latter has been included to add information about the \emph{not}GCs classification.}\label{fig:PERFORMANCES}
\end{figure}

Panels in Fig.~\ref{fig:PERFORMANCES} show the performances of the GNG model as a function of the used band filters, probing, as expected, the effectiveness of the complete spectrum dataset in order to identify GCs with ground-based imaging. 

The \emph{ugri} case has the best trade-off between purity and completeness (related to the GCs) in the three classification experiments, i.e. the model is able to correctly identify the GCs with an acceptable level of precision and sensitivity. Since the purity and the completeness are referred to GCs, the average efficiencies have been displayed in Fig.~\ref{fig:PERFORMANCES}, in order to include classification information about the \textit{not}GCs. Concerning the incomplete spectrum datasets, the experiment involving the \emph{U}-band improves the purity by $\sim0.6\%$ to $\sim6.3\%$, the completeness by $\sim3.1\%$ to $\sim7.7\%$ and the average efficiency by $\sim3.8\%$ to $\sim9.4\%$. Therefore, there is a significant performance gap between experiments with or without the \emph{U}-band. 

Comparing the accuracy reached by the GNG network on the HST and the single band VST data, the experiment based on the \emph{u} band is the only one outcoming comparable results. Only using the complete spectrum the GNG achieves better performance than single-band HST GNG experiments, improving purity, completeness and efficiency (by $\sim8.3\%$, $\sim5.3\%$ and $\sim1.8\%$, respectively, for the \textit{GCALL} experiment).  

\cite{brescia:2012} probed the capability of several ML methods to disentangle GCs from background and foreground sources using single band, high-quality, deep photometry. To achieve similar results on ground-based VST data, it is necessary to use at least two filters, where one of them has to be the \emph{u}-band. 

Regarding the feature selection procedure, it is possible to compare results obtained by \cite{angora2017} with HST data, which added a feature selection procedure provided by the Random Forest model, finding a set of relevant features composed by only photometric quantities, from which the \textit{Kron radius} and the \textit{ellipticity} were also rejected. Despite some differences between the datasets, for instance, the magnitude coverage, depth and number of bands, the FS results are similar to those performed through $\Phi$Lab. 
In fact, in both cases the \textit{Kron radius} is rejected, showing a negligible informative contribution ($<1\%$). Concerning the \textit{ellipticity}, defined as $1-\mathit{B\_WORLD}/\mathit{A\_WORLD}$, it is related to another $\Phi$LAB rejected feature, the \textit{ELONGATION}, connected to the \textit{ELLIPTICITY} through: $\mathit{ELLIPTICITY}=1-(1/\mathit{ELONGATION})$, \cite{bertin:1996}. Therefore, its rejection is motivated by the informative contribution already carried by the \textit{ELONGATION} feature.

\subsection{Comparison with other techniques}\label{ss:tradcomp}

\begin{table*}
\centering
\caption[Bivariate bimodal Gaussian parameters fitted through the GMM method]{Bivariate bimodal Gaussian parameters fitted through the GMM method compared with parameters fitted by \cite{dabrusco:2016} and by \cite{cantiello2018}. Values in the last row of the table refer to the discrepancies of distribution peaks that are consistent within $(0.8, 3.1)\sigma$. These discrepancies have been estimated as $t=|\mu_1-\mu_2|/\sigma$ \citep{taylor1986}.}\label{tab:GMM:fit}
\begin{tabular}{cccccccccccccc}
\multicolumn{1}{c}{} & \phantom{ciao}& \multicolumn{4}{c}{GMM} & \phantom{ciao} &\multicolumn{2}{c}{$\,\,$\citeauthor{dabrusco:2016}}& \phantom{ciao}&\multicolumn{4}{c}{$\qquad$\citeauthor{cantiello2018}}\\
\multicolumn{1}{c}{} && \multicolumn{2}{c}{blue} & \multicolumn{2}{c}{red}&& blue & red && \multicolumn{2}{c}{blue} & \multicolumn{2}{c}{red}\\
\cline{3-14}
&& \textit{u-i} & \textit{g-i}& \textit{u-i} & \textit{g-i} && \multicolumn{2}{c}{\textit{g-i}} &&\textit{u-i} & \textit{g-i} & \textit{u-i} & \textit{g-i}\\\hline
p && \multicolumn{2}{c}{0.63} & \multicolumn{2}{c}{0.37} && 0.63 & 0.37 && 0.52 & 0.53 & 0.48 & 0.47 \\\hline
N && \multicolumn{2}{c}{446} & \multicolumn{2}{c}{273} && $1853$ & $1095$ && 78& 79 & 71 & 70 \\\hline
$\mu$ && 2.14 & 0.75 & 2.79 & 0.98 && 0.74 & 0.95 && 2.08 & 0.78 & 2.74 & 1.06\\\hline
$\sigma$ && 0.32 & 0.13 & 0.35 & 0.15 && 0.08 & 0.12 &&0.34& 0.11& 0.55& 0.17\\\hline\hline
&\multicolumn{6}{l}{$\sigma$'s discrepancy between peaks} & 1.1 & 3.0 && 1.6 & 1.1 & 0.8 & 3.1\\
\hline
\end{tabular}
\end{table*}

With the term \emph{other} we refer to those methodologies that do not exploit ML, and use instead several combinations of cuts in a more or less complex parameter space, to separate GCs from background/foreground sources. 
We have compared our results with two works, respectively, \cite{dabrusco:2016} and \cite{cantiello2018}, which analyze the Fornax region with the same VST images. 

\cite{dabrusco:2016} applied a Principal Component Analysis (PCA, \citealt{Bishop:2006}) on the natural colours, identifying a \textit{locus} in the PC space dominated by the presence of GCs, excluding brighter and fainter sources (i.e. cuts on \textit{G}-band) and using the SExtractor \textit{CLASS\_STAR} parameter. 

\cite{cantiello2018} introduced a \textit{morpho-photometric} approach: in order to analyze the properties of the GC sample (bimodality, density maps, radial profiles), they use a statistical background decontamination method. Here, for simplicity, we compare our results to the catalogue provided by \citealt{cantiello2018}, being aware that this is oversimplified. \cite{cantiello2018} used a spectroscopic set of sources in order to find the parameter space occupied by GCs and applied a set of cuts on the features: $\Delta X$, \textit{CLASS\_STAR}$_X$, \textit{FWHM}$_X$, \textit{FLUX\_RADIUS}$_X$, \textit{KRON\_RADIUS}$_X$, \textit{PETRO\_RADIUS}$_I$, \textit{ELONGATION}$_X$, \textit{u-i}, \textit{g-i}, $\Delta(\mathit{u-i})$, $\Delta(\mathit{g-i})$, $\Delta m_X$, $m_X$. $X$ labels the \textit{g} and \textit{i} band, while $\Delta$ indicates the difference between two apertures (with respect to magnitudes and colours), respectively, $6$ and $12$. In addition they used a selection on the colour-colour plane: $|(\mathit{g-i})-[0.362(\mathit{u-i})-0.0205]|\leq0.2$. The high number of cuts derives from the need to reduce contamination introduced by peculiar sources. In order to study the colour bimodality, they selected GCs inside several annular regions, concentric on \textit{NGC1399}. For each GC set within the annular region they applied a GMM in order to fit a bimodal univariate Gaussian distribution using the \textit{u-i} and \textit{g-i} colour. Moreover, they compared the bimodal and the unimodal distributions, statistically validating the colour bimodality.    

Regarding the split between blue and red GCs, the blue GCs of \cite{dabrusco:2016} are the sources whose \textit{g-i} colours are less than $0.85$, while the blue GCs of \cite{cantiello2018} are those sources with \textit{u-i}$<2.5$. These thresholds are stated by authors in their respective works.

Table~\ref{tab:GMM:fit} shows the bimodal Gaussian best fit together with the parameters estimated by \cite{dabrusco:2016}, which applied a GMM method only for the colour \textit{g-i} of a larger GC set (their VST catalogue covers $\sim 8.4 deg^2$), and by \cite{cantiello2018}, which refer to an annular region whose radii are $2.5$ and $5$ $\arcmin$, respectively. 

Using the criterion of acceptability \citep{taylor1986}, we have estimated the number of standard deviations for which our measure ($\mu_1$) differs from that of \cite{dabrusco:2016} and (\citealt{cantiello2018}, $\mu_2$), i.e. $t = |\mu_1-\mu_2|/\sigma$. The resulting values are reported in the last row of Table~\ref{tab:GMM:fit}. The discrepancies with the peak of the blue sub-populations are less than $1.6\sigma$, i.e. the measures are comparable. 
Concerning the red sub-populations, only the peaks related to \textit{u-i} are compatible below $1\sigma$; the \textit{g-i} peaks differ from each other by at least $3\sigma$.
However, we point out that in absolute terms the observed differences are small ($<0.1$ mag) and they could be easily explained by noticing that the different studies sample different galactocentric distances; in particular the value of \citealt{cantiello2018} refers to an annulus within 5' from NGC1399 where the red GC component usually peaks at redder colours, and they did not use the \textit{r} band, so the samples are inherently different in terms of possible contamination.

\begin{table*}
\centering\caption[Comparing our predicted GCs]{Intersection between our prediction and GCs identified by \cite{dabrusco:2016} and \cite{cantiello2018}. Their prediction has been intersected with our set of spectroscopic confirmed GCs (column TRAIN) and with our predicted GCs (column PRED). The percentage refers to the amount of training and predicted sources, i.e. common sources divided by training (or predicted) GCs. The total number of common GCs, together with the fraction (i.e. common GCs divided by all GCs in our work), is reported in column TRAIN+TEST. Columns BLUE and RED are the common blue and red GCs; the percentage refers to the amount of blue and red GCs predicted in this work. The last column of the table (FP) indicates the intersection with the false positives (sources excluded from our prediction), i.e. the number of our removed GCs that \cite{dabrusco:2016} and \cite{cantiello2018} classified as GCs; the percentage refers to the amount of GCs labeled as FPs.}\label{tab:COMP}
\begin{tabular}{ccccccc}
& \multicolumn{6}{c}{Common sources identified as GCs in our work}\\\cline{2-7}
& TRAIN & PRED & TRAIN + PRED & BLUE & RED & FP\\\hline
\citeauthor{dabrusco:2016} &329& 322 & 651 & 322 &227 & 2\\
$[\%]$ & 92.2 & 88.9 & 90.5 &72.2 &83.2 & 2.3 \\\hline
\citeauthor{cantiello2018} &286 & 291 & 577 & 358 & 180 &29\\
$[\%]$ & 80.1 & 80.4 & 80.3 & 80.3 & 66.0 & 25.7\\\hline
\end{tabular}
\end{table*}

\begin{figure}\centering
\includegraphics[width=\columnwidth]{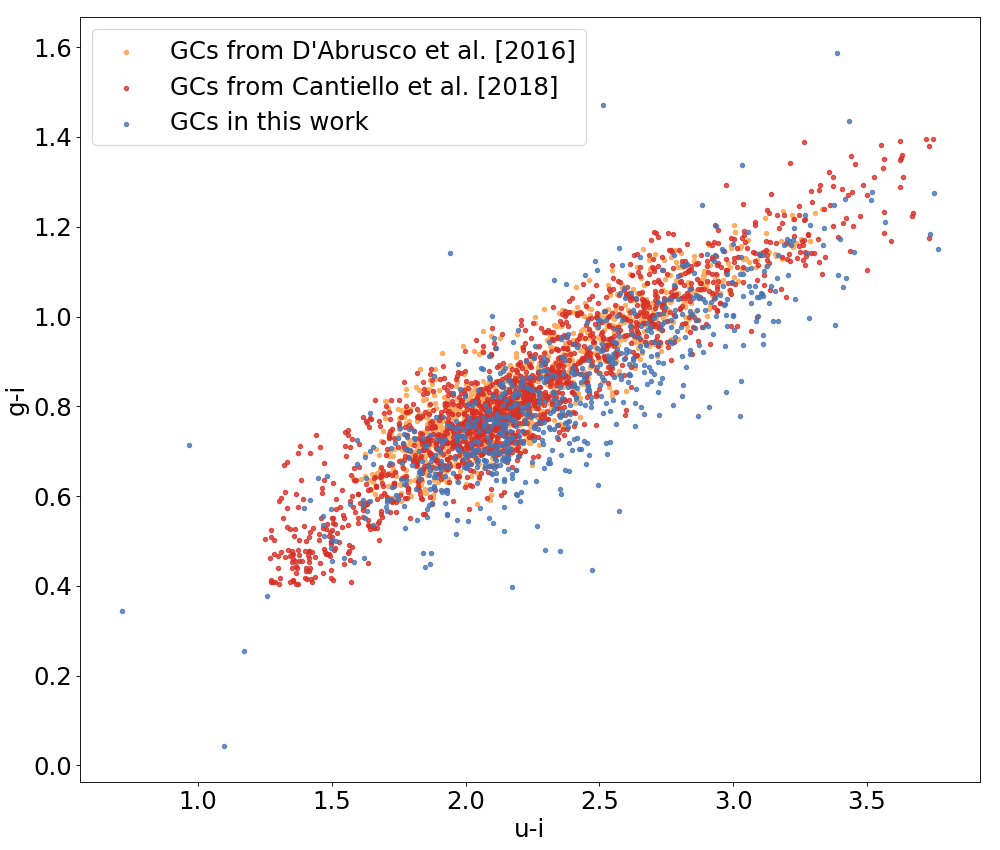}
\caption[\textit{G-I} vs \textit{U-I} diagram compares our predicted GC with the literature]{\textit{U-I} vs \textit{G-I} diagrams related to our predicted GC (blue) and those predicted by (\citealt{dabrusco:2016}, orange) and (\citealt{cantiello2018}, red).}\label{fig:COMP:colours}
\end{figure}

Fig.~\ref{fig:COMP:colours} shows the \textit{U-I} vs \textit{G-I} diagram related to the three sets of candidate GCs, and shows a large overlap between the sets. In Table \ref{tab:COMP} the common sources between our GC catalogue and the candidate GCs provided by \cite{dabrusco:2016} and \cite{cantiello2018} are reported. There are, respectively, $91\%$ and $80\%$ GCs in common, of which $72\%$ and $80\%$ are blue, $83\%$ and $66\%$ are red. This result confirms the capability of our method to predict the GC class type. Furthermore, among the excluded FPs, only $2$ and $29$ sources are candidate GCs, strengthening the FPs selection robustness, based on a GMM approach. 

The resulting common sources reflect the difference between the selection approaches. 
All the cited works, included this one, use a GMM best fit to model a bimodal Gaussian underlying the GC populations, but through a different colour fitting: \cite{dabrusco:2016} performed the fit with the \textit{G-I} colour; \cite{cantiello2018} uses both \textit{U-I} and \textit{G-I} colours separately, in this work we used both colours by modeling a bivariate bimodal Gaussian distribution and producing a less sharp cut on the colour distributions. 

Furthermore, we intersected the GCs provided by \cite{dabrusco:2016} and \cite{cantiello2018} with both our training and predicted \textit{not}GCs, i.e. stars and galaxies present in our training set together with those predicted by our method. From the intersection with the training stars, resulted $86$ and $126$ sources, respectively for \cite{dabrusco:2016} and \cite{cantiello2018}, while only $2$ and $1$ sources resulted from the intersection with the training galaxies. Since the training \textit{not}GCs are spectroscopically confirmed, these sources represent a set of candidate FPs for \cite{dabrusco:2016} and \cite{cantiello2018}.   

\begin{figure*}
\centering
\includegraphics[width=2.\columnwidth]{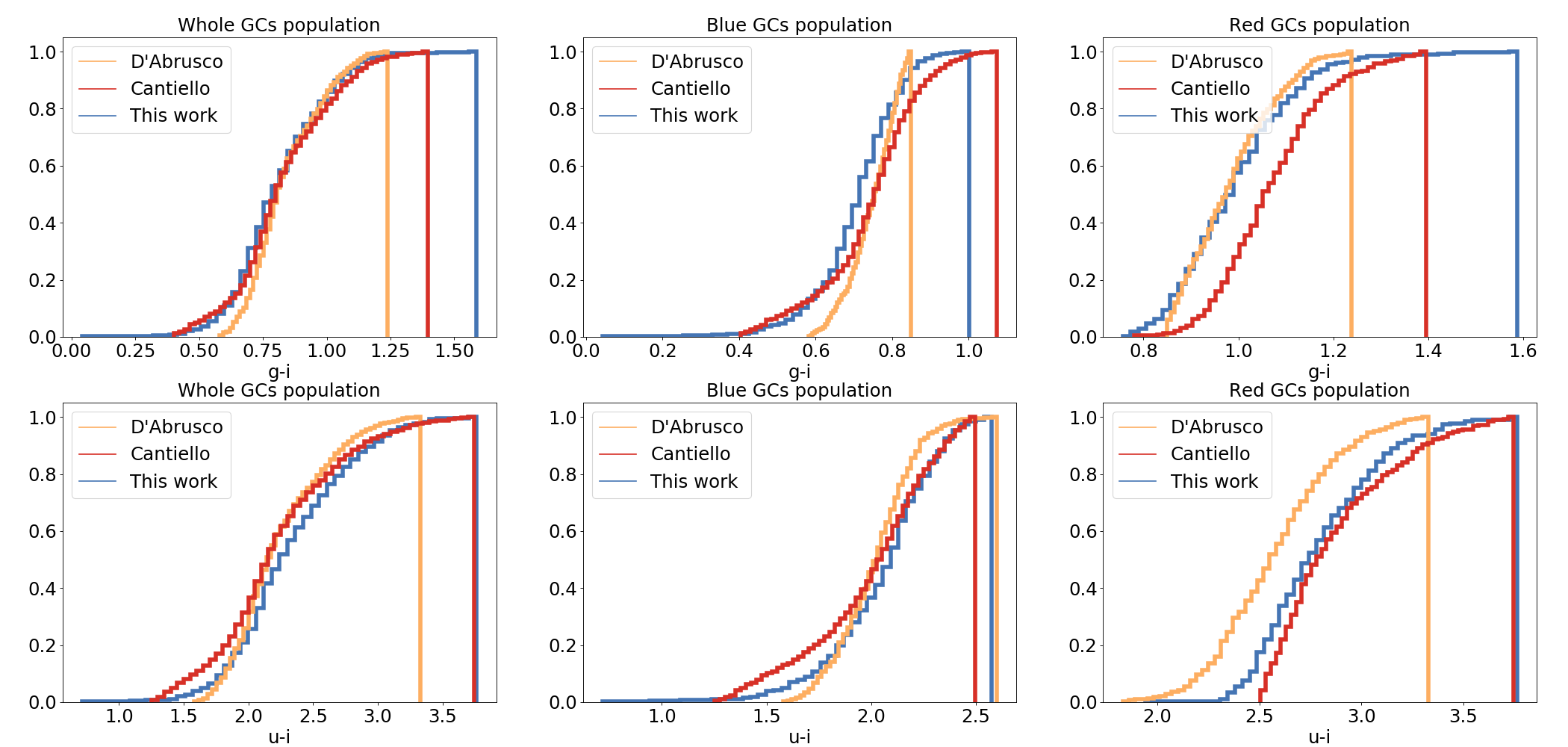}
\caption[]{Cumulative distribution functions (CDFs) related to colours \textit{G-I} (first row) and \textit{U-I} (second row) for the whole GC population (first column), for the blue GC subpopulation (second column) and for the red GC subpopulation (third column). In all panels our CDFs are in blue, \citealt{dabrusco:2016} CDFs are in orange and \citealt{cantiello2018} CDFs are in red.}\label{fig:cdfs}
\end{figure*}

In order to explore the differences between the GC populations and sub-populations, we estimated the Cumulative Distribution Functions (CDFs) related to the \textit{u-i} and \textit{g-i} colours, illustrated in Fig.~\ref{fig:cdfs} for the whole GC population and for both the red and blue GC sub-populations, comparing our selected GCs with the selection performed by \citeauthor{cantiello2018} and \citeauthor{dabrusco:2016}. Furthermore we applied a Kolmogorov-Smirnov test (KS test, \citealt{peacock1983, fasano1987}) to estimate whether two samples have been extracted from the same distribution (here after null hypothesis). Concerning the whole GC systems, the largest difference is found comparing the \textit{u-i} distribution with \citeauthor{cantiello2018} (bottom left panel in Fig.~\ref{fig:cdfs}), for which the p-value is $<10^{-7}$; this could imply the rejection of the null hypothesis, nevertheless the same GCs distributions in \textit{g-i} colours (top left panel in Fig.~\ref{fig:cdfs}) show a very similar CDFs with a p-value $\sim4\%$, i.e. the null hypothesis can not be rejected. This discrepancy could be due to the different magnitude spanning range of colours, for instance $~1.5mag$ for $g-i$ and $3.5mag$ for $u-i$, which could cause a higher relevance of outliers in the \textit{u-i} case. Regarding the sub-populations (center and right panels in Fig.~\ref{fig:cdfs}), the distributions are affected by the difference between the GC sets (due to the different GC identification techniques) and by the not uniform blue-red selection criteria. Nevertheless the KS test between our and \citeauthor{dabrusco:2016} red GCs returns a p-value of $11.7\%$, so we can not reject the null hypothesis. 

\section{Analysis of density maps of GC spatial distribution}\label{ss:densitymap}
\begin{figure}\centering
\includegraphics[width=\columnwidth]{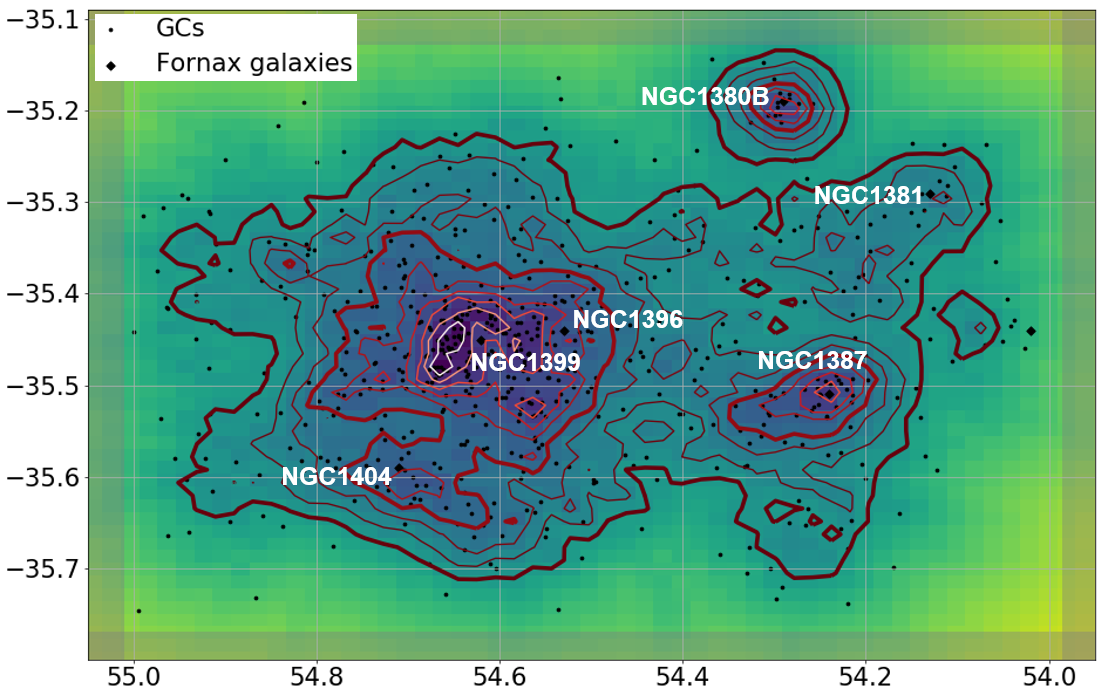}
\includegraphics[width=\columnwidth]{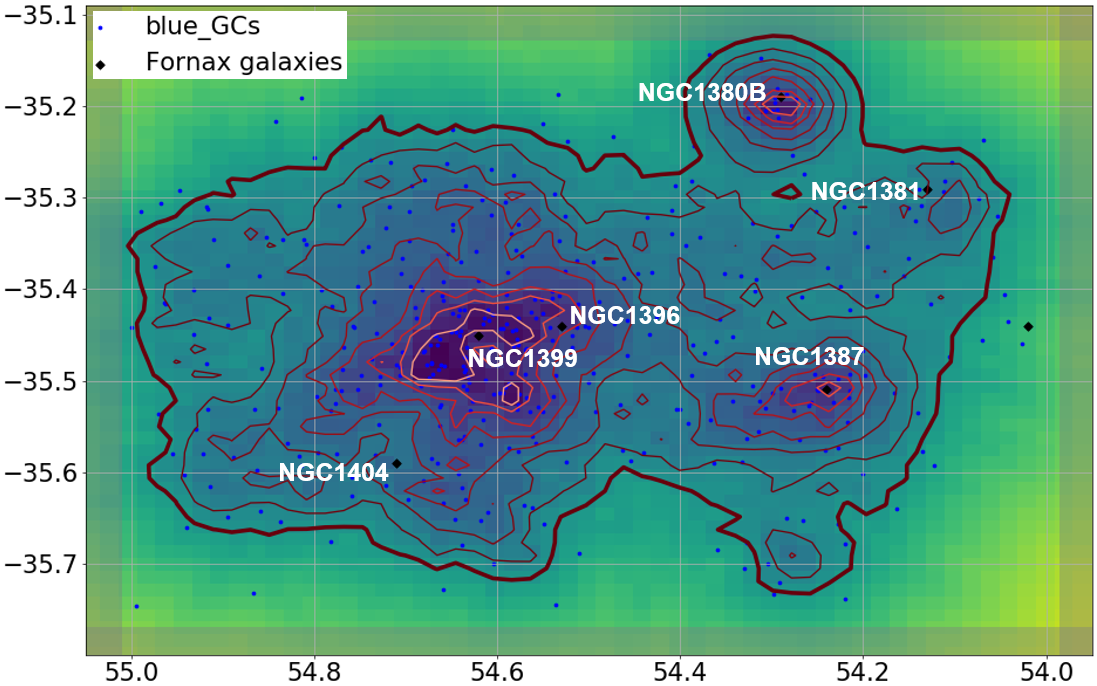}
\includegraphics[width=\columnwidth]{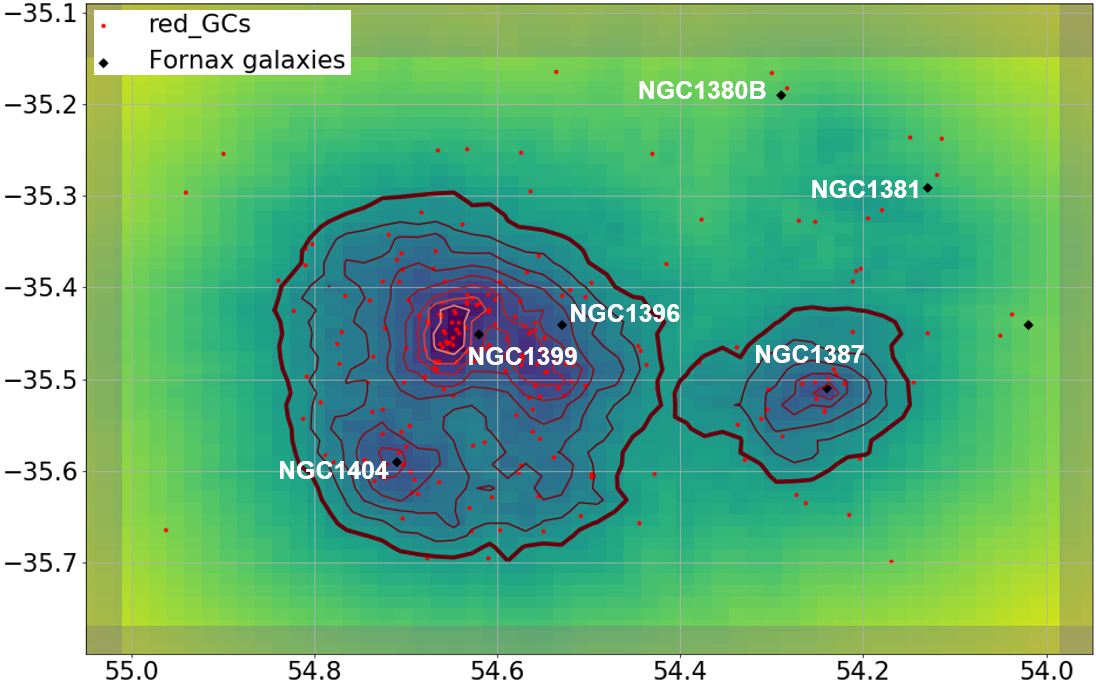}
\caption[Density Maps]{Density maps of GC spatial distribution. Top panel: whole GC population. Middle panel: blue GC sub-population. Bottom panel: red GC sub-population. The contours indicate $10$ ($9$ for the sub-populations) log-spaced density levels. The main Fornax galaxies are marked with black filled diamonds.}\label{fig:MAPS:densitiesmaps}
\end{figure}

Finally, as further validation method, we present the density maps of the spatial distribution from the GC sky locations, estimated both for the whole population and for the blue and red sub-populations separately. The extracted density maps can be directly compared with those of \citealt{dabrusco:2016} and \citealt{cantiello2018}.

The density maps are related to the core of Fornax cluster, with right ascension $\in (54.0, 55.0)$ and declination $\in (-35.75, -35.13)$, and have been estimated from the GCs sky coordinates, applying a K-Nearest Neighbour (KNN) method \citep{duda:2000} on a regular squared grid covering the sky region, following the same process presented in \citealt{dabrusco2015, dabrusco:2016}. Each knot in the grid has a density defined as $d=K/(\pi\cdot r_k^2)$, i.e. the ratio between the $K$ neighbour GCs used to estimate the density, and the (projected) area of the circle whose radius is equal to the distance of the $K$th nearest neighbour. 
The value of $K$ shapes the densities map: small $K$-values imply maps with compact density structures, while high $K$-values lead to large structures losing spatial information \citep{dabrusco2015,dabrusco:2016}. In the following, $K$ is taken equals to $9$, adopting the same strategy proposed by \cite{dabrusco:2016} which focused on the study of large spatial scale GC distribution. The GCs used in this process ($719$) are those used as training set for our GNG model and derived from the \textit{run} executions after the GMM exclusion process. 
It is worth to say that the density is underestimated at the edges of the selected region, due to the lack of sources beyond those edges.  Panels in Fig.~\ref{fig:MAPS:densitiesmaps} show the extracted density maps for the whole GC population (top panel), and for the blue and red sub-populations (middle and bottom panel, respectively). In these figures the gray areas represent the region in which the density is underestimated.

\begin{figure}\centering
\includegraphics[width=\columnwidth]{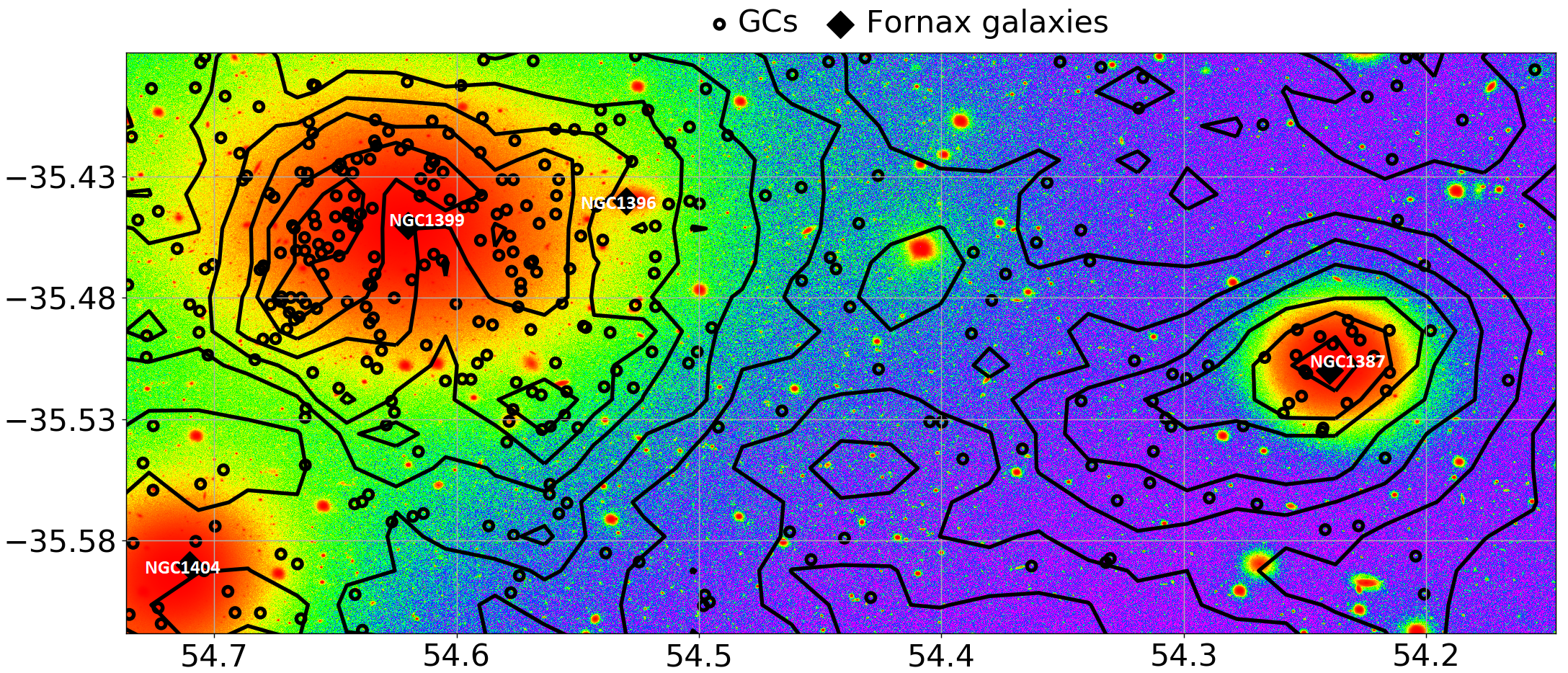}
\caption[Detail of the density map overlapped to the FDS \textit{G}-band]{Detail of the density map overlapped to the FDS \textit{G}-band in the region that includes \textit{NGC1399, NGC1396, NGC1404, NGC1387}, with limits RA $\in (54.146, 54.44)$ and DEC $\in (-35.38, -35.62)$.}\label{fig:MAPS:detail}
\end{figure}

Looking at the top panel in Fig.~\ref{fig:MAPS:densitiesmaps}, the irregular shape of the region designed as \textit{A} clearly shows a structure stretched in the W-E direction, due to the gravitational interaction between the giant elliptical galaxy \textit{NGC1399} (around which the density is maximum) and the nearby galaxies (\textit{NGC1396, NGC1404, NGC1387, NGC1381}). This region contains $85\%$ of the involved GCs in the density map estimation. Within such region it is possible to distinguish an overdensity associated with the \textit{NGC1399-NGC1396-NGC1404} complex region (B, $46\%$), where a bridge is connecting \textit{NGC1399} and \textit{NGC1404} in the SE-NW direction (discovered by \citealt{bassino2006} and emphasized by \citealt{dabrusco:2016}). On a larger scale this complex region stretches to the west, combining densities related to \textit{NGC1387} and \textit{NGC1381}. At the north there is an isolated density region centered on \textit{NGC1380B} (C, $2.5\%$). \citealt{iodice2017} have detected a previously unknown region of intracluster light (ICL). This overdensity of ICL is located in between the three bright galaxies in the core, \textit{NGC1387}, \textit{NGC1379}, and \textit{NGC1381}. They also show that the ICL is the counterpart in the diffuse light of the known over-density in the population of blue globular clusters. A detail of the connection between \textit{NGC1387} and the complex region \textit{NGC1399-NGC1396-NGC1404} is illustrated in Fig.~\ref{fig:MAPS:detail}, where the iso-density contours and the distribution of GCs are overlapped onto the FDS \textit{G}-band image of the region whose limits are: RA $\in (54.146, 54.440)$ and DEC $\in (-35.380, -35.620)$. In this figure is shown the connection between \textit{NGC1399} and \textit{NGC1387} which reflects the ``bridge-like'' stellar steam between the two galaxies made by several filamentary structures, detected by \cite{iodice2016}, whose existence had initially been proposed by \cite{bassino2006} and confirmed by \cite{dabrusco:2016}. This low surface brightness structure seems to be confirmed by the GCs distribution, whose iso-density contour forms a connection between the two galaxies. This suggests an ongoing interaction between the two galaxies where a fraction of GCs, originally belonging to \textit{NGC1387}, may have been stripped by the more massive \textit{NGC1399}. 

This interaction between the central structure (\textit{NGC1399-NGC1404-NGC1396}) and \textit{NGC1387} is particularly evident in the density map of red GCs (bottom panel in Fig.~\ref{fig:MAPS:densitiesmaps}). Although \textit{NG1396} is a dwarf galaxy separated by $500 km/s$ from the central galaxies, thus inducing a projection effect, it has been included in analogy with \citealt{dabrusco:2016}. Most of the red GCs are concentrated in two regions, E ($75\%$) and F ($10\%$). The inner contours of \textit{NGC1399} are characterized by a tail that stretches toward east (i.e. toward \textit{NGC1387}), maybe due to the interaction between the ellipticals, although it could be a projection effect. Indeed, unlike isolated systems (e.g. the blue GC distribution around \textit{NGC1380B}), the shape of the contours is strongly irregular, although the incompleteness of GC detections in the center of giant galaxies could contribute to the observed irregular density structures. The blue GCs density (middle panel in Fig.~\ref{fig:MAPS:densitiesmaps}) shows a large and stretched complex region (D, $90\%$), which connects \textit{NGC1381} with the mean central overdensity. This suggests that the GC stripping is not confined to \textit{NGC1387}, but acts on a broader scale.

The fact that no overdensity connected to \textit{NGC1379}, located at $(54.02, -35.44)$, is detected is likely due to the fact that the field covered in this work is limited to the range $(54.02, 55.38)$, and the density is underestimated in direction of  \textit{NGC1379}.

\section{Conclusions}\label{sec:conc}
In this work we have presented an approach based on Astroinformatics methodologies to the identification of GCs from ground-based data. The models under investigation were the GNG, fully implemented using the GPU-oriented Theano library \citep{theano:2016}, and the $\Phi$Lab feature selection method \citep{brescia:2018b}.
\begin{itemize}
    \item[-] The difference with the standard GNG model is the \textit{batch} sample extraction, which not only allows a faster convergence towards the minimum of the cost function, but also improves scalability, together with the capability to fully exploit the computing resources of the host machine. 
    \item[-] We probed the efficiency of the feature selection method $\Phi$Lab to individuate the complete set of relevant features, by excluding features whose informative contribution was negligible. We have also shown how the relevant set of features found is essentially in agreement with the physics of the problem, since the distribution and projections of the selected hyperspace allow the separation between the class types.
    \item[-] Comparing the GNG performance with one of the widely used method in Astrophysics, the MLPQNA \citep{brescia:2012}, we confirmed the capability of GNG to separate GCs from background and foreground sources, reaching a satisfying trade-off between purity and completeness, comparable to the MLPQNA results, particularly when the full set of bands was used (i.e. \textit{ugri}).
    \item[-] Furthermore, we classified an unlabeled set of sources, extracted from the whole catalogue and validated through the limited amount of HST detected sources as \textit{ground truth}. Having evidence for candidate false positives, we have applied a GMM in order to exclude them. The bimodal bivariate Gaussian fit returned a set of parameters fully comparable with the literature, thus validating our results.
\end{itemize}

In order to investigate the prediction capabilities of our methods, the model performances and the identified set of GCs have been compared with other similar works:
\begin{itemize}
    \item[-] by comparing our multi-band ground-based results with those obtained with the single-band HST photometry (see \citealt{brescia:2012, angora2017}), we showed that, only using all ground-based photometry, the classifiers reach levels of accuracy comparable with those obtained with HST single-band photometry. In particular, by introducing the information carried by \textit{u}-band we reached comparable results to HST experiments, although the different efficiencies of the instruments limit the ground-based analysis to brighter sources..
    \item[-] the matching with the results obtained by \cite{dabrusco:2016} and \cite{cantiello2018} probed the robustness of our method, fully comparable with other techniques which exploit different approaches;
    \item[-] finally, the density maps for the red, blue and whole GC populations showed the usefulness of our prediction method and underlined some interesting features of the Fornax core, comparable to other studies \citep{dabrusco:2016, cantiello2018, iodice2016, bassino2006}.
\end{itemize}
Although our approach requires a spectroscopic knowledge in order to build a broad and pure Knowledge Base, indispensable for training ML models, the  method avoids the introduction of arbitrary photometric cuts, which, although plausible, are bound to a maximum of three-dimensional viewpoints of the phenomenology, thus unavoidably originating contamination effects. It is important to underline that our results are constrained by the exiguous number of labeled sources available in the catalogue and by the unavoidable inhomogeneities among the filters, particularly concerning the presence of $\sim 82\%$ missing data among the \textit{u}-band samples, which we proved to be crucial to effectively separate the GCs from different type of sources. In future works, an improvement could be obtained by a reduction of VST data tailored to compact sources (in progress) or by including external photometry such as, e.g. DECam \textit{u}-band \citep{decam2018}. The ongoing reduction of the full FDS survey data, will allow to extend these results to the whole Fornax cluster out to the virial radius.

\section*{Acknowledgements}
The authors thank the anonymous referee for all very helpful comments and suggestions that improved the scientific quality of the presented work.
MB acknowledges the \textit{INAF PRIN-SKA 2017 program 1.05.01.88.04} and the funding from \textit{MIUR Premiale 2016: MITIC}. MP acknowledges support from PRIN INAF 2014 ``Fornax Cluster Imaging and Spectroscopic Deep Survey''. MP and SC acknowledge support from the project ``Quasars at high redshift: physics and Cosmology'' financed by the ASI/INAF agreement 2017-14-H.0. GL, RP and NRN acknowledge support rom the European Union’s Horizon 2020 Sundial Innovative Training Network, grant n.721463. NRN acknowledges support from the 100 Top Talent Program of the Sun Yat-sen University, Guandong Province. MS and EI acknowledge financial support from the VST project. R.D'A. is supported by NASA contract NAS8-03060 (Chandra X-ray Center). G.D. acknowledges support from CONICYT project Basal AFB-170002. DAMEWARE has been used for machine learning experiments \citep{DAMEWARE}. Topcat has been used for this work \citep{Taylor2005}. C$^3$ has been used for efficient catalogue cross-matching \citep{Riccio:2017}.

\bibliographystyle{mnras}
\bibliography{main} 

\begin{thebibliography}{}
\makeatletter
\relax
\def\mn@urlcharsother{\let\do\@makeother \do\$\do\&\do\#\do\^\do\_\do\%\do\~}
\def\mn@doi{\begingroup\mn@urlcharsother \@ifnextchar [ {\mn@doi@}
  {\mn@doi@[]}}
\def\mn@doi@[#1]#2{\def\@tempa{#1}\ifx\@tempa\@empty \href
  {http://dx.doi.org/#2} {doi:#2}\else \href {http://dx.doi.org/#2} {#1}\fi
  \endgroup}
\def\mn@eprint#1#2{\mn@eprint@#1:#2::\@nil}
\def\mn@eprint@arXiv#1{\href {http://arxiv.org/abs/#1} {{\tt arXiv:#1}}}
\def\mn@eprint@dblp#1{\href {http://dblp.uni-trier.de/rec/bibtex/#1.xml}
  {dblp:#1}}
\def\mn@eprint@#1:#2:#3:#4\@nil{\def\@tempa {#1}\def\@tempb {#2}\def\@tempc
  {#3}\ifx \@tempc \@empty \let \@tempc \@tempb \let \@tempb \@tempa \fi \ifx
  \@tempb \@empty \def\@tempb {arXiv}\fi \@ifundefined
  {mn@eprint@\@tempb}{\@tempb:\@tempc}{\expandafter \expandafter \csname
  mn@eprint@\@tempb\endcsname \expandafter{\@tempc}}}

\bibitem[\protect\citeauthoryear{{Abbott} et~al.,}{{Abbott}
  et~al.}{2018}]{decam2018}
{Abbott} T.~M.~C.,  et~al., 2018, preprint (\mn@eprint {arXiv} {1801.03181})

\bibitem[\protect\citeauthoryear{{Angora}, {Brescia}, {Riccio}, {Cavuoti},
  {Paolillo}  \& {Puzia}}{{Angora} et~al.}{2017}]{angora2017}
{Angora} G.,  {Brescia} M.,  {Riccio} G.,  {Cavuoti} S.,  {Paolillo} M.,
  {Puzia} T.~H.,  2017, CEUR Workshop Proceedings, 2022, 381

\bibitem[\protect\citeauthoryear{Ashman \& Zepf}{Ashman \&
  Zepf}{2008}]{ashman:2008}
Ashman K.,  Zepf S.,  2008, Globular Cluster Systems.
Cambridge Astrophysics, Cambridge University Press

\bibitem[\protect\citeauthoryear{{Ashman}, {Bird}  \& {Zepf}}{{Ashman}
  et~al.}{1994}]{KMM1994}
{Ashman} K.~M.,  {Bird} C.~M.,   {Zepf} S.~E.,  1994, \mn@doi [Astronomical
  Journal] {10.1086/117248}, 108, 2348

\bibitem[\protect\citeauthoryear{{Baron}}{{Baron}}{2019}]{baron2019}
{Baron} D.,  2019, arXiv e-prints, \href
  {https://ui.adsabs.harvard.edu/abs/2019arXiv190407248B} {p. arXiv:1904.07248}

\bibitem[\protect\citeauthoryear{{Bassino}, {Faifer}, {Forte}, {Dirsch},
  {Richtler}, {Geisler}  \& {Schuberth}}{{Bassino} et~al.}{2006}]{bassino2006}
{Bassino} L.~P.,  {Faifer} F.~R.,  {Forte} J.~C.,  {Dirsch} B.,  {Richtler} T.,
   {Geisler} D.,   {Schuberth} Y.,  2006, \mn@doi [Astronomy and Astrophysics]
  {10.1051/0004-6361:20054563}, 451, 789

\bibitem[\protect\citeauthoryear{Batista \& Monard}{Batista \&
  Monard}{2003}]{BatistaMonard2003}
Batista G. E. A. P.~A.,  Monard M.~C.,  2003, \mn@doi [Applied Artificial
  Intelligence] {10.1080/713827181}, 17, 519

\bibitem[\protect\citeauthoryear{{Bertin} \& {Arnouts}}{{Bertin} \&
  {Arnouts}}{1996}]{bertin:1996}
{Bertin} E.,  {Arnouts} S.,  1996, \mn@doi [Astrophysics and Space Science]
  {10.1051/aas:1996164}, 117, 393

\bibitem[\protect\citeauthoryear{Bishop}{Bishop}{2006}]{Bishop:2006}
Bishop C.~M.,  2006, Pattern Recognition and Machine Learning (Information
  Science and Statistics).
Springer-Verlag New York, Inc., Secaucus, NJ, USA

\bibitem[\protect\citeauthoryear{{Borne} et~al.,}{{Borne}
  et~al.}{2009}]{Borne2009}
{Borne} K.,  et~al., 2009, in astro2010: The Astronomy and Astrophysics Decadal
  Survey.  (\mn@eprint {arXiv} {0909.3892})

\bibitem[\protect\citeauthoryear{{Bortoletti}, {Di Fiore}, {Fanelli}  \&
  {Zellini}}{{Bortoletti} et~al.}{2003}]{bortoletti}
{Bortoletti} A.,  {Di Fiore} C.,  {Fanelli} S.,   {Zellini} P.,  2003, \mn@doi
  [IEEE Transactions on Neural Networks] {10.1109/TNN.2003.809425}, 14, 263

\bibitem[\protect\citeauthoryear{Breiman}{Breiman}{2001}]{Breiman2001}
Breiman L.,  2001, \mn@doi [Machine Learning] {10.1023/A:1010933404324}, 45, 5

\bibitem[\protect\citeauthoryear{Breiman, Last  \& Rice}{Breiman
  et~al.}{2003}]{breiman:2003}
Breiman L.,  Last M.,   Rice J.,  2003, in Statistical Challenges in Astronomy.
  Springer New York, New York, NY, pp 243--254

\bibitem[\protect\citeauthoryear{{Brescia} \& {Longo}}{{Brescia} \&
  {Longo}}{2013}]{brescia:2013}
{Brescia} M.,  {Longo} G.,  2013, \mn@doi [Nuclear Instruments and Methods in
  Physics Research A] {10.1016/j.nima.2012.12.027}, 720, 92

\bibitem[\protect\citeauthoryear{{Brescia}, {Cavuoti}, {Paolillo}, {Longo}  \&
  {Puzia}}{{Brescia} et~al.}{2012}]{brescia:2012}
{Brescia} M.,  {Cavuoti} S.,  {Paolillo} M.,  {Longo} G.,   {Puzia} T.,  2012,
  \mn@doi [Monthly Notices of the Royal Astronomical Society]
  {10.1111/j.1365-2966.2011.20375.x}, 421, 1155

\bibitem[\protect\citeauthoryear{{Brescia}, {Cavuoti}, {D'Abrusco}, {Longo}  \&
  {Mercurio}}{{Brescia} et~al.}{2013}]{Brescia2013}
{Brescia} M.,  {Cavuoti} S.,  {D'Abrusco} R.,  {Longo} G.,   {Mercurio} A.,
  2013, \mn@doi [\apj] {10.1088/0004-637X/772/2/140}, \href
  {http://adsabs.harvard.edu/abs/2013ApJ...772..140B} {772, 140}

\bibitem[\protect\citeauthoryear{Brescia et~al.,}{Brescia
  et~al.}{2014}]{DAMEWARE}
Brescia M.,  et~al., 2014, Publications of the Astronomical Society of the
  Pacific, 126, 783

\bibitem[\protect\citeauthoryear{{Brescia}, {Cavuoti}, {Amaro}, {Riccio},
  {Angora}, {Vellucci}  \& {Longo}}{{Brescia} et~al.}{2018}]{Brescia2018}
{Brescia} M.,  {Cavuoti} S.,  {Amaro} V.,  {Riccio} G.,  {Angora} G.,
  {Vellucci} C.,   {Longo} G.,  2018, Communications in Computer and
  Information Science, 822

\bibitem[\protect\citeauthoryear{Brescia, Salvato, Cavuoti, Ananna, Riccio,
  LaMassa, Urry  \& Longo}{Brescia et~al.}{2019}]{brescia:2018b}
Brescia M.,  Salvato M.,  Cavuoti S.,  Ananna T.~T.,  Riccio G.,  LaMassa
  S.~M.,  Urry C.~M.,   Longo G.,  2019, \mn@doi [Monthly Notices of the Royal
  Astronomical Society] {10.1093/mnras/stz2159}, 489, 663

\bibitem[\protect\citeauthoryear{{Brodie} \& {Strader}}{{Brodie} \&
  {Strader}}{2006}]{brodie2006}
{Brodie} J.~P.,  {Strader} J.,  2006, \mn@doi [\araa]
  {10.1146/annurev.astro.44.051905.092441}, 44, 193

\bibitem[\protect\citeauthoryear{Byrd, Nocedal  \& Schnabel}{Byrd
  et~al.}{1994}]{Byrd1994}
Byrd R.,  Nocedal J.,   Schnabel R.,  1994, \mn@doi [Mathematical Programming]
  {10.1007/BF01582063}, 63, 129

\bibitem[\protect\citeauthoryear{{Camino}, {Hammerschmidt}  \&
  {State}}{{Camino} et~al.}{2019}]{camino:2019}
{Camino} R.~D.,  {Hammerschmidt} C.~A.,   {State} R.,  2019, arXiv e-prints,
  \href {https://ui.adsabs.harvard.edu/abs/2019arXiv190210666C} {p.
  arXiv:1902.10666}

\bibitem[\protect\citeauthoryear{{Cantiello}, {Grado}, {Rejkuba}, {Arnaboldi},
  {Capaccioli}, {Greggio}, {Iodice}  \& {Limatola}}{{Cantiello}
  et~al.}{2018a}]{can2018}
{Cantiello} M.,  {Grado} A.,  {Rejkuba} M.,  {Arnaboldi} M.,  {Capaccioli} M.,
  {Greggio} L.,  {Iodice} E.,   {Limatola} L.,  2018a, \mn@doi [\aap]
  {10.1051/0004-6361/201731325}, \href
  {http://adsabs.harvard.edu/abs/2018A%26A...611A..21C} {611, A21}

\bibitem[\protect\citeauthoryear{{Cantiello} et~al.,}{{Cantiello}
  et~al.}{2018b}]{cantiello2018}
{Cantiello} M.,  et~al., 2018b, \mn@doi [\aap] {10.1051/0004-6361/201730649},
  \href {http://adsabs.harvard.edu/abs/2018A%26A...611A..93C} {611, A93}

\bibitem[\protect\citeauthoryear{{Cantiello} et~al.}{{Cantiello}
  et~al.}{2019}]{cant2019}
{Cantiello} M.,  et~al., 2019, in preparation

\bibitem[\protect\citeauthoryear{{Cavuoti}, {Brescia}, {Longo}  \&
  {Mercurio}}{{Cavuoti} et~al.}{2012}]{Cavuoti:2012}
{Cavuoti} S.,  {Brescia} M.,  {Longo} G.,   {Mercurio} A.,  2012, \mn@doi
  [A\&A] {10.1051/0004-6361/201219755}, 546, A13

\bibitem[\protect\citeauthoryear{Cavuoti, Garofalo, Brescia, Pescape, Longo  \&
  Ventre}{Cavuoti et~al.}{2013}]{cavuoti2013b}
Cavuoti S.,  Garofalo M.,  Brescia M.,  Pescape A.,  Longo G.,   Ventre G.,
  2013, \mn@doi [Smart Innovation, Systems and Technologies]
  {10.1007/978-3-642-35467-0_4}, 19, 29

\bibitem[\protect\citeauthoryear{{Cavuoti}, {Brescia}, {De Stefano}  \&
  {Longo}}{{Cavuoti} et~al.}{2015}]{Cavuoti:2015}
{Cavuoti} S.,  {Brescia} M.,  {De Stefano} V.,   {Longo} G.,  2015, \mn@doi
  [Experimental Astronomy] {10.1007/s10686-015-9443-4}, 39, 45

\bibitem[\protect\citeauthoryear{{D'Abrusco}, {Fabbiano}  \&
  {Zezas}}{{D'Abrusco} et~al.}{2015}]{dabrusco2015}
{D'Abrusco} R.,  {Fabbiano} G.,   {Zezas} A.,  2015, The Astrophysical Journal,
  805, 26

\bibitem[\protect\citeauthoryear{{D'Abrusco} et~al.,}{{D'Abrusco}
  et~al.}{2016}]{dabrusco:2016}
{D'Abrusco} R.,  et~al., 2016, \mn@doi [Astrophysical Journal, Letters]
  {10.3847/2041-8205/819/2/L31}, 819, L31

\bibitem[\protect\citeauthoryear{{D'Ago} et~al.}{{D'Ago}
  et~al.}{2019}]{dago2019}
{D'Ago} G.,  et~al., 2019, in preparation

\bibitem[\protect\citeauthoryear{D'Isanto, Cavuoti, Brescia, Donalek, Longo,
  Riccio  \& Djorgovski}{D'Isanto et~al.}{2016}]{D'Isanto20163119}
D'Isanto A.,  Cavuoti S.,  Brescia M.,  Donalek C.,  Longo G.,  Riccio G.,
  Djorgovski S.,  2016, \mn@doi [Monthly Notices of the Royal Astronomical
  Society] {10.1093/mnras/stw157}, 457, 3119

\bibitem[\protect\citeauthoryear{Delli~Veneri, Cavuoti, Brescia, Longo  \&
  Riccio}{Delli~Veneri et~al.}{2019}]{Delliveneri:2019}
Delli~Veneri M.,  Cavuoti S.,  Brescia M.,  Longo G.,   Riccio G.,  2019,
  \mn@doi [Monthly Notices of the Royal Astronomical Society]
  {10.1093/mnras/stz856}, 486, 1377

\bibitem[\protect\citeauthoryear{Duda, Hart  \& Stork}{Duda
  et~al.}{2000}]{duda:2000}
Duda R.~O.,  Hart P.~E.,   Stork D.~G.,  2000, Pattern Classification (2Nd
  Edition).
Wiley-Interscience

\bibitem[\protect\citeauthoryear{{Fasano} \& {Franceschini}}{{Fasano} \&
  {Franceschini}}{1987}]{fasano1987}
{Fasano} G.,  {Franceschini} A.,  1987, \mn@doi [Monthly Notices of the RAS]
  {10.1093/mnras/225.1.155}, 225, 155

\bibitem[\protect\citeauthoryear{Fawcett}{Fawcett}{2006}]{fawcett:2006}
Fawcett T.,  2006, Pattern Recognition Letters, 27, 861

\bibitem[\protect\citeauthoryear{{Feigelson} \& {Hilbe}}{{Feigelson} \&
  {Hilbe}}{2014}]{Feigelson2014}
{Feigelson} E.,  {Hilbe} J.~M.,  2014, in American Astronomical Society Meeting
  Abstracts \#223. p. 253.02

\bibitem[\protect\citeauthoryear{Floudas \& Pardalos}{Floudas \&
  Pardalos}{2006}]{Floudas:2006}
Floudas C. C.~A.,  Pardalos P.~M.,  2006, Encyclopedia of Optimization.
Springer-Verlag, Berlin, Heidelberg

\bibitem[\protect\citeauthoryear{Fritzke}{Fritzke}{1994}]{fritzke:1994}
Fritzke B.,  1994, \mn@doi [Neural Networks] {10.1016/0893-6080(94)90091-4}, 7,
  1441

\bibitem[\protect\citeauthoryear{Fritzke}{Fritzke}{1995}]{fritzke:1995}
Fritzke B.,  1995, in Tesauro G.,  Touretzky D.~S.,   Leen T.~K.,  eds, ,
  Advances in Neural Information Processing Systems 7.
MIT Press, pp 625--632

\bibitem[\protect\citeauthoryear{{Geisler} \& {Forte}}{{Geisler} \&
  {Forte}}{1990}]{geisler1990}
{Geisler} D.,  {Forte} J.~C.,  1990, \mn@doi [Astrophysical Journal, Letters]
  {10.1086/185654}, 350, L5

\bibitem[\protect\citeauthoryear{Gheyas \& Smith}{Gheyas \&
  Smith}{2010}]{Gheyas2010}
Gheyas I.~A.,  Smith L.~S.,  2010, \mn@doi [Pattern Recognition]
  {https://doi.org/10.1016/j.patcog.2009.06.009}, 43, 5

\bibitem[\protect\citeauthoryear{Guyon \& Elisseeff}{Guyon \&
  Elisseeff}{2003}]{Guyon2003}
Guyon I.,  Elisseeff A.,  2003, J. Mach. Learn. Res., 3, 1157

\bibitem[\protect\citeauthoryear{Guyon, Gunn, Nikravesh  \& Zadeh}{Guyon
  et~al.}{2006}]{Guyon:2006}
Guyon I.,  Gunn S.,  Nikravesh M.,   Zadeh L.~A.,  2006, Feature Extraction:
  Foundations and Applications (Studies in Fuzziness and Soft Computing).
Springer-Verlag, Berlin, Heidelberg

\bibitem[\protect\citeauthoryear{Hanley \& McNeil}{Hanley \&
  McNeil}{1982}]{hanley:1982}
Hanley J.~A.,  McNeil B.~J.,  1982, \mn@doi [Radiology]
  {10.1148/radiology.143.1.7063747}, 143, 29

\bibitem[\protect\citeauthoryear{{Hara} \& {Maehara}}{{Hara} \&
  {Maehara}}{2016}]{Hara2016}
{Hara} S.,  {Maehara} T.,  2016, preprint (\mn@eprint {arXiv} {1611.05940})

\bibitem[\protect\citeauthoryear{Hara \& Maehara}{Hara \&
  Maehara}{2017}]{Hara2017}
Hara S.,  Maehara T.,  2017, in 31st AAAI Conference on Artificial
  Intelligence, AAAI 2017. pp 1985--1991

\bibitem[\protect\citeauthoryear{Hastie, Tibshirani  \& Friedman}{Hastie
  et~al.}{2001}]{hastie2012}
Hastie T.,  Tibshirani R.,   Friedman J.,  2001, The Elements of Statistical
  Learning.
Springer Series in Statistics, Springer New York Inc., New York, NY, USA

\bibitem[\protect\citeauthoryear{Hastie, Tibshirani  \& Friedman}{Hastie
  et~al.}{2009}]{hastie2009}
Hastie T.,  Tibshirani R.,   Friedman J.,  2009, The Elements of Statistical
  Learning: Data Mining, Inference, and Prediction, Second Edition.
Springer Series in Statistics, Springer New York

\bibitem[\protect\citeauthoryear{{Iodice} et~al.,}{{Iodice}
  et~al.}{2016}]{iodice2016}
{Iodice} E.,  et~al., 2016, \mn@doi [Astrophysical Journal]
  {10.3847/0004-637X/820/1/42}, 820, 42

\bibitem[\protect\citeauthoryear{{Iodice} et~al.,}{{Iodice}
  et~al.}{2017}]{iodice2017}
{Iodice} E.,  et~al., 2017, \mn@doi [\apj] {10.3847/1538-4357/aa6846}, 839, 21

\bibitem[\protect\citeauthoryear{Jain \& Zongker}{Jain \&
  Zongker}{1997}]{Jain:1997}
Jain A.,  Zongker D.,  1997, \mn@doi [IEEE Trans. Pattern Anal. Mach. Intell.]
  {10.1109/34.574797}, 19, 153

\bibitem[\protect\citeauthoryear{Jolliffe}{Jolliffe}{2002}]{Jolliffe2002}
Jolliffe I.~T.,  2002, Principal Component Analysis.
Springer

\bibitem[\protect\citeauthoryear{Kohavi}{Kohavi}{1995}]{Kohavi1995}
Kohavi R.,  1995, in Proceedings of the 14th International Joint Conference on
  Artificial Intelligence - Volume 2. IJCAI'95.
Morgan Kaufmann Publishers Inc., San Francisco, CA, USA, pp 1137--1143, \url
  {http://dl.acm.org/citation.cfm?id=1643031.1643047}

\bibitem[\protect\citeauthoryear{Kohavi \& John}{Kohavi \&
  John}{1997}]{Kohavi1997}
Kohavi R.,  John G.~H.,  1997, \mn@doi [Artificial Intelligence]
  {https://doi.org/10.1016/S0004-3702(97)00043-X}, 97, 273

\bibitem[\protect\citeauthoryear{{Kuijken}}{{Kuijken}}{2011}]{Kuijken2011}
{Kuijken} K.,  2011, The Messenger, 146, 8

\bibitem[\protect\citeauthoryear{Kullback \& Leibler}{Kullback \&
  Leibler}{1951}]{Kullback}
Kullback S.,  Leibler R.~A.,  1951, \mn@doi [Ann. Math. Statist.]
  {10.1214/aoms/1177729694}, 22, 79

\bibitem[\protect\citeauthoryear{{Kundu} \& {Whitmore}}{{Kundu} \&
  {Whitmore}}{1998}]{kundu1998}
{Kundu} A.,  {Whitmore} B.~C.,  1998, \mn@doi [\aj] {10.1086/300643}, 116, 2841

\bibitem[\protect\citeauthoryear{Kursa \& Rudnicki}{Kursa \&
  Rudnicki}{2010}]{Kursa2010}
Kursa M.,  Rudnicki W.,  2010, \mn@doi [Journal of Statistical Software,
  Articles] {10.18637/jss.v036.i11}, 36, 1

\bibitem[\protect\citeauthoryear{Lal, Chapelle, Weston  \& Elisseeff}{Lal
  et~al.}{2006}]{Lal2006}
Lal T.,  Chapelle O.,  Weston J.,   Elisseeff A.,  2006, Embedded methods.
Springer, Berlin, Germany, pp 137--165

\bibitem[\protect\citeauthoryear{Marlin}{Marlin}{2008}]{marlin2008}
Marlin B.,  2008, PhD thesis, Department of Computer Science, University of
  Toronto

\bibitem[\protect\citeauthoryear{Martinetz \& Schulten}{Martinetz \&
  Schulten}{1991}]{martinez:1991}
Martinetz T.~M.,  Schulten K.~J.,  1991, in Kohonen T.,  M{\"a}kisara K.,
  Simula O.,   Kangas J.,  eds, {P}roceedings of the International Conference
  on Artificial Neural Networks 1991 {\rm ({E}spoo, {F}inland)}. Amsterdam; New
  York: North-Holland, pp 397--402

\bibitem[\protect\citeauthoryear{Martinetz, Berkovich  \& Schulten}{Martinetz
  et~al.}{1993}]{martinez:1993}
Martinetz T.~M.,  Berkovich S.~G.,   Schulten K.~J.,  1993, \mn@doi [Trans.
  Neur. Netw.] {10.1109/72.238311}, 4, 558

\bibitem[\protect\citeauthoryear{Montoro \& Abascal}{Montoro \&
  Abascal}{1993}]{montoro:1993}
Montoro J. C.~G.,  Abascal J. L.~F.,  1993, \mn@doi [The Journal of Physical
  Chemistry] {10.1021/j100118a044}, 97, 4211

\bibitem[\protect\citeauthoryear{{Mu{\~n}oz} et~al.,}{{Mu{\~n}oz}
  et~al.}{2014}]{munoz:2014}
{Mu{\~n}oz} R.~P.,  et~al., 2014, \mn@doi [Astrophysical Journal, Supplement]
  {10.1088/0067-0049/210/1/4}, 210, 4

\bibitem[\protect\citeauthoryear{{Muratov} \& {Gnedin}}{{Muratov} \&
  {Gnedin}}{2010}]{muratov2010}
{Muratov} A.~L.,  {Gnedin} O.~Y.,  2010, \mn@doi [Astrophysical Journal]
  {10.1088/0004-637X/718/2/1266}, 718, 1266

\bibitem[\protect\citeauthoryear{Murtagh \& Legendre}{Murtagh \&
  Legendre}{2014}]{Murtagh2014}
Murtagh F.,  Legendre P.,  2014, \mn@doi [Journal of Classification]
  {10.1007/s00357-014-9161-z}, 31, 274

\bibitem[\protect\citeauthoryear{{Nakoneczny}, {Bilicki}, {Solarz}, {Pollo},
  {Maddox}, {Spiniello}, {Brescia}  \& {Napolitano}}{{Nakoneczny}
  et~al.}{2018}]{Nakoneczny:2019}
{Nakoneczny} S.,  {Bilicki} M.,  {Solarz} A.,  {Pollo} A.,  {Maddox} N.,
  {Spiniello} C.,  {Brescia} M.,   {Napolitano} N.~R.,  2018, arXiv e-prints

\bibitem[\protect\citeauthoryear{Parker}{Parker}{2010}]{parker2010}
Parker R.,  2010, Missing Data Problems in Machine Learning.
VDM Verlag

\bibitem[\protect\citeauthoryear{{Peacock}}{{Peacock}}{1983}]{peacock1983}
{Peacock} J.~A.,  1983, \mn@doi [Monthly Notices of the RAS]
  {10.1093/mnras/202.3.615}, 202, 615

\bibitem[\protect\citeauthoryear{Pedregosa et~al.,}{Pedregosa
  et~al.}{2011}]{sklearn:2011}
Pedregosa F.,  et~al., 2011, J. Mach. Learn. Res., 12, 2825

\bibitem[\protect\citeauthoryear{{Pota} et~al.,}{{Pota}
  et~al.}{2013}]{pota:2013}
{Pota} V.,  et~al., 2013, \mn@doi [Monthly Notices of the RAS]
  {10.1093/mnras/sts029}, 428, 389

\bibitem[\protect\citeauthoryear{{Pota} et~al.,}{{Pota}
  et~al.}{2018}]{pota:2018}
{Pota} V.,  et~al., 2018, preprint (\mn@eprint {arXiv} {1803.03275})

\bibitem[\protect\citeauthoryear{{Poulos} \& {Valle}}{{Poulos} \&
  {Valle}}{2016}]{poulos:2016}
{Poulos} J.,  {Valle} R.,  2016, arXiv e-prints, \href
  {https://ui.adsabs.harvard.edu/abs/2016arXiv161009075P} {p. arXiv:1610.09075}

\bibitem[\protect\citeauthoryear{{Puzia}, {Paolillo}, {Goudfrooij},
  {Maccarone}, {Fabbiano}  \& {Angelini}}{{Puzia} et~al.}{2014}]{puzia2014}
{Puzia} T.~H.,  {Paolillo} M.,  {Goudfrooij} P.,  {Maccarone} T.~J.,
  {Fabbiano} G.,   {Angelini} L.,  2014, \mn@doi [Astrophysical Journal]
  {10.1088/0004-637X/786/2/78}, 786, 78

\bibitem[\protect\citeauthoryear{{Riccio}, {Brescia}, {Cavuoti}, {Mercurio},
  {di Giorgio}  \& {Molinari}}{{Riccio} et~al.}{2017}]{Riccio:2017}
{Riccio} G.,  {Brescia} M.,  {Cavuoti} S.,  {Mercurio} A.,  {di Giorgio} A.~M.,
    {Molinari} S.,  2017, \mn@doi [Publications of the ASP]
  {10.1088/1538-3873/129/972/024005}, 129, 024005

\bibitem[\protect\citeauthoryear{Russell \& Norvig}{Russell \&
  Norvig}{2010}]{Russell2010}
Russell S.~J.,  Norvig P.,  c2010., Artificial intelligence :, 3rd ed. edn.
Prentice Hall series in artificial intelligence, Prentice Hall,, Upper Saddle
  River :

\bibitem[\protect\citeauthoryear{{Schipani} et~al.,}{{Schipani}
  et~al.}{2012}]{schipani:2012}
{Schipani} P.,  et~al., 2012, Memorie della Societa Astronomica I\-ta\-lia\-na
  Supplementi, 19, 393

\bibitem[\protect\citeauthoryear{{Schuberth}, {Richtler}, {Hilker}, {Dirsch},
  {Bassino}, {Romanowsky}  \& {Infante}}{{Schuberth} et~al.}{2010}]{schub:2010}
{Schuberth} Y.,  {Richtler} T.,  {Hilker} M.,  {Dirsch} B.,  {Bassino} L.~P.,
  {Romanowsky} A.~J.,   {Infante} L.,  2010, \mn@doi [Astronomy and
  Astrophysics] {10.1051/0004-6361/200912482}, 513, A52

\bibitem[\protect\citeauthoryear{Stehman}{Stehman}{1997}]{stehman:1997}
Stehman S.~V.,  1997, Remote Sensing of Environment, 62, 77

\bibitem[\protect\citeauthoryear{{Tangaro} et~al.,}{{Tangaro}
  et~al.}{2015}]{Tangaro:2015}
{Tangaro} S.,  et~al., 2015, \mn@doi [Computational and Mathematical Methods in
  Medicine, vol.~2015, Article ID 814104, 10 pages, 2015]
  {10.1155/2015/814104}, \href
  {http://adsabs.harvard.edu/abs/2015CMMM.201514104T} {2015}

\bibitem[\protect\citeauthoryear{Taylor}{Taylor}{1996}]{taylor1986}
Taylor J.~R.,  1996, An Introduction to Error Analysis: The Study of
  Uncertainties in Physical Measurements, 2 sub edn.
University Science Books

\bibitem[\protect\citeauthoryear{{Taylor}}{{Taylor}}{2005}]{Taylor2005}
{Taylor} M.~B.,  2005, in {Shopbell} P.,  {Britton} M.,   {Ebert} R.,  eds,
  Astronomical Society of the Pacific Conference Series Vol. 347, Astronomical
  Data Analysis Software and Systems XIV. p.~29

\bibitem[\protect\citeauthoryear{{The Theano Development Team} et~al.,}{{The
  Theano Development Team} et~al.}{2016}]{theano:2016}
{The Theano Development Team} et~al., 2016, preprint (\mn@eprint {arXiv}
  {1605.02688})

\bibitem[\protect\citeauthoryear{Tibshirani}{Tibshirani}{2013}]{Tibshirani2012}
Tibshirani R.~J.,  2013, \mn@doi [Electron. J. Statist.] {10.1214/13-EJS815},
  7, 1456

\bibitem[\protect\citeauthoryear{Tikhonov}{Tikhonov}{1998}]{Tikhonov1998}
Tikhonov A.,  1998, Nonlinear Ill-Posed Problems.
Springer Netherlands

\bibitem[\protect\citeauthoryear{Van Der~Maaten}{Van
  Der~Maaten}{2014}]{VanDerMaaten:2014}
Van Der~Maaten L.,  2014, J. Mach. Learn. Res., 15, 3221

\bibitem[\protect\citeauthoryear{{Wittmann}, {Lisker}, {Pasquali}, {Hilker}  \&
  {Grebel}}{{Wittmann} et~al.}{2016}]{Wittmann:2016}
{Wittmann} C.,  {Lisker} T.,  {Pasquali} A.,  {Hilker} M.,   {Grebel} E.~K.,
  2016, \mn@doi [\mnras] {10.1093/mnras/stw827}, \href
  {http://adsabs.harvard.edu/abs/2016MNRAS.459.4450W} {459, 4450}

\bibitem[\protect\citeauthoryear{{Yoon}, {Jordon}  \& {van der Schaar}}{{Yoon}
  et~al.}{2018}]{yoon:2018}
{Yoon} J.,  {Jordon} J.,   {van der Schaar} M.,  2018, arXiv e-prints, \href
  {https://ui.adsabs.harvard.edu/abs/2018arXiv180602920Y} {p. arXiv:1806.02920}

\bibitem[\protect\citeauthoryear{{Zepf}, {Ashman}  \& {Geisler}}{{Zepf}
  et~al.}{1995}]{ashman:1995}
{Zepf} S.~E.,  {Ashman} K.~M.,   {Geisler} D.,  1995, \mn@doi [Astrophysical
  Journal] {10.1086/175549}, 443, 570

\bibitem[\protect\citeauthoryear{{Zhang}, {Xie}  \& {Xing}}{{Zhang}
  et~al.}{2018}]{zhang:2018}
{Zhang} H.,  {Xie} P.,   {Xing} E.,  2018, arXiv e-prints, \href
  {https://ui.adsabs.harvard.edu/abs/2018arXiv180801684Z} {p. arXiv:1808.01684}

\bibitem[\protect\citeauthoryear{van~der Maaten \& Hinton}{van~der Maaten \&
  Hinton}{2008}]{vanDerMaaten2008}
van~der Maaten L.,  Hinton G.,  2008, Journal of Machine Learning Research, 9,
  2579

\makeatother
\end{thebibliography}

\appendix
\section{Feature importances}\label{app:FI}
In this section we report five tables regarding the feature selection process. Table~\ref{tab:FS} shows the importance values estimated by $\Phi$LAB to select the parameter space used in the six classification experiments. The top panel encloses the selected features, while bottom panel contains the rejected ones. Table~\ref{tab:PSs} specifies the involved features for the four experiments chosen to validate the feature selection performed by $\Phi$LAB (see Sec.~\ref{ss:FS}). The results achieved by the GNG and RF are respectively shown in Table~\ref{tab:PSs:performance} and Table~\ref{tab:PSs:RFperformance}. Finally, Table~\ref{tab:roleofu} illustrates the results achieved by the GNG and RF on a dataset composed by: (\textit{i}) the same \textit{ugri} samples after having removed the information regarding the \textit{u}-band; (\textit{ii}) a dataset whose features represent only the information carried by the \textit{u}-band, (\textit{iii}) the whole \textit{ugri} informative contribution (taken from Table~\ref{tab:results} and from Table~\ref{tab:PSs:RFperformance}).

\begin{table*}
\centering\caption[Selected features informative contributions]{Feature importance values for the features selected by $\Phi$LAB (top table) and for the features rejected (bottom table), related to the six performed experiments.}\label{tab:FS}
  \resizebox{.645\textwidth}{!}{  
\begin{tabular}{lcccccc}
\textbf{SELECTED} & \multicolumn{3}{c}{\bf ugri} & \multicolumn{3}{c}{\bf gri}\\\hline
FEATURE & 3CLASS & GCALL & GCSTAR & 3CLASS & GCALL & GCSTAR \\\hline
\textit{u FWHM} & 0.0261 & 0.0085 & 0.0094 & & & \\\hline
\textit{g FWHM} & 0.0811 & 0.0281 & 0.0240 & 0.0639 & 0.0294 & 0.0203 \\\hline
\textit{r FWHM} & 0.0676 & 0.0407 & 0.0221 & 0.0848 & 0.0272 & 0.0257 \\\hline
\textit{i FWHM} & 0.0924 & 0.0225 & 0.0181 & 0.0609 & 0.0257 & 0.0391 \\\hline
\textit{u FLUX RADIUS} & 0.0290 & 0.0084 & 0.0060 & & & \\\hline
\textit{g FLUX RADIUS} & 0.0431 & 0.0321 & 0.0120 & 0.0782 & 0.0231 & 0.0139\\\hline
\textit{r FLUX RADIUS} & 0.0349 & 0.0188 & 0.0137 & 0.0485 & 0.0312 & 0.0098\\\hline
\textit{i FLUX RADIUS} & 0.0390 & 0.0210 & 0.0069 & 0.0804 & 0.0162 & 0.0071\\\hline
\textit{u MAG AUTO} & 0.0106 & 0.0320 & 0.0191 &  &  & \\\hline
\textit{g MAG APER4} & 0.0057 & 0.0130 & 0.0174 &  &  & \\\hline
\textit{r MAG APER6} & 0.0107 & 0.0080 & 0.0068 &  &  & \\\hline
\textit{i MAG APER8} & 0.0063 & 0.0087 & 0.0149 &  &  & \\\hline
\textit{g MAG AUTO} & 0.0108 & 0.0115 & 0.0145 & 0.0120 & 0.0105 & 0.0150\\\hline
\textit{g MAG APER4} & 0.0199 & 0.0046 & 0.0132 & 0.0090 & 0.0082 & 0.0162\\\hline
\textit{g MAG APER6} & 0.0188 & 0.0111 & 0.0179 & 0.0101 & 0.0075 & 0.0118\\\hline
\textit{g MAG APER8} & 0.0066 & 0.0051 & 0.0245 & 0.0089 & 0.0128 & 0.0212\\\hline
\textit{r MAG AUTO} & 0.0127 & 0.0247 & 0.0205 & 0.0190 & 0.0184 & 0.0213\\\hline
\textit{r MAG APER4} & 0.0148 & 0.0107 & 0.0600 & 0.0124 & 0.0137 & 0.0144\\\hline
\textit{r MAG APER6} & 0.0195 & 0.0106 & 0.0432 & 0.0083 & 0.0124 & 0.0232\\\hline
\textit{r MAG APER8} & 0.0163 & 0.0145 & 0.0359 & 0.0089 & 0.0167 & 0.0228\\\hline
\textit{i MAG AUTO} & 0.0058 & 0.0099 & 0.0081 & 0.0155 & 0.0274 & 0.0373\\\hline
\textit{i MAG APER4} & 0.0087 & 0.0131 & 0.0231 & 0.0149 & 0.0211 & 0.0469\\\hline
\textit{i MAG APER6} & 0.0063 & 0.0123 & 0.0200 & 0.0190 & 0.0208 & 0.0414\\\hline
\textit{i MAG APER8} & 0.0072 & 0.0100 & 0.0148 & 0.0176 & 0.0162 & 0.0399\\\hline
\textit{u-g AUTO} & 0.0092 & 0.0244 & 0.0120 &  &  & \\\hline
\textit{u-g APER4} & 0.0301 & 0.0220 & 0.0096 &  &  & \\\hline
\textit{u-g APER6} & 0.0120 & 0.0260 & 0.0103 &  &  & \\\hline
\textit{u-g APER8} & 0.0132 & 0.0393 & 0.0109 &  &  & \\\hline
\textit{g-r AUTO} & 0.0087 & 0.0183 & 0.0108 & 0.0173 & 0.0350 & 0.0302\\\hline
\textit{g-r APER4} & 0.0054 & 0.0130 & 0.0100 & 0.0154 & 0.0239 & 0.0299\\\hline
\textit{g-r APER6} & 0.0074 & 0.0185 & 0.0090 & 0.0160 & 0.0302 & 0.0283\\\hline
\textit{g-r APER8} & 0.0094 & 0.0185 & 0.0108 & 0.0146 & 0.0263 & 0.0452\\\hline
\textit{r-i AUTO} & 0.0047 & 0.0065 & 0.0078 & 0.0083 & 0.0142 & 0.0150\\\hline
\textit{r-i APER4} & 0.0057 & 0.0067 & 0.0131 & 0.0130 & 0.0154 & 0.0198\\\hline
\textit{r-i APER6} & 0.0065 & 0.0059 & 0.0107 & 0.0135 & 0.0171 & 0.0191\\\hline
\textit{r-i APER8} & 0.0053 & 0.0074 & 0.0092 & 0.0089 & 0.0198 & 0.0235\\\hline
\textit{u MU MAX} & 0.0090 & 0.0085 & 0.0195 & & & \\\hline
\textit{g MU MAX} & 0.0190 & 0.0118 & 0.0383 & 0.0121 & 0.0110 & 0.0179\\\hline
\textit{r MU MAX} & 0.0206 & 0.0082 & 0.0683 & 0.0152 & 0.0127 & 0.0222\\\hline
\textit{i MU MAX} & 0.0120 & 0.0102 & 0.0145 & 0.0178 & 0.0146 & 0.0552\\\hline
\textit{u A WORLD} & 0.0085 & 0.0327 & 0.0037 & & &\\\hline
\textit{g A WORLD} & 0.0105 & 0.0287 & 0.0044 & 0.0148 & 0.0492 & 0.0058\\\hline
\textit{r A WORLD} & 0.0129 & 0.0248 & 0.0047 & 0.0190 & 0.0292 & 0.0050\\\hline
\textit{i A WORLD} & 0.0116 & 0.0247 & 0.0052 & 0.0258 & 0.0420 & 0.0081\\\hline
\textit{u B WORLD} & 0.0043 & 0.0206 & 0.0079 & & & \\\hline
\textit{g B WORLD} & 0.0088 & 0.0263 & 0.0048 & 0.0273 & 0.0449 & 0.0109\\\hline
\textit{r B WORLD} & 0.0084 & 0.0217 & 0.0046 & 0.0207 & 0.0492 & 0.0094\\\hline
\textit{i B WORLD} & 0.0082 & 0.0360 & 0.0062 & 0.0227 & 0.0499 & 0.0298\\\hline
\textit{i PETRO RADIUS} & 0.0112 & 0.0021 & 0.0042 & 0.0069 & 0.0033 & 0.0090\\\hline
\hline
\textbf{REJECTED} & \multicolumn{3}{c}{\bf ugri} & \multicolumn{3}{c}{\bf gri}\\\hline
FEATURE & 3CLASS & GCALL & GCSTAR & 3CLASS & GCALL & GCSTAR \\\hline
\textit{u PETRO RADIUS} & 0.0018 & 0.0027 & 0.0012 & & & \\\hline
\textit{g PETRO RADIUS} & 0.0156 & 0.0011 & 0.0014 & 0.0086 & 0.0030 & 0.0034\\\hline
\textit{r PETRO RADIUS} & 0.0084 & 0.0012 & 0.0054 & 0.0063 & 0.0032 & 0.0046\\\hline
\textit{u KRON RADIUS} & 0.0028 & 0.0034 & 0.0051 & & & \\\hline
\textit{g KRON RADIUS} & 0.0002 & 0.0002 & 0.0001 & 0.0004 & 0.0010 & 0.0006\\\hline
\textit{r KRON RADIUS} & 0.0003 & 0.0003 & 0.0001 & 0.0009 & 0.0010 & 0.0009\\\hline
\textit{i KRON RADIUS} & 0.0006 & 0.0005 & 0.0006 & 0.0014 & 0.0025 & 0.0023\\\hline
\textit{u ELONG} & 0.0032 & 0.0047 & 0.0027 & & & \\\hline
\textit{g ELONG} & 0.0038 & 0.0028 & 0.0031 & 0.0049 & 0.0049 & 0.0038\\\hline
\textit{r ELONG} & 0.0029 & 0.0034 & 0.0048 & 0.0058 & 0.0050 & 0.0047\\\hline
\textit{i ELONG} & 0.0024 & 0.0029 & 0.0044 & 0.0050 & 0.0049 & 0.0059\\\hline
\textit{u THETA} & 0.0018 & 0.0029 & 0.0025 & & & \\\hline
\textit{g THETA} & 0.0013 & 0.0034 & 0.0041 & 0.0023 & 0.0040 & 0.0041\\\hline
\textit{r THETA} & 0.0017 & 0.0026 & 0.0024 & 0.0021 & 0.0041 & 0.0041\\\hline
\textit{i THETA} & 0.0018 & 0.0028 & 0.0030 & 0.0025 & 0.0036 & 0.0043\\\hline
\end{tabular}
}
\end{table*}

\begin{table*}\centering\caption[Parameter spaces]{List of features composing the four parameter spaces used for validation of the feature selection process with the method $\Phi$LAB: column FULL refers to the original parameter space, including all available features; column BEST refers to the best solution obtained by the FS method; column MIXED refers to a variant of the best PS obtained by replacing all $15$ features rejected by $\Phi$LAB to a subset randomly extracted from the best solution; finally, column BEST+REJECTED is another variant of the BEST PS, where the $15$ rejected features were inserted in place of the least significant features of the BEST PS.}\label{tab:PSs}
  \resizebox{.55\textwidth}{!}{  
\begin{tabular}{lcccc}
\textbf{FEATURE}           & \textbf{FULL} & \textbf{BEST} & \textbf{MIXED} & \textbf{BEST+REJECTED} \\\hline\hline
\textit{u FWHM}            & $\times$ & $\times$ & $\times$ & $\times$ \\\hline
\textit{g FWHM}            & $\times$ & $\times$ &          & $\times$ \\\hline
\textit{r FWHM}            & $\times$ & $\times$ & $\times$ & $\times$ \\\hline
\textit{i FWHM}            & $\times$ & $\times$ & $\times$ & $\times$ \\\hline
\textit{u FLUX RADIUS}     & $\times$ & $\times$ &          & $\times$ \\\hline
\textit{g FLUX RADIUS}     & $\times$ & $\times$ & $\times$ & $\times$ \\\hline
\textit{r FLUX RADIUS}     & $\times$ & $\times$ & $\times$ & $\times$ \\\hline
\textit{i FLUX RADIUS}     & $\times$ & $\times$ & $\times$ &          \\\hline
\textit{u MAG AUTO}        & $\times$ & $\times$ &          & $\times$ \\\hline
\textit{u MAG APER4}       & $\times$ & $\times$ & $\times$ & $\times$ \\\hline
\textit{u MAG APER6}       & $\times$ & $\times$ &          & $\times$ \\\hline
\textit{u MAG APER8}       & $\times$ & $\times$ &          & $\times$ \\\hline
\textit{g MAG AUTO}        & $\times$ & $\times$ & $\times$ &          \\\hline
\textit{g MAG APER4}       & $\times$ & $\times$ & $\times$ & $\times$ \\\hline
\textit{g MAG APER6}       & $\times$ & $\times$ &          &          \\\hline
\textit{g MAG APER8}       & $\times$ & $\times$ & $\times$ &          \\\hline
\textit{r MAG AUTO}        & $\times$ & $\times$ &          & $\times$ \\\hline
\textit{r MAG APER4}       & $\times$ & $\times$ & $\times$ &          \\\hline
\textit{r MAG APER6}       & $\times$ & $\times$ & $\times$ & $\times$ \\\hline
\textit{r MAG APER8}       & $\times$ & $\times$ & $\times$ & $\times$ \\\hline
\textit{i MAG AUTO}        & $\times$ & $\times$ & $\times$ &          \\\hline
\textit{i MAG APER4}       & $\times$ & $\times$ & $\times$ & $\times$ \\\hline
\textit{i MAG APER6}       & $\times$ & $\times$ & $\times$ & $\times$ \\\hline
\textit{i MAG APER8}       & $\times$ & $\times$ & $\times$ &          \\\hline
\textit{u-g AUTO}          & $\times$ & $\times$ & $\times$ & $\times$ \\\hline
\textit{u-g APER4}         & $\times$ & $\times$ &          & $\times$ \\\hline
\textit{u-g APER6}         & $\times$ & $\times$ & $\times$ & $\times$ \\\hline
\textit{u-g APER8}         & $\times$ & $\times$ &          & $\times$ \\\hline
\textit{g-r AUTO}          & $\times$ & $\times$ & $\times$ & $\times$ \\\hline
\textit{g-r APER4}         & $\times$ & $\times$ & $\times$ & $\times$ \\\hline
\textit{g-r APER6}         & $\times$ & $\times$ & $\times$ & $\times$ \\\hline
\textit{g-r APER8}         & $\times$ & $\times$ &          & $\times$ \\\hline
\textit{r-i AUTO}          & $\times$ & $\times$ & $\times$ &          \\\hline
\textit{r-i APER4}         & $\times$ & $\times$ &          &          \\\hline
\textit{r-i APER6}         & $\times$ & $\times$ & $\times$ & $\times$ \\\hline
\textit{r-i APER8}         & $\times$ & $\times$ & $\times$ &          \\\hline
\textit{u MU MAX}          & $\times$ & $\times$ &          & $\times$ \\\hline
\textit{g MU MAX}          & $\times$ & $\times$ & $\times$ &          \\\hline
\textit{r MU MAX}          & $\times$ & $\times$ & $\times$ & $\times$ \\\hline
\textit{i MU MAX}          & $\times$ & $\times$ &          & $\times$ \\\hline
\textit{u A WORLD}         & $\times$ & $\times$ & $\times$ &          \\\hline
\textit{g A WORLD}         & $\times$ & $\times$ &          & $\times$ \\\hline
\textit{r A WORLD}         & $\times$ & $\times$ & $\times$ & $\times$ \\\hline
\textit{i A WORLD}         & $\times$ & $\times$ & $\times$ & $\times$ \\\hline
\textit{u B WORLD}         & $\times$ & $\times$ & $\times$ & $\times$ \\\hline
\textit{g B WORLD}         & $\times$ & $\times$ & $\times$ &          \\\hline
\textit{r B WORLD}         & $\times$ & $\times$ & $\times$ &          \\\hline
\textit{i B WORLD}         & $\times$ & $\times$ &          & $\times$ \\\hline
\textit{i PETRO RADIUS}    & $\times$ & $\times$ & $\times$ &          \\\hline
\textit{u PETRO RADIUS}    & $\times$ &          & $\times$ & $\times$ \\\hline
\textit{g PETRO RADIUS}    & $\times$ &          & $\times$ & $\times$ \\\hline
\textit{r PETRO RADIUS}    & $\times$ &          & $\times$ & $\times$ \\\hline
\textit{u KRON RADIUS}     & $\times$ &          & $\times$ & $\times$ \\\hline
\textit{g KRON RADIUS}     & $\times$ &          & $\times$ & $\times$ \\\hline
\textit{r KRON RADIUS}     & $\times$ &          & $\times$ & $\times$ \\\hline
\textit{i KRON RADIUS}     & $\times$ &          & $\times$ & $\times$ \\\hline
\textit{u ELONG}           & $\times$ &          & $\times$ & $\times$ \\\hline
\textit{g ELONG}           & $\times$ &          & $\times$ & $\times$ \\\hline
\textit{r ELONG}           & $\times$ &          & $\times$ & $\times$ \\\hline
\textit{i ELONG}           & $\times$ &          & $\times$ & $\times$ \\\hline
\textit{u THETA}           & $\times$ &          & $\times$ & $\times$ \\\hline
\textit{g THETA}           & $\times$ &          & $\times$ & $\times$ \\\hline
\textit{r THETA}           & $\times$ &          & $\times$ & $\times$ \\\hline
\textit{i THETA}           & $\times$ &          & $\times$ & $\times$ \\\hline
\textbf{TOTAL}             & \textbf{64} & \textbf{49} & \textbf{49} & \textbf{49} \\\hline\hline
\end{tabular}
}
\end{table*}

\begin{table*}\centering\caption{Classification results in terms of statistical estimators (the same used in Table~\ref{tab:results}) achieved by GNG and K-menas on \textit{ugri} and \textit{gri} datasets, using the \textit{BEST} parameter space. The results refer to the three described classification problems: \textit{3CLASS} (top table), \textit{GCs vs ALL} (middle table), \textit{GCs vs STARS} (bottom table).}\label{tab:kmeanscompare}
\begin{tabular}{lcccc}
\hline
\multicolumn{1}{c}{3CLASS} & \multicolumn{2}{c}{\emph{ugri}} & \multicolumn{2}{c}{\emph{gri}} \\
\hline
ESTIMATOR [\%] & GNG  & K-means & GNG  & K-means \\\hline
AE         & 85.5 & 82.9   & 79.4 & 82.1 \\\hline\hline
pur STAR   & 89.4 & 81.3   & 77.9 & 79.9 \\\hline
compl STAR & 76.8 & 70.6   & 65.4 & 62.1 \\\hline
F1 STAR    & 83.1 & 76.0   & 71.6 & 71.0 \\\hline\hline
pur GCs    & 77.7 & 70.4   & 75.3 & 75.8 \\\hline
compl GCs  & 86.9 & 80.6   & 82.0 & 81.6 \\\hline
F1 GCs     & 82.3 & 75.5   & 78.5 & 78.7 \\\hline\hline
pur gal    & 93.5 & 93.6   & 88.9 & 89.0 \\\hline
compl gal  & 92.1 & 92.0   & 91.1 & 90.6 \\\hline
F1 gal     & 92.8 & 92.8   & 90.0 & 89.8 \\\hline\hline
\end{tabular}\\
\begin{tabular}{lcccc}
\hline
\multicolumn{1}{c}{GCs vs ALL} & \multicolumn{2}{c}{\emph{ugri}} & \multicolumn{2}{c}{\emph{gri}} \\
\hline
ESTIMATOR [\%] & GNG & K-means & GNG & K-means \\\hline
AE          & 87.8 & 86.0 & 82.8 & 80.4 \\\hline\hline
pur notGC   & 82.2 & 75.4 & 81.1 & 78.5 \\\hline
compl notGC & 89.6 & 83.5 & 82.8 & 75.2 \\\hline
F1 notGC    & 85.9 & 79.4 & 81.9 & 76.9 \\\hline\hline
pur GCs     & 91.5 & 90.8 & 87.1 & 81.7 \\\hline
compl GCs   & 87.4 & 87.2 & 80.3 & 80.3 \\\hline
F1 GCs      & 89.4 & 89.0 & 83.7 & 81.0 \\\hline\hline
\end{tabular}\\
\begin{tabular}{lcccc}
\hline
\multicolumn{1}{c}{GCs vs STARs} & \multicolumn{2}{c}{\emph{ugri}} & \multicolumn{2}{c}{\emph{gri}} \\
\hline
ESTIMATOR [\%] & GNG & K-means & GNG & K-means \\\hline
AE         & 83.8 & 85.7 & 77.2 & 80.5 \\\hline\hline
pur STAR   & 83.8 & 87.4 & 75.5 & 81.6 \\\hline
compl STAR & 88.3 & 93.6 & 82.9 & 90.8 \\\hline
F1 STAR    & 86.0 & 90.5 & 79.2 & 86.2 \\\hline\hline
pur GCs    & 88.3 & 80.0 & 81.3 & 81.4 \\\hline
compl GCs  & 81.0 & 65.7 & 71.0 & 59.4 \\\hline
F1 GCs     & 84.6 & 72.8 & 76.1 & 70.4 \\\hline\hline
\end{tabular}
\end{table*}

\begin{figure*}\centering
\includegraphics[scale=1]{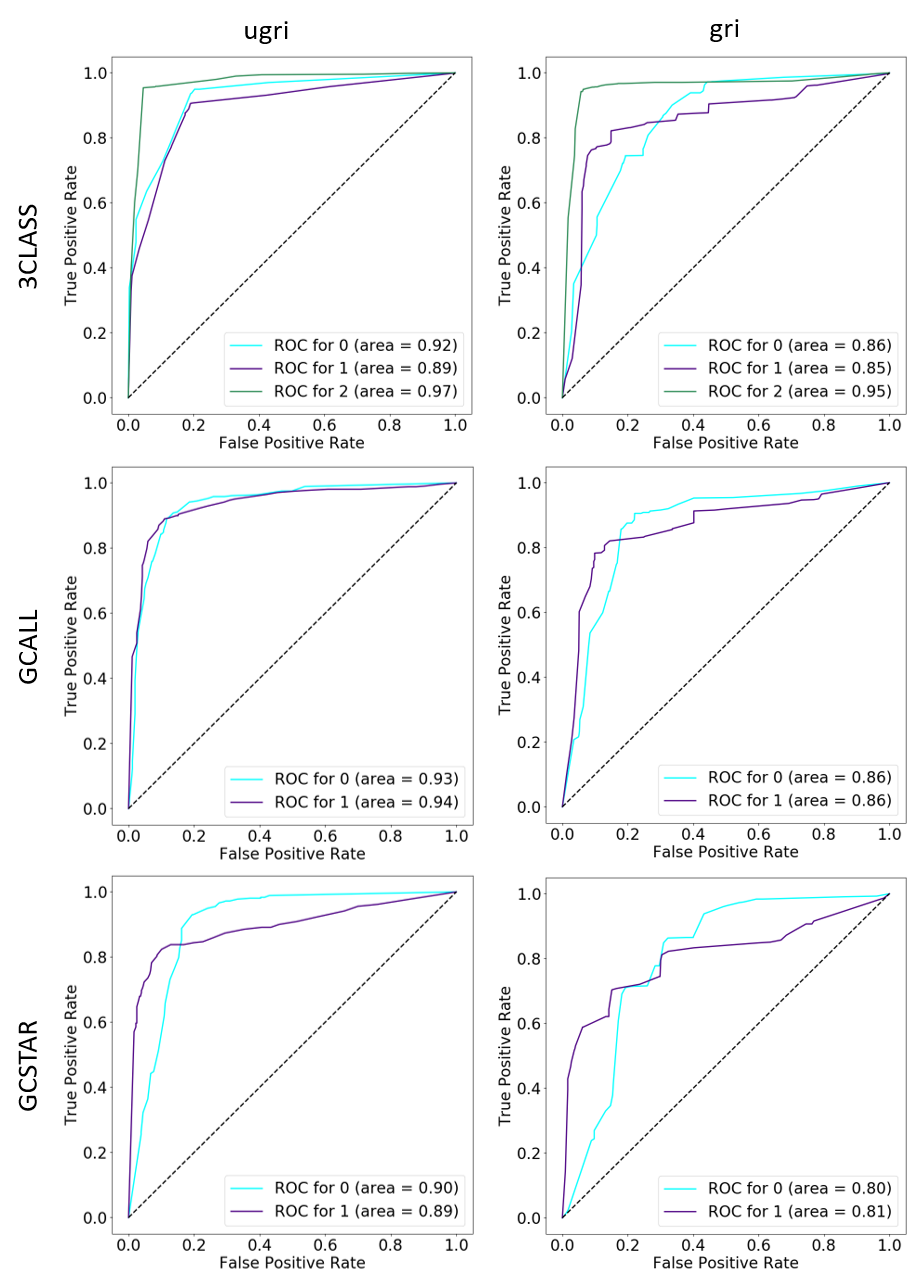}
\caption{ROC curves related to the GNG performances for the six experiments. GCs are labeled as $1$ (purple), stars are labeled with $0$ (light blue) in the \textit{3CLASS} and \textit{GCSTAR} experiments, while galaxies are labeled with $2$ in the \textit{3CLASS} experiments (green), in the \textit{GCALL} experiments the label $0$ refers to the \textit{not}GCs. In all the panels is reported the area and the curve and the non-discrimination line (dotted).}\label{fig:GNG:ROC}
\end{figure*}

\begin{table*}\centering\caption[]{GNG classification results in terms of statistical estimators (the same used in Table~\ref{tab:results}) for both \emph{ugri} and \emph{gri} dataset types. Top Table reports the results for the $3$-class experiment, middle Table for the $2$-class experiment between GCs and not GCs (stars + galaxies), and bottom Table shows the results concerning the $2$-class experiment between GCs and stars. The columns BEST, FULL, MIXED and BEST+REJECTED are related to the four parameter spaces, described in Table~\ref{tab:PSs}.}\label{tab:PSs:performance}
\begin{tabular}{lcccccccc}
\hline
\multicolumn{1}{c}{\textbf{3CLASS}} & \multicolumn{4}{c}{\emph{\bf ugri}} & \multicolumn{4}{c}{\emph{\bf gri}} \\
\hline
\multirow{ 2}{*}{ESTIMATOR [\%]}  & \multirow{ 2}{*}{BEST} & \multirow{ 2}{*}{FULL} & \multirow{ 2}{*}{MIXED} 
& BEST +  & \multirow{ 2}{*}{BEST} & \multirow{ 2}{*}{FULL} & \multirow{ 2}{*}{MIXED} & BEST +  \\
  &  &  &  &  REJECTED &  &  &  &  REJECTED \\\hline

AE         & 86.5 & 51.9 & 40.0  & 55.3  & 79.4 & 55.8 & 30.8 & 30.0 \\\hline\hline
pur STAR   & 85.8 & 33.3 & 37.7  & 39.0  & 71.9 & 29.4 & 0    & 0 \\\hline
compl STAR & 80.3 & 9.7  & 7.3   & 8.8   & 66.9 & 28.1 & 0    & 0 \\\hline
F1 STAR    & 83.0 & 15.0 & 12.2  & 14.4  & 69.3 & 28.8 & 0    & 0 \\\hline\hline
pur GCs    & 80.0 & 42.3 & 43.0  & 45.0  & 78.2 & 51.7 & 43.2 & 43.2 \\\hline
compl GCs  & 90.8 & 31.7 & 37.5  & 79.6  & 79.6 & 53.2 & 56.8 & 50.2 \\\hline
F1 GCs     & 85.1 & 36.2 & 40.0  & 57.5  & 78.9 & 52.4 & 49.1 & 46.5 \\\hline\hline
pur gal    & 92.5 & 44.3 & 44.5  & 44.6  & 88.3 & 56.1 & 28.3 & 28.3 \\\hline
compl gal  & 95.4 & 78.5 & 78.4  & 80.8  & 92.1 & 56.1 & 50.0 & 49.9 \\\hline
F1 gal     & 93.9 & 56.9 & 56.7  & 57.4  & 90.2 & 56.1 & 36.1 & 36.1 \\\hline\hline
\hline
\multicolumn{1}{c}{\textbf{GCs vs ALL}} & \multicolumn{4}{c}{\emph{\bf ugri}} & \multicolumn{4}{c}{\emph{\bf gri}} \\
\hline
\multirow{ 2}{*}{ESTIMATOR [\%]}  & \multirow{ 2}{*}{BEST} & \multirow{ 2}{*}{FULL} & \multirow{ 2}{*}{MIXED} 
& BEST +  & \multirow{ 2}{*}{BEST} & \multirow{ 2}{*}{FULL} & \multirow{ 2}{*}{MIXED} & BEST +  \\
  &  &  &  &  REJECTED &  &  &  &  REJECTED \\\hline
  AE        & 88.7 & 59.6 & 59.1 & 59.3 & 84.0 & 60.0 & 66.4 & 66.3 \\\hline\hline
pur notGC   & 85.1 & 51.3 & 50.7 & 51.1 & 81.3 & 62.4 & 71.4 & 71.4 \\\hline
compl notGC & 88.0 & 44.0 & 40.3 & 40.6 & 85.5 & 42.3 & 48.9 & 48.9 \\\hline
F1 notGC    & 86.5 & 47.4 & 45.0 & 45.2 & 83.4 & 52.3 & 60.1 & 58.1 \\\hline\hline
pur GCs     & 91.3 & 64.1 & 63.2 & 63.4 & 86.2 & 60.3 & 63.9 & 63.9 \\\hline
compl GCs   & 89.1 & 70.5 & 72.3 & 72.5 & 82.1 & 76.2 & 82.2 & 82.2 \\\hline
F1 GCs      & 90.2 & 67.2 & 67.5 & 67.7 & 84.2 & 68.2 & 71.9 & 71.9 \\\hline\hline
\hline
\multicolumn{1}{c}{\textbf{GCs vs STARs}} & \multicolumn{4}{c}{\emph{\bf ugri}} & \multicolumn{4}{c}{\emph{\bf gri}} \\
\hline
\multirow{ 2}{*}{ESTIMATOR [\%]}  & \multirow{ 2}{*}{BEST} & \multirow{ 2}{*}{FULL} & \multirow{ 2}{*}{MIXED} 
& BEST +  & \multirow{ 2}{*}{BEST} & \multirow{ 2}{*}{FULL} & \multirow{ 2}{*}{MIXED} & BEST +  \\
  &  &  &  &  REJECTED &  &  &  &  REJECTED \\\hline
AE         & 86.8 & 55.0 & 52.9 & 50.8 & 78.2 & 50.0 & 50.0 & 50.0 \\\hline\hline
pur STAR   & 87.0 & 56.5 & 54.5 & 52.8 & 77.3 & 54.2 & 54.2 & 54.2 \\\hline
compl STAR & 83.2 & 52.1 & 49.3 & 47.1 & 84.9 & 50.1 & 50.0 & 49.6 \\\hline
F1 STAR    & 85.1 & 54.2 & 51.7 & 49.5 & 81.1 & 52.1 & 52.1 & 51.9 \\\hline\hline
pur GCs    & 91.6 & 53.6 & 51.6 & 49.6 & 79.7 & 45.6 & 45.6 & 45.4 \\\hline
compl GCs  & 80.3 & 57.9 & 56.8 & 54.7 & 70.3 & 49.8 & 49.8 & 49.4 \\\hline
F1 GCs     & 85.6 & 55.7 & 54.1 & 52.0 & 75.0 & 47.6 & 47.6 & 47.4 \\\hline\hline
\end{tabular}
\end{table*}

\begin{table*}\centering\caption{Classification results in terms of statistical estimators (the same used in Table~\ref{tab:results}) achieved by GNG and RF on different datasets: (i) composed only by the information carried by the \textit{u}-band (first two columns), (ii) composed by same \textit{ugri} samples without the information carried by the \textit{u}-band (named as \textit{gri*}, third and fourth columns), (iii) the whole \textit{ugri} informative contribution (last two columns, the performances are takes from Table~\ref{tab:results} and from \ref{tab:PSs:RFperformance}). The results refer to the three described classification problems: \textit{3CLASS} (top table), \textit{GCs vs ALL} (middle table), \textit{GCs vs STARS} (bottom table).}\label{tab:roleofu}
\begin{tabular}{p{1.9cm}p{0.8cm}p{1.3cm}p{0.8cm}p{1.3cm}p{0.8cm}p{1.3cm}}
\hline
\multicolumn{1}{c}{3CLASS} & \multicolumn{2}{c}{\emph{u}} & \multicolumn{2}{c}{\emph{gri*}} & \multicolumn{2}{c}{\emph{ugri}} \\
\hline
ESTIMATOR [\%] & GNG  & RF & GNG  & RF 	& GNG  & RF			\\\hline
AE         & 85.5 & 84.1   & 79.4 & 82.4 & 86.5	& 94.4	\\\hline\hline
pur STAR   & 89.4 & 80.4   & 77.9 & 81.1 & 85.8	& 85.4	\\\hline
compl STAR & 76.8 & 81.7   & 65.4 & 79.8 & 80.3	& 85.5	\\\hline
F1 STAR    & 83.1 & 81.0   & 71.6 & 80.4 & 83.0	& 85.5	\\\hline\hline
pur GCs    & 77.7 & 79.3   & 75.3 & 80.3 & 80.0	& 86.9	\\\hline
compl GCs  & 86.9 & 80.8   & 82.0 & 80.2 & 90.8	& 93.3	\\\hline
F1 GCs     & 82.3 & 80.1   & 78.5 & 80.2 & 85.1	& 90.0	\\\hline\hline
pur gal    & 93.5 & 93.3   & 88.9 & 90.7 & 92.5	& 95.6	\\\hline
compl gal  & 92.1 & 90.1   & 91.1 & 85.2 & 95.4	& 97.7	\\\hline
F1 gal     & 92.8 & 91.7   & 90.0 & 87.9 & 93.9	& 96.6	\\\hline\hline
\end{tabular}\\
\begin{tabular}{p{1.9cm}p{0.8cm}p{1.3cm}p{0.8cm}p{1.3cm}p{0.8cm}p{1.3cm}}
\hline
\multicolumn{1}{c}{GCs vs ALL} & \multicolumn{2}{c}{\emph{u}} & \multicolumn{2}{c}{\emph{gri*}} & \multicolumn{2}{c}{\emph{ugri}} \\
\hline
ESTIMATOR [\%] & GNG & RF & GNG & RF & GNG & RF 	\\\hline
AE          & 87.8 & 86.6 & 82.8 & 85.0 & 88.7 & 92.2		\\\hline\hline
pur notGC   & 82.2 & 81.2 & 81.1 & 80.2 & 85.1 & 91.0		\\\hline
compl notGC & 89.6 & 80.8 & 82.8 & 83.8 & 88.0 & 89.9		\\\hline
F1 notGC    & 85.9 & 81.0 & 81.9 & 82.0 & 86.5 & 90.4		\\\hline\hline
pur GCs     & 91.5 & 90.8 & 87.1 & 88.0 & 91.3 & 92.8		\\\hline
compl GCs   & 87.4 & 90.1 & 80.3 & 84.1 & 89.1 & 92.6		\\\hline
F1 GCs      & 89.4 & 90.5 & 83.7 & 86.0 & 90.2 & 92.7		\\\hline\hline
\end{tabular}\\
\begin{tabular}{p{1.9cm}p{0.8cm}p{1.3cm}p{0.8cm}p{1.3cm}p{0.8cm}p{1.3cm}}
\hline
\multicolumn{1}{c}{GCs vs STARs} & \multicolumn{2}{c}{\emph{u}} & \multicolumn{2}{c}{\emph{gri*}} & \multicolumn{2}{c}{\emph{ugri}} \\
\hline
ESTIMATOR [\%] & GNG & RF & GNG & RF & GNG & RF		\\\hline
AE         & 83.8 & 90.8 & 77.2 & 87.0 & 86.8 & 88.2	\\\hline\hline
pur STAR   & 83.8 & 90.7 & 75.5 & 84.8 & 87.0 & 85.9	\\\hline
compl STAR & 88.3 & 97.1 & 82.9 & 86.0 & 83.2 & 92.2	\\\hline
F1 STAR    & 86.0 & 93.9 & 79.2 & 85.4 & 85.1 & 88.9	\\\hline\hline
pur GCs    & 88.3 & 92.6 & 81.3 & 89.1 & 91.6 & 90.8	\\\hline
compl GCs  & 81.0 & 80.4 & 71.0 & 80.4 & 80.3 & 95.2	\\\hline
F1 GCs     & 84.6 & 86.5 & 76.1 & 84.7 & 85.6 & 92.9	\\\hline\hline
\end{tabular}
\end{table*}

\begin{table*}\centering\caption[]{Random Forest classification results in terms of statistical estimators (the same used in Table~\ref{tab:results}) for both \emph{ugri} and \emph{gri} dataset types. Top Table reports the results for the $3$-class experiment, middle Table for the $2$-class experiment between GCs and not GCs (stars + galaxies), and bottom Table shows the results concerning the $2$-class experiment between GCs and stars. The columns BEST, FULL, MIXED and BEST+REJECTED are related to the four parameter spaces, described in Table~\ref{tab:PSs}.}\label{tab:PSs:RFperformance}
\begin{tabular}{lcccccccc}
\hline
\multicolumn{1}{c}{\textbf{3CLASS}} & \multicolumn{4}{c}{\emph{\bf ugri}} & \multicolumn{4}{c}{\emph{\bf gri}} \\
\hline
\multirow{ 2}{*}{ESTIMATOR [\%]}  & \multirow{ 2}{*}{BEST} & \multirow{ 2}{*}{FULL} & \multirow{ 2}{*}{MIXED} 
& BEST +  & \multirow{ 2}{*}{BEST} & \multirow{ 2}{*}{FULL} & \multirow{ 2}{*}{MIXED} & BEST +  \\
  &  &  &  &  REJECTED &  &  &  &  REJECTED \\\hline
  AE       & 94.4 & 92.6 & 93.0  & 92.7  & 92.6 & 91.1 & 90.2 & 90.2 \\\hline\hline
pur STAR   & 85.4 & 84.6 & 85.1  & 84.7  & 83.7 & 80.5 & 77.1 & 78.9 \\\hline
compl STAR & 85.7 & 86.1 & 85.7  & 86.1  & 84.1 & 80.6 & 80.5 & 80.2 \\\hline
F1 STAR    & 85.5 & 85.3 & 85.4  & 85.4  & 83.9 & 80.5 & 78.8 & 79.5 \\\hline\hline
pur GCs    & 86.9 & 87.2 & 88.0  & 86.9  & 88.9 & 86.5 & 86.0 & 86.7 \\\hline
compl GCs  & 93.3 & 85.5 & 85.1  & 85.1  & 91.2 & 90.2 & 88.8 & 89.6 \\\hline
F1 GCs     & 90.0 & 86.3 & 86.5  & 86.0  & 90.0 & 88.3 & 87.4 & 88.1 \\\hline\hline
pur gal    & 95.6 & 97.1 & 96.8  & 95.8  & 94.6 & 94.6 & 95.1 & 95.0 \\\hline
compl gal  & 97.7 & 97.0 & 96.1  & 97.8  & 96.6 & 89.9 & 88.8 & 89.1 \\\hline
F1 gal     & 96.6 & 97.0 & 96.4  & 96.8  & 95.6 & 92.2 & 91.5 & 92.0 \\\hline\hline
\hline
\multicolumn{1}{c}{\textbf{GCs vs ALL}} & \multicolumn{4}{c}{\emph{\bf ugri}} & \multicolumn{4}{c}{\emph{\bf gri}} \\
\hline
\multirow{ 2}{*}{ESTIMATOR [\%]}  & \multirow{ 2}{*}{BEST} & \multirow{ 2}{*}{FULL} & \multirow{ 2}{*}{MIXED} 
& BEST +  & \multirow{ 2}{*}{BEST} & \multirow{ 2}{*}{FULL} & \multirow{ 2}{*}{MIXED} & BEST +  \\
  &  &  &  &  REJECTED &  &  &  &  REJECTED \\\hline
AE          & 92.2 & 92.0 & 91.7 & 92.3 & 88.1 & 87.4 & 86.1 & 87.1 \\\hline\hline
pur notGC   & 91.0 & 90.1 & 90.2 & 90.3 & 84.6 & 83.0 & 83.4 & 83.2 \\\hline
compl notGC & 89.9 & 90.2 & 89.7 & 89.6 & 91.3 & 91.3 & 90.8 & 91.2 \\\hline
F1 notGC    & 90.4 & 90.1 & 89.9 & 89.9 & 87.8 & 87.0 & 86.9 & 87.0 \\\hline\hline
pur GCs     & 92.8 & 93.0 & 92.7 & 92.7 & 92.2 & 91.3 & 90.7 & 92.1 \\\hline
compl GCs   & 92.6 & 91.1 & 90.4 & 92.8 & 84.9 & 83.1 & 82.0 & 82.6 \\\hline
F1 GCs      & 92.7 & 92.0 & 91.5 & 92.7 & 88.4 & 86.8 & 86.1 & 87.1 \\\hline\hline
\hline
\multicolumn{1}{c}{\textbf{GCs vs STARs}} & \multicolumn{4}{c}{\emph{\bf ugri}} & \multicolumn{4}{c}{\emph{\bf gri}} \\
\hline
\multirow{ 2}{*}{ESTIMATOR [\%]}  & \multirow{ 2}{*}{BEST} & \multirow{ 2}{*}{FULL} & \multirow{ 2}{*}{MIXED} 
& BEST +  & \multirow{ 2}{*}{BEST} & \multirow{ 2}{*}{FULL} & \multirow{ 2}{*}{MIXED} & BEST +  \\
  &  &  &  &  REJECTED &  &  &  &  REJECTED \\\hline
AE         & 88.2 & 88.0 & 87.8 & 87.6 & 88.1 & 88.1 & 87.3 & 87.7 \\\hline\hline
pur STAR   & 85.9 & 85.9 & 86.3 & 85.7 & 85.9 & 85.7 & 83.7 & 85.9 \\\hline
compl STAR & 92.2 & 90.4 & 91.7 & 91.6 & 92.7 & 93.8 & 92.8 & 92.7 \\\hline
F1 STAR    & 88.9 & 88.1 & 88.9 & 88.6 & 89.2 & 88.7 & 88.0 & 89.2 \\\hline\hline
pur GCs    & 90.8 & 89.7 & 90.7 & 91.0 & 91.1 & 91.7 & 90.2 & 91.1 \\\hline
compl GCs  & 95.2 & 84.2 & 84.1 & 85.1 & 83.4 & 83.9 & 83.8 & 81.4 \\\hline
F1 GCs     & 92.9 & 86.9 & 87.3 & 88.0 & 87.1 & 87.6 & 86.9 & 86.0 \\\hline\hline
\end{tabular}
\end{table*}








\clearpage

\bsp	
\label{lastpage}
\end{document}